\RequirePackage{lineno}
\documentclass[aps,prd,twocolumn,superscriptaddress,showpacs]{revtex4}
\usepackage{multirow}
\usepackage{graphicx}% Include figure files
\usepackage{dcolumn}% Align table columns on decimal point
\usepackage{setspace}
\usepackage{subfigure}
\usepackage{slashbox}

\begin{document}
\newpage

%\begin{flushright}
%  METbb PRD\
%  \today \\
%  Version 2.6\\
  % {\scriptsize(preliminary cross section, uncorrected energies)}\\
%\end{flushright}

\newcommand{\ttbar}{\mbox{$t\bar{t}$}}
\newcommand{\thetastar}{\mbox{$\theta^*$}}
\newcommand{\costhetastar}{\mbox{$\cos\thetastar$}}
\newcommand{\Mlb}{\mbox{$M_{lb}$}}
\newcommand{\Et}{\mbox{$E_T$}}
\newcommand{\Pt}{\mbox{$p_T$}}
\newcommand{\Ht}{\mbox{$H_T$}}
\newcommand{\Flong}{\mbox{$F_0$}}
\newcommand{\Fplus}{\mbox{$F_+$}}
\newcommand{\met}{\mbox{$\protect \raisebox{0.3ex}{$\not$}\Et$}}
\newcommand{\mpt}{\mbox{$\protect \raisebox{0.3ex}{$\not$}\Pt$}}
\newcommand{\mht}{\mbox{$\protect \raisebox{0.3ex}{$\not$}\Ht$}}
\newcommand{\gevcc}{\mbox{GeV/$c^2$}}
\newcommand{\jetprob}{{\sc jetprob}}
\newcommand{\secvtx}{{\sc secvtx}}
\newcommand{\gray}{}

% You should use BibTeX and revtex.bst for references
\bibliographystyle{revtex}

\vspace*{1.5cm}

%%%%% uncomment for the PRD
\title{Search for single top quark production in \boldmath{$p\bar p$} collisions at 
\boldmath{${\sqrt{s}=1.96}$}\,TeV in the missing transverse energy plus jets topology}

\affiliation{Institute of Physics, Academia Sinica, Taipei, Taiwan 11529, Republic of China} 
\affiliation{Argonne National Laboratory, Argonne, Illinois 60439} 
\affiliation{University of Athens, 157 71 Athens, Greece} 
\affiliation{Institut de Fisica d'Altes Energies, Universitat Autonoma de Barcelona, E-08193, Bellaterra (Barcelona), Spain} 
\affiliation{Baylor University, Waco, Texas  76798} 
\affiliation{Istituto Nazionale di Fisica Nucleare Bologna, $^{dd}$University of Bologna, I-40127 Bologna, Italy} 
\affiliation{Brandeis University, Waltham, Massachusetts 02254} 
\affiliation{University of California, Davis, Davis, California  95616} 
\affiliation{University of California, Los Angeles, Los Angeles, California  90024} 
\affiliation{University of California, San Diego, La Jolla, California  92093} 
\affiliation{University of California, Santa Barbara, Santa Barbara, California 93106} 
\affiliation{Instituto de Fisica de Cantabria, CSIC-University of Cantabria, 39005 Santander, Spain} 
\affiliation{Carnegie Mellon University, Pittsburgh, PA  15213} 
\affiliation{Enrico Fermi Institute, University of Chicago, Chicago, Illinois 60637}
\affiliation{Comenius University, 842 48 Bratislava, Slovakia; Institute of Experimental Physics, 040 01 Kosice, Slovakia} 
\affiliation{Joint Institute for Nuclear Research, RU-141980 Dubna, Russia} 
\affiliation{Duke University, Durham, North Carolina  27708} 
\affiliation{Fermi National Accelerator Laboratory, Batavia, Illinois 60510} 
\affiliation{University of Florida, Gainesville, Florida  32611} 
\affiliation{Laboratori Nazionali di Frascati, Istituto Nazionale di Fisica Nucleare, I-00044 Frascati, Italy} 
\affiliation{University of Geneva, CH-1211 Geneva 4, Switzerland} 
\affiliation{Glasgow University, Glasgow G12 8QQ, United Kingdom} 
\affiliation{Harvard University, Cambridge, Massachusetts 02138} 
\affiliation{Division of High Energy Physics, Department of Physics, University of Helsinki and Helsinki Institute of Physics, FIN-00014, Helsinki, Finland} 
\affiliation{University of Illinois, Urbana, Illinois 61801} 
\affiliation{The Johns Hopkins University, Baltimore, Maryland 21218} 
\affiliation{Institut f\"{u}r Experimentelle Kernphysik, Karlsruhe Institute of Technology, D-76131 Karlsruhe, Germany} 
\affiliation{Center for High Energy Physics: Kyungpook National University, Daegu 702-701, Korea; Seoul National University, Seoul 151-742, Korea; Sungkyunkwan University, Suwon 440-746, Korea; Korea Institute of Science and Technology Information, Daejeon 305-806, Korea; Chonnam National University, Gwangju 500-757, Korea; Chonbuk National University, Jeonju 561-756, Korea} 
\affiliation{Ernest Orlando Lawrence Berkeley National Laboratory, Berkeley, California 94720} 
\affiliation{University of Liverpool, Liverpool L69 7ZE, United Kingdom} 
\affiliation{University College London, London WC1E 6BT, United Kingdom} 
\affiliation{Centro de Investigaciones Energeticas Medioambientales y Tecnologicas, E-28040 Madrid, Spain} 
\affiliation{Massachusetts Institute of Technology, Cambridge, Massachusetts  02139} 
\affiliation{Institute of Particle Physics: McGill University, Montr\'{e}al, Qu\'{e}bec, Canada H3A~2T8; Simon Fraser University, Burnaby, British Columbia, Canada V5A~1S6; University of Toronto, Toronto, Ontario, Canada M5S~1A7; and TRIUMF, Vancouver, British Columbia, Canada V6T~2A3} 
\affiliation{University of Michigan, Ann Arbor, Michigan 48109} 
\affiliation{Michigan State University, East Lansing, Michigan  48824}
\affiliation{Institution for Theoretical and Experimental Physics, ITEP, Moscow 117259, Russia} 
\affiliation{University of New Mexico, Albuquerque, New Mexico 87131} 
\affiliation{Northwestern University, Evanston, Illinois  60208} 
\affiliation{The Ohio State University, Columbus, Ohio  43210} 
\affiliation{Okayama University, Okayama 700-8530, Japan} 
\affiliation{Osaka City University, Osaka 588, Japan} 
\affiliation{University of Oxford, Oxford OX1 3RH, United Kingdom} 
\affiliation{Istituto Nazionale di Fisica Nucleare, Sezione di Padova-Trento, $^{ee}$University of Padova, I-35131 Padova, Italy} 
\affiliation{LPNHE, Universite Pierre et Marie Curie/IN2P3-CNRS, UMR7585, Paris, F-75252 France} 
\affiliation{University of Pennsylvania, Philadelphia, Pennsylvania 19104}
\affiliation{Istituto Nazionale di Fisica Nucleare Pisa, $^{ff}$University of Pisa, $^{gg}$University of Siena and $^{hh}$Scuola Normale Superiore, I-56127 Pisa, Italy} 
\affiliation{University of Pittsburgh, Pittsburgh, Pennsylvania 15260} 
\affiliation{Purdue University, West Lafayette, Indiana 47907} 
\affiliation{University of Rochester, Rochester, New York 14627} 
\affiliation{The Rockefeller University, New York, New York 10021} 
\affiliation{Istituto Nazionale di Fisica Nucleare, Sezione di Roma 1, $^{ii}$Sapienza Universit\`{a} di Roma, I-00185 Roma, Italy} 

\affiliation{Rutgers University, Piscataway, New Jersey 08855} 
\affiliation{Texas A\&M University, College Station, Texas 77843} 
\affiliation{Istituto Nazionale di Fisica Nucleare Trieste/Udine, I-34100 Trieste, $^{jj}$University of Trieste/Udine, I-33100 Udine, Italy} 
\affiliation{University of Tsukuba, Tsukuba, Ibaraki 305, Japan} 
\affiliation{Tufts University, Medford, Massachusetts 02155} 
\affiliation{Waseda University, Tokyo 169, Japan} 
\affiliation{Wayne State University, Detroit, Michigan  48201} 
\affiliation{University of Wisconsin, Madison, Wisconsin 53706} 
\affiliation{Yale University, New Haven, Connecticut 06520} 
\author{T.~Aaltonen}
\affiliation{Division of High Energy Physics, Department of Physics, University of Helsinki and Helsinki Institute of Physics, FIN-00014, Helsinki, Finland}
\author{J.~Adelman}
\affiliation{Enrico Fermi Institute, University of Chicago, Chicago, Illinois 60637}
\author{B.~\'{A}lvarez~Gonz\'{a}lez$^w$}
\affiliation{Instituto de Fisica de Cantabria, CSIC-University of Cantabria, 39005 Santander, Spain}
\author{S.~Amerio$^{ee}$}
\affiliation{Istituto Nazionale di Fisica Nucleare, Sezione di Padova-Trento, $^{ee}$University of Padova, I-35131 Padova, Italy} 

\author{D.~Amidei}
\affiliation{University of Michigan, Ann Arbor, Michigan 48109}
\author{A.~Anastassov}
\affiliation{Northwestern University, Evanston, Illinois  60208}
\author{A.~Annovi}
\affiliation{Laboratori Nazionali di Frascati, Istituto Nazionale di Fisica Nucleare, I-00044 Frascati, Italy}
\author{J.~Antos}
\affiliation{Comenius University, 842 48 Bratislava, Slovakia; Institute of Experimental Physics, 040 01 Kosice, Slovakia}
\author{G.~Apollinari}
\affiliation{Fermi National Accelerator Laboratory, Batavia, Illinois 60510}
\author{J.~Appel}
\affiliation{Fermi National Accelerator Laboratory, Batavia, Illinois 60510}
% ...

\collaboration{CDF Collaboration\footnote{With visitors from $^a$University of Massachusetts Amherst, Amherst, Massachusetts 01003,
$^b$Universiteit Antwerpen, B-2610 Antwerp, Belgium, 
$^c$University of Bristol, Bristol BS8 1TL, United Kingdom,
$^d$Chinese Academy of Sciences, Beijing 100864, China, 
$^e$Istituto Nazionale di Fisica Nucleare, Sezione di Cagliari, 09042 Monserrato (Cagliari), Italy,
$^f$University of California Irvine, Irvine, CA  92697, 
$^g$University of California Santa Cruz, Santa Cruz, CA  95064, 
$^h$Cornell University, Ithaca, NY  14853, 
$^i$University of Cyprus, Nicosia CY-1678, Cyprus, 
$^j$University College Dublin, Dublin 4, Ireland,
$^k$University of Edinburgh, Edinburgh EH9 3JZ, United Kingdom, 
$^l$University of Fukui, Fukui City, Fukui Prefecture, Japan 910-0017,
$^m$Kinki University, Higashi-Osaka City, Japan 577-8502,
$^n$Universidad Iberoamericana, Mexico D.F., Mexico,
$^o$University of Iowa, Iowa City, IA  52242,
$^p$Kansas State University, Manhattan, KS 66506,
$^q$Queen Mary, University of London, London, E1 4NS, England,
$^r$University of Manchester, Manchester M13 9PL, England,
$^s$Muons, Inc., Batavia, IL 60510, 
$^t$Nagasaki Institute of Applied Science, Nagasaki, Japan, 
$^u$University of Notre Dame, Notre Dame, IN 46556,
$^v$Obninsk State University, Obninsk, Russia,
$^w$University de Oviedo, E-33007 Oviedo, Spain, 
$^x$Texas Tech University, Lubbock, TX  79609, 
$^y$IFIC(CSIC-Universitat de Valencia), 56071 Valencia, Spain,
$^z$Universidad Tecnica Federico Santa Maria, 110v Valparaiso, Chile,
$^{aa}$University of Virginia, Charlottesville, VA  22906,
$^{bb}$Bergische Universit\"at Wuppertal, 42097 Wuppertal, Germany,
$^{cc}$Yarmouk University, Irbid 211-63, Jordan,
$^{kk}$On leave from J.~Stefan Institute, Ljubljana, Slovenia, 
}}
\noaffiliation

\begin{abstract}
We report a search for single top quark production with the CDF\, II detector 
using 2.1~fb$^{-1}$ of integrated luminosity of $p\bar p$ collisions at $\sqrt{s}=1.96$~TeV. The data selected consist of events characterized by large energy imbalance in the transverse plane and hadronic jets, and no identified electrons and muons, so the sample is enriched in $W \to \tau \nu$ decays.
In order to suppress backgrounds, additional kinematic and topological requirements are imposed through a neural network, and at least one of the jets must be identified as a $b$~quark jet.
We measure an excess of signal-like events in agreement with the standard model prediction, but inconsistent with 
a model without single top quark production by 2.1 standard deviations ($\sigma$), with a median expected sensitivity of 1.4\,$\sigma$. Assuming a top quark mass of 175\,\gevcc ~and ascribing the excess to single top quark production, the cross section is measured to be $4.9^{ +2.5}_{-2.2}$\,(stat+syst)\,pb, consistent with measurements performed in independent datasets and with the standard model prediction. 
\end{abstract}

%uncomment for the PRD
\maketitle

\section{\label{sec:Intro}Introduction}
At the Tevatron the dominant standard model (SM) mechanism for top quark production in $p\bar p$ collisions is the production through strong interactions of $\ttbar$ pairs with a cross section of approximately 
%$7.0$\,pb\,\cite{Kidonakis:2008mu}\,\cite{Cacciari:2008zb}\,\cite{Moch:2008ai}.
$7.0$\,pb\,\cite{cacciari-2004-0404}.
Top quarks can also be produced singly through electroweak processes, which are interesting in their own right. The single top quark production cross section is directly proportional to the square of the magnitude of the $|V_{tb}|$ element of the Cabibbo-Kobayashi-Maskawa (CKM) matrix\,\cite{Cabibbo,KM}. A measurement of the single top quark production cross section thus constrains the value of the latter. A value of $| V_{tb} |$ smaller than unity could thus indicate the presence of a fourth family of quarks\,\cite{Alwall:2006bx}, while on the other hand an apparent $| V_{tb} |$ value significantly greater than one could point, for instance, to the existence of a heavy $W$-like boson enhancing the cross section. A review of new physics models affecting the single top quark production cross section is given in Ref.\,\cite{Tait:2000sh}. \par
\begin{figure}[t]
\includegraphics[width=9.5cm]{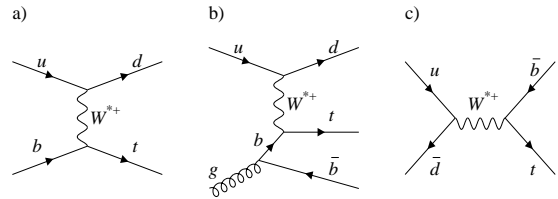}
\caption{Feynman diagrams for single top quark production. Represented are the LO (a) and NLO (b) $t$-channel processes, and the LO $s$-channel process (c).}
\label{fig:sttfeyn}
\end{figure}

At the Tevatron, a single top quark can be produced at leading order (LO) together with a $b$~quark in the $s$-channel, or paired with a light quark in the $t$-channel (the charge conjugated process is assumed throughout). The $t$-channel process also has a large next-to-leading-order (NLO) contribution which gives an additional $\bar b$~quark in the final state. The Feynman diagrams for the above processes are shown in Fig.\,\ref{fig:sttfeyn}. The SM NLO calculations predict the single top quark production cross section to be $\sigma_s = 0.88 \pm 0.11$\,pb for the $s$-channel and $\sigma_t =1.98 \pm 0.25$\,pb for the $t$-channel\,\cite{singletops, singletopt,Campbell:2009ss}, for an assumed  top quark mass of  175~GeV/$c^2$.

The top quark has a predicted lifetime of roughly 10$^{-25}$s, and the SM predicts it decays into a $W$ boson and a $b$~quark almost 100\% of the time, assuming $|V_{tb}|^2 \gg |V_{ts}|^2+|V_{td}|^2$. The $W$ boson subsequently decays to either a quark-antiquark pair or a lepton pair. Events with decays $W \rightarrow e \nu$ and  $W \rightarrow \mu \nu$ are the favored identification modes at a hadron collider due to the presence of the charged lepton and large missing transverse energy from the neutrino. The identification of both objects suppresses the otherwise large QCD background. In single top quark searches, $W \rightarrow \tau \nu$ decays have been studied until now only in events in which the tau decays to $e$ or $\mu$.
Electroweak production of top quarks is difficult to isolate due to the low cross section and large backgrounds which can be estimated only with large uncertainties on their rates.

The first evidence of electroweak top quark production has been achieved by the D0 collaboration with the charged lepton plus missing energy plus jets signature using 0.9\,fb$^{-1}$ of integrated luminosity\,\cite{Abazov:2006gd,Abazov:2008kt}. 
A more recent measurement in the same decay mode with an observed significance of $3.7\sigma$ has been released by the CDF collaboration using 2.2\,fb$^{-1}$ of integrated luminosity\,\cite{Aaltonen:2008sy}. 
%These D0 and CDF collaborations published measurements reached an observed significance of $3.4\sigma$ and $3.7\sigma$ respectively. 
In March 2009, both collaborations achieved the observation ($5.0\,\sigma$ significance) level, using 2.3\,fb$^{-1}$  and up to 3.2\,fb$^{-1}$ of integrated luminosity for the D0 collaboration\,\cite{Abazov:2009ii} and the CDF collaboration\,\cite{Aaltonen:2009jj}, respectively. 
To add acceptance to the dataset with identified $e$ or $\mu$, the latter measurement uses for the first time events containing jets, large missing transverse energy, and no reconstructed electrons or muons. The analysis of these events is described in this paper. 

This signature comprises events with $W \rightarrow \tau \nu$ decays where the hadronic $\tau$ decays are dominant, and with $W \rightarrow e \nu$ or $W \rightarrow \mu \nu$ decays where the $e, \mu$ are unidentified. Because the event selection vetoes on the presence of a reconstructed $e$ or $\mu$ this measurement is statistically independent from the one in Ref.\,\cite{Aaltonen:2008sy}. The single top quark candidate events analyzed share the same signature as events where the SM Higgs boson is produced in association with a $W$ or $Z$ boson, where the $W$ decays leptonically in either hadronic $\tau$s or unidentified $e$ or $\mu$, or the $Z$ decays to neutrinos. The techniques used in this single top quark search are shared with the ones deployed in the SM Higgs boson search in the same signature\,\cite{HiggsPRL}. Since the single top quark events are a background to the SM Higgs boson search, measuring the single top quark production cross section in this sample means getting a step closer to reaching sensitivity to the SM Higgs boson signal.

%The results of this measurement are combined with the analyses performed by CDF in the sample with reconstructed leptons to increase the overall sensitivity to the signal.

With respect to the single top quark search in the sample with identified leptons\,\cite{Aaltonen:2008sy} this analysis has the challenge of much larger backgrounds masking the presence of the signal. In fact, the sample under study is dominated by QCD multijet production, where a mismeasurement of one or more of the jet energies yields large energy imbalance in the detector. The QCD background dominates the signal by 4 orders of magnitude after the application of the online trigger selection.
%This analysis has the challenge with respect to the one with identified leptons\,\cite{Aaltonen:2008sy} of having much larger backgrounds masking the presence of the signal. 
Also, the absence of reconstructed charged leptons and the presence of the neutrino result in underconstrained kinematics, leaving us with the impossibility to reconstruct the top quark invariant mass from its decay products.
To improve the signal-to-background ratio ($s/b$), we select jets identified as originating from $b$~quarks using $b$-tagging algorithms. Even after these requirements, the $s/b$ is still too low to achieve sensitivity to single top quark production. We further exploit the kinematic and topological characteristics of single top quark events using  neural networks to isolate the signal from the dominant QCD background and subsequently from the remaining backgrounds. 

%The results of this measurement are combined with the analyses performed by CDF in the sample with reconstructed leptons to increase the overall sensitivity to the signal.
%The single top quark candidate events analyzed share the same signature as events where the Higgs boson is produced in association with a $W$ or $Z$ boson, and the $Z$ decays to neutrinos, or the $W$ decays leptonically in either hadronic $\tau$s or unidentified $e$ or $\mu$. The techniques used in this single top quark search are shared with the ones deployed in the Higgs boson search\,\cite{HiggsPRL}. Since the single top quark events are a background to the Higgs boson search, measuring the single top quark production cross section in this sample means getting a step closer to reaching sensitivity to the Higgs boson signal.

We report results based on data taken with the CDF\, II detector between July 2002 and August 2007, corresponding to an integrated luminosity of $2.1$~fb$^{-1}$. The paper is organized as follows. Section\,\ref{sec:detector} contains a brief description of the CDF\,II detector. The analyzed dataset is described in Sec.\,\ref{sec:trigger}. Section\,\ref{sec:reco} contains the definition of the reconstructed objects used in this analysis. In Sec.\,\ref{sec:SandBmodel} we discuss the Monte Carlo simulation and data-based techniques we use to model the signal and the backgrounds. Section\,\ref{sec:evsel} describes the preselection and the neural network event selection used to suppress the dominant QCD background. Section\,\ref{sec:NNdisc} describes the distribution used to scan for a signal, while Sec.\,\ref{sec:sys} lists the sources of systematic uncertainties that affect the final result. Section\,\ref{sec:likelihood} presents the likelihood procedure used to measure the cross section and the $|V_{tb}|$ matrix element. The single top quark production cross section measured in missing transverse energy plus jets events is presented in Sec.\,\ref{sec:results}. Finally, results are  summarized in Sec.\,\ref{sec:summary}.

\section{\label{sec:detector}The CDF\, II Detector}

The CDF\,II detector\,\cite{CDFdetector} is an azimuthally and forward-backward symmetric apparatus designed to study $p\bar p$ collisions at the Fermilab Tevatron. It uses a cylindrical coordinate system as described in\,\cite{coordinate}.
 It consists of a magnetic spectrometer surrounded by calorimeters and muon detectors. The charged particle tracking system is contained in a 1.4~T solenoid in which the magnetic field is parallel to the beam. A set of silicon microstrip detectors provides charged particle tracking in the radial distance from 1.5 to 28~cm\,\cite{L00,SVXII,ISL}. A 3.1~m long open-cell drift chamber, the central outer tracker (COT)\,\cite{COT}, covers the radial distance from 40 to 137~cm. The COT provides up to 96 measurements of the track position with alternating axial and $\pm$2$^{\circ}$-stereo superlayers of 12-wire layers each. The fiducial region of the silicon detector extends in pseudorapidity $|\eta|$ up to $|\eta| \le 2$, while the COT provides full radial coverage up to $|\eta|\le 1$. Segmented electromagnetic and hadronic calorimeters surround the tracking system and measure the energy of interacting particles. 
 
The electromagnetic and hadronic calorimeters are lead-scintillator and iron-scintillator sampling devices respectively, covering the range $|\eta|\le 3.6$. They are segmented in the central region ($|\eta|<1.1$) in towers of  0.1 in $\eta$ and 15$^\circ$ in azimuthal angle $\phi$, and in the forward region ($1.1<|\eta|<3.6$) in towers of  0.1 to 0.64 units of $\eta$ (corresponding to a nearly constant $2.7^{\circ}$ change in polar angle) and 7.5$^\circ$ in azimuthal angle for $1.1<|\eta|<2.11$ and 15$^\circ$ for $|\eta|>2.11$. The electromagnetic calorimeters\,\cite{ecal,pem} are instrumented with proportional and scintillating strip detectors that measure the transverse profile of electromagnetic showers at a depth corresponding to the expected shower maximum.
The measured energy resolution for electrons in the electromagnetic calorimeters are $13.5\%/\sqrt{E_T} \oplus 2\% $ in the central region, and $16\%/\sqrt{E_T} \oplus 1\%$ in the forward region\,\cite{cha} where the units of $E_T$ are in GeV.
We also measure the single-particle (pion) energy resolution in the hadronic calorimeters to be $50\%/\sqrt{E_T} \oplus 
3\%$ for the central and $80\%/\sqrt{E_T} \oplus 5\%$, for the forward detector\,\cite{TDR}.

Drift chambers located outside the central hadronic calorimeters and behind a 60~cm thick iron shield detect muons with $|\eta| \le 0.6$\,\cite{CMU}. Additional drift chambers and scintillation  counters detect muons in the region $0.6<|\eta|<1.5$\,\cite{CMX}. Gas Cherenkov counters with a coverage of $3.7<|\eta|<4.7$ measure the average number of inelastic $p\bar p$ collisions and thereby determine the luminosity\,\cite{WZprl}. 
 
\section{\label{sec:dataset} $\met$ plus jets Dataset}
\label{sec:trigger}

The events of interest are those with a top quark produced in association with one or more jets, where the top quark decays to a $b$~quark and a $W$ boson, and the $W$ boson decays leptonically. Since we are looking at events with non-identified electrons and muons, or hadronically decaying taus, we use a trigger devised to select events on the presence of two calorimetric jets and large missing transverse energy. The missing transverse energy $\vec{\met}$ is calculated as the negative vector sum of the energy in each calorimeter tower multiplied by a unit vector in the azimuthal direction of the tower. The $\met$ symbol is used for the magnitude of $\vec{\met}$.

CDF uses a three-level trigger system, the first two consisting of special purpose electronics and the third level consisting of conventional computer processors. For triggering purposes the calorimeter granularity is simplified to a $24\times 24$ grid in $\eta-\phi$ space and each trigger tower spans approximately 15$^\circ$ in $\phi$ and 0.2 in $\eta$ covering one or two physical towers. At level\,1, $\met \ge 25$~GeV is required, while at level\,2 we require the presence of two calorimetric clusters, each with transverse energy greater than 10~GeV. Finally, at level\,3 $\met \ge 35$~GeV is required.  With increasing instantaneous luminosity delivered by the Tevatron collider, tighter constraints at trigger level were needed. Starting from March 2005, the level\,2 trigger definition changed by demanding that one of the two calorimeter energy clusters be central, {\it i.e.} $|\eta| \le 1.1$. The steadily increasing Tevatron performance required an additional change in November 2006, where the trigger path was turned off as soon as the initial luminosity exceeded $1.90\times10^{32} $cm$^{-2}$ s$^{-1}$. Starting from April 2007, for instantaneous luminosity above $2\times 10^{32}\,{\rm cm}^{-2}$ s$^{-1}$ events were randomly discarded based on a scaling factor between 1 and 40 to keep the trigger rate at a reasonable level. 

%The trigger rate varies with the instantaneous luminosity, and corresponds to an effective cross section between 5 and 50~nb.  
Overall, 14\,963\,805 events pass the online trigger requirements, corresponding to an average effective cross section for the data collected by the trigger around 10\,nb. The single top quark production cross section times $W \to \ell \nu$ branching ratio is about 1\,pb, further reduced by the fact that this analysis is devised to collect mostly $W \to \tau \nu$ decays and only a fraction of the $W \to e/\mu ~ \nu$ decays. The $s/b$ ratio for events surviving the trigger level selection is thus of the order of 1/10\,000.

The $\met$ plus jets trigger efficiency is computed using data collected with a high $\Pt$ muon, and with data collected with a trigger requiring the presence of a jet with $E_T > 20$\,GeV.
%A high pT inclusive jet sample (JET50) was also used as a cross-check. The jet samples suer from small
%statistics, since these events are dominated by dijet production with fake
%ET . The muon sample is richer in events with real ET (and is the most similar to our signal),
%and has sucient statistics to determine the ET eciency. 
Trigger efficiencies are calculated for all three levels of the CDF trigger, and are then parametrized as a function of $\met$ and $E_T$ of the jets.
%corrected ET for L1 and L3 triggers: (L1(ET ) and L3(ET )) corrected jet ET for L2 trigger: (L2(jet ET )).
The systematic uncertainties originating from the choice of the samples used in the efficiency
calculations are large at small $\met$, and therefore we require every event to have $\met > $50\,GeV.  
The trigger is nearly 100\% efficient if the jets with the highest and second highest transverse energies satisfy the conditions $E^{j_1}_T > $35\,GeV and $E^{j_2}_T > $25\,GeV respectively.
Additionally, we require the spatial separation between the two leading $E_T$ jets to be greater than $\Delta R = 1$, where $\Delta R = \sqrt{\Delta \eta^2 + \Delta \phi^2}$ is the distance in the $\eta - \phi$ space between the two jet centroids, in order to avoid jet merging performed by the level\,2 jet clustering algorithm.
%Events passing the trigger
%are required to comply with the following additional requirements. 
%We exclude runs where the detector was not fully operational.
%ranges 217990-220272 and 245448-246231 due to the presence of hot towers biasing in this
%periods of data-taking.
A small fraction of events do not pass the level\,1 requirements due to a hardware problem, and are recovered with a trigger on inclusive jets\,\cite{artur}.

\section{Event Reconstruction}
\label{sec:reco}

Events are considered whenever the primary event vertex is reconstructed inside the luminous region ($|z|<60$~cm) along the beam axis.
Jets are identified using a fixed-cone algorithm which loops over calorimetric towers, with a cone radius of 0.4 in $\eta-\phi$ space. 
The jet energies are corrected for variations in calorimeter response and the presence of multiple $p \bar p$ collisions.
First, we take into account calorimeter response variations in $\eta$ and over time, and energy loss in the uninstrumented regions.
After a small correction for the extra energy deposited inside the jet cone by multiple collisions in the same accelerator bunch crossing, a correction for calorimeter non-linearity is applied so that the jet energies correspond to the most probable in-cone hadronic energy. 
%H1
Each of these steps has an individual systematic uncertainty that is added in quadrature to derive the total uncertainty which decreases from 8\% for jet transverse energies around 15\,GeV down to 3\% for jet energies above 60\,GeV.

After these corrections the jet energy provides a good estimate of the initial parton energy. 
This is verified by transverse momentum balance in events with a single jet recoiling against a well measured probe object such as a prompt photon or a $Z \to \ell^+ \ell^-$\,\cite{Bhatti:2005ai}.  Jet energies are further corrected using the algorithm developed by the H1 collaboration\,\cite{H1}, which combines the measurement of the momentum of charged particles in the spectrometer with the calorimeter energy measurement. 
%This approach provides an improvement of $\sim 10\%$ in jet energy resolution\,\cite{artur}.

In order to improve the $s/b$ ratio, we exploit the heavy flavor content of single top quark events using a $b$-tagging algorithm based on secondary vertex reconstruction (\secvtx) as described in detail in Ref.\,\cite{secvtx}. The algorithm aims at the identification of jets containing a $b$-hadron by reconstructing its decay vertex with at least two good quality tracks with hits in the silicon vertex  detector. A $b$-tagged jet must have a secondary vertex displaced from the primary vertex by more than $7.5 \,\sigma_{VTX}$ in the transverse plane, where $\sigma_{VTX}=190\,\mu$m is the typical secondary vertex spacial resolution in the transverse plane.
%NEW
A second algorithm, \jetprob, is also used to identify jets originating from $b$~quarks. This algorithm computes the probability that all tracks associated with a jet come from the primary $p \bar p$ interaction vertex. The probability calculation is based on the impact parameters of the tracks in the jet, and their uncertainties\,\cite{jetprob}. We consider a jet to be \jetprob-tagged when the probability of all jet tracks to come from the primary vertex is less than 0.05.
Electrons are reconstructed as charged particles in the tracking system that leave the majority of their energy in the electromagnetic section of the calorimeter. Muons are identified as charged particles in the tracker that leave hits in the muon chambers located outside the calorimeter. If isolated high momentum muons are found in the event, $\vec{\met}$ is corrected by subtracting the average muon ionization energy released in the calorimeter and adding the muon \Pt\ to the vector sum.
%The exact definition of $e, \mu$ can be found in\,\cite{PRDobs}. [CHECK] 
No specific tau identification algorithm is used in this analysis and in Ref.\,\cite{Aaltonen:2008sy}. Events with $\tau$ leptons decaying leptonically are sometimes collected in the dataset analyzed in Ref.\,\cite{Aaltonen:2008sy} by identifying their $e\,$ or $\mu$ decay products, while the event selection described here collects hadronic $\tau$ decays whenever the decay products are reconstructed as jets. We veto events with reconstructed electrons and muons in order to keep this sample statistically independent from the one analyzed in Ref.\,\cite{Aaltonen:2008sy}.
\par
The critical part of this analysis is the requirement of the $\met$ signature. The $\met$ in the event can stem not only from neutrinos, but also from various instrumental and detector effects. Events containing large $\met$ could have originated from non-collision sources, such as cosmic or beam-halo muons passing through the detector or noisy or dead calorimeter cells causing an energy imbalance. These types of events are removed by requiring that the event observables indicate an inelastic collision with large momentum transfer, such as the presence of at least one high quality primary vertex in the collision and at least one central jet with $E_T > 10$\,GeV. Additional requirements are also imposed to remove events consistent with beam-halo muons traversing the detector or those caused by noisy calorimeter cells. After these requirements, the leading source of $\met$ is jet energy mismeasurement due to either jets pointing to non-instrumented regions of the calorimeter, or to calorimeter resolution effects. Both categories of events are 
characterized by $\vec{\met}$ often being aligned with the projection of one of the jet three-momenta ($\vec{j}$) in the azimuthal plane. Other characteristic properties of this instrumental background will be described in Sec.\,\ref{sec:evsel}, together with the strategy devised to suppress it.

%ENDNEW

\section{Signal and background modeling}
\label{sec:SandBmodel}

\subsection{\label{sec:Signal} Signal modeling}

% USE TOP MASS 175 GeV

The single top quark production is simulated assuming a top quark mass of 175\,GeV/$c^2$ using the {\sc madevent}\,\cite{madevent} matrix element generator, interfaced to the {\sc cteq5l}~\cite{Lai:1999wy} parametrization of the parton distribution functions (PDF). The {\sc madevent} generator models the polarization of the top quark and the distributions of the final state decay products accordingly. The transition from final state colored particles to colorless objects is done through the  {\sc pythia}~\cite{PYTHIA} parton showering and hadronization routines.

It has been shown that the inclusion of the next-to-leading-order diagrams results in an increase in the cross section for $s$-channel production mode, but does not change significantly its kinematics\,\cite{singletops}.
The $s$-channel events are thus generated at leading order and the cross section is scaled to the next-to-leading-order rate\,\cite{singletops}.

For the $t$-channel, the leading-order process for single top quark production is a 2 $\rightarrow$ 2 process with a $b$~quark in the initial state: $b + u  \rightarrow d + t$ or $b + \bar d  \rightarrow  \bar u + t$. Single anti-top quark production implies the conjugate processes.  As several authors have pointed out\,\cite{Boos:2006af,singletops}, the distribution of observable jets is not adequately represented by the LO contribution to the $t$-channel production of single top quark and it is better predicted by next-to-leading-order calculations.  In the latter, the $b$~quark stems from a gluon splitting into a $b \bar b$ pair. The $\bar b$~quark required by the flavor conservation of the strong interaction is created by LO parton shower programs through backward evolution following the {\sc dglap} scheme\,\cite{DGLAP}.
The high-\Pt\ tail of the transverse momentum distribution of the $\bar b$~quark is not well modeled by this scheme.
The mismodeling is estimated by comparing with a NLO calculation\,\cite{singletops}.
%\cite{Boos:2006af}. 
%Besides, the pseudorapidity distributions of the $\bar b$~quark simulated at LO are biased towards higher pseudorapidities compared to the NLO predictions.
The modeling of the $t$-channel single top quark process can be improved by producing simulated events with a matrix element generator, followed by the simulation of the production of observable particles by {\sc pythia}. For this, two samples are used: one for the leading 2 $\rightarrow$ 2 process,
%$b + q  \rightarrow q + t$
 and one for the 2  $\rightarrow$ 3 process with a gluon in the initial state $g + q  \rightarrow q + t + \bar b$. In the latter process, the $\bar b$~quark is directly produced in the hard scattering described by the matrix element. It also describes the important high-\Pt\ tail of the $\bar b$~quark \Pt\ distribution. The construction of a Monte Carlo simulated sample following the NLO predictions is done by matching the 2 $\rightarrow$ 2 and 2 $\rightarrow$ 3 processes as described in\,\cite{Jan}.

\subsection{\label{sec:backgrounds} Background modeling}

There are numerous standard model processes besides single top quark production that can produce the signature  characterized by large $\met$, relatively low jet multiplicity, and no reconstructed charged leptons. The most significant background at the first stage of the analysis is the QCD multijet production.
%with a cross section of the order of $\mu$b, which is about 9 orders of magnitude greater than the signal before requiring the presence of $b$-jets. 
Although these processes generally do not produce neutrinos, mismeasured jet energies do
result in imbalance in the measured transverse energy by which the QCD events
can pass the basic selection. Furthermore, QCD $b$~quark pair production yields neutrinos whenever one
$b$-hadron decays semi-leptonically, thus giving additional $\met$. 
The background sources for this final state are due mainly to QCD production of heavy-quark pairs ($b\bar b$ and $c\bar c$) and jets falsely tagged as $b$-jets.

Because of the high production rate for QCD at a hadron collider and the large statistics needed in order to describe this process adequately in an analysis looking for a very small signal, the Monte Carlo simulation of an acceptable amount of QCD events is prohibitive. Moreover, the systematic uncertainties associated with the Monte Carlo simulation of QCD jet production are high. For these reasons, we estimate the QCD background solely from data.

%CHANGE Given the theoretical uncertainties related to the production cross section for the generation of N-parton events, it is important to have a method for the background estimate that does not require any Monte Carlo information, and thus, based solely on data. 
Events collected by the $\met$+jets trigger are expected to be composed mostly of QCD production of light flavor jets. We model the heavy flavor jets QCD rate and distributions by weighting events without any $b$-tagging requirement by the probability to tag a jet as a $b$-jet. This probability is extracted from events depleted in single top quark signal, {\it i.e.} events with $50 \le \met \le 70$~GeV and $\Delta \phi(\vec{\met}, \vec{j_2})<0.4$, and two or three jets\,\cite{artur} 
The tag rate per jet is evaluated as a ratio of $b-$tagged to fiducial jets, where the fiducial jets are the ones in the kinematic region where the secondary vertex detection efficiency is nonzero. The tag rate is parametrized in terms of variables sensitive to both the efficiency of the identification of true heavy-flavored objects and the rate of false tags. These variables are the jet $\Et$, the absolute value of the jet $\eta$, the scalar sum of the transverse energies of the jets in the event ($\Ht$), and the fraction of jet $\Pt$ carried by the charged particles inside the jet which are significantly displaced from the collision point. To compute the last quantity, all charged particles satisfying $0.5 \le \Pt \le 200$~GeV/$c$ are used, and they are required to have the distance of closest approach to the beamline ($d_0$) significantly displaced from the beamline, {\it i.e.} $|d_0 / \sigma_{d_0}|>2.5$ where $\sigma_{d_0}$ is the uncertainty on $d_0$.

%This method has the advantage of including the prediction of false tags, which hence requires no additional estimation.
%The tag rates per jet as a function of these variables are shown in Fig.\,\ref{fig:tagrate} for jets within the vertex detector acceptance (fiducial jets).

%%%%%%%%%%%%%%%%%
%\begin{figure}[htbp]
%\begin{center}
%\centering
%\includegraphics[width=5.0cm]{Rate_et-3.eps}
%\hfill
%\includegraphics[width=5.0cm]{Rate_eta-3.eps}
%\hfill
%\includegraphics[width=5.0cm]{Rate_Ht-3.eps}
%\hfill
%\caption{Tag rate for fiducial jets as a function of jet $\Et$, $N_{\rm trk}^{jet}$ and $N_{\rm vert}$.} \label{fig:tagrate}
%\end{center}
%\end{figure}
%%%%%%%%%%%%%%%%%%

The tag rate parametrization is then used to estimate the probability that a fiducial jet in the signal candidate sample is tagged. We construct three independent parametrizations to estimate the background in events with exactly one \secvtx-tagged jet (1S), two \secvtx-tagged jets (2S category), and one \secvtx-tagged and one \jetprob-tagged jet (SJ category). Events which belong to both 2S and SJ categories are assigned to the 2S subsample. Events with three $b$-tagged jets are discarded. In this way the three selections are orthogonal by construction so that an event can belong to only one category.

\begin{figure}
  \centering
\subfigure
%[$\met$]
{\includegraphics[width=8cm]{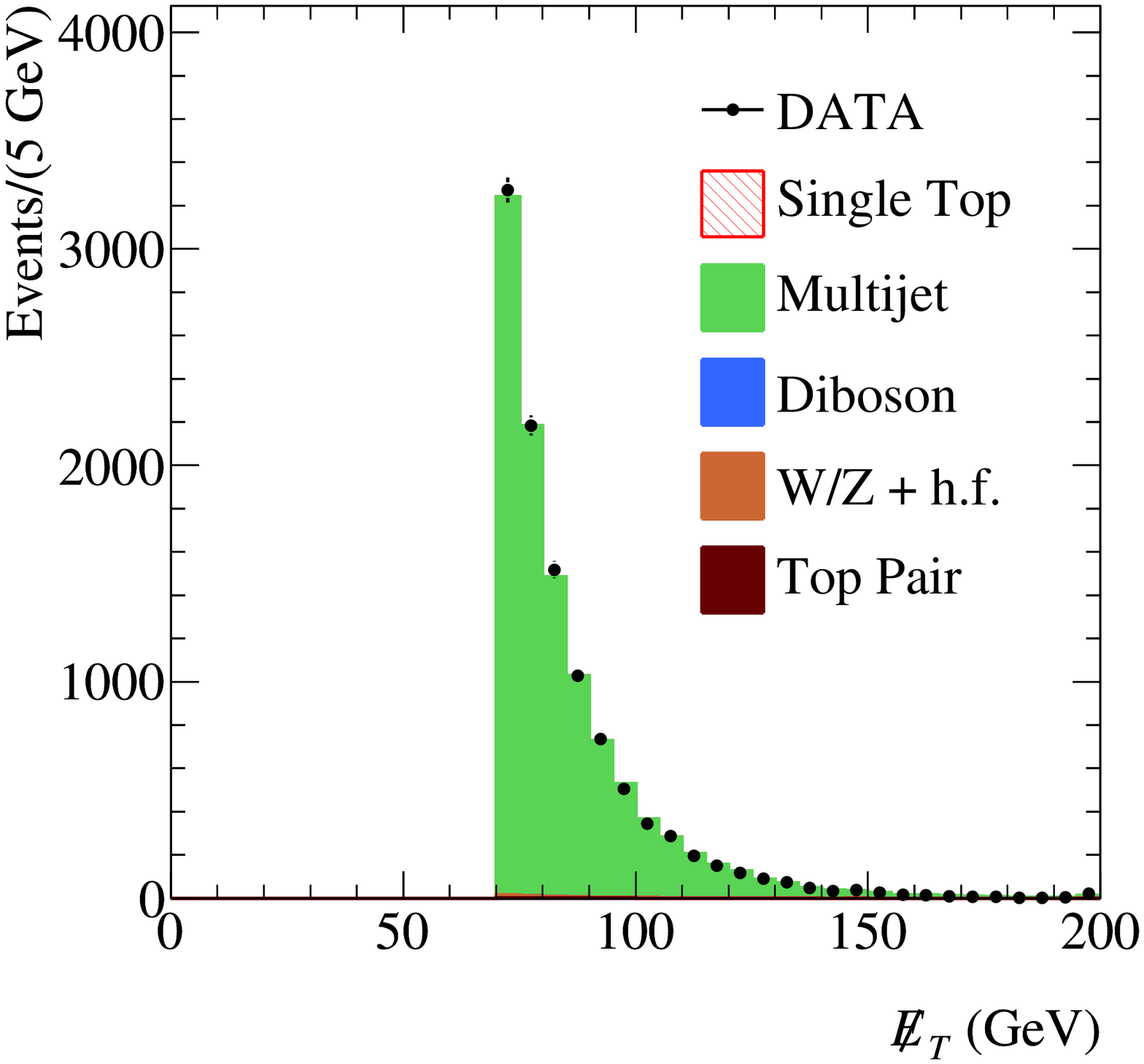}}
%\subfigure[$\Delta\phi(\mpt,\met)$]{\includegraphics[width=.45\linewidth]{finalPlotss/QCD_CR/PRDPlots_DPhiMET_TrkMPT_All_Validation\gray}}
\subfigure
%[Invariant mass of $\met$ , $jet_1$ and $jet_2$]
{\includegraphics[width=8cm]{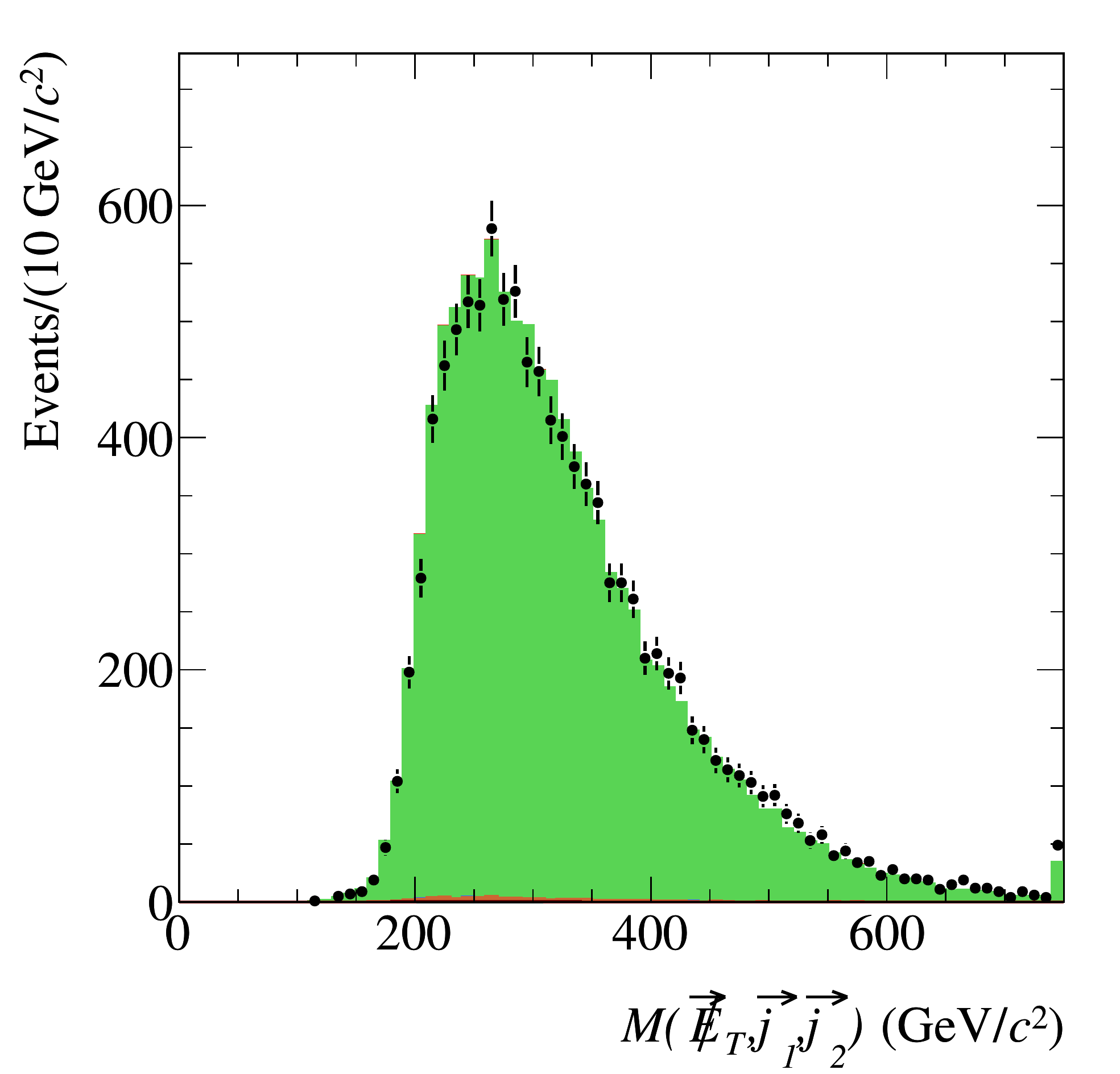}}
%\subfigure[Min($\Delta\phi(\met,jet_i)$)]{\includegraphics[width=.45\linewidth]{finalPlotss/QCD_CR/PRDPlots_MinDPhiMET_Ji_All_Validation\gray}}
    \caption{Comparison of the background modeling to data in a QCD-dominated control region. The green area represents the multijet background model, while the points represent the data. The bin at the right end of the $x$-axis represents the overflow bin. The distributions of the observables under study show good agreement between data and the background model.}%~\ref{sec:NNselection}.}
     \label{fig:CR1}
\end{figure}

By summing the probability of $b$-tagging each fiducial jet in each $b$-tag subsample and weighting the rate and distributions of data events before any $b-$tagging requirement (pretag sample), we predict the rate of QCD $b$-tagged jet multijet production background events and its kinematic distributions.
We predict the kinematic properties of events with one \secvtx-tagged jet QCD background from the pretag sample,
%and shapes of events with two \secvtx-tagged jets, and one \secvtx-tagged jet plus a \jetprob-tagged jet QCD background from the sample with one \secvtx-tagged jet. 
and the kinematic properties of QCD background events in the 2S and SJ categories from events with one \secvtx-tagged jet. 
The parametrizations do not completely account for the fact that events with pair production of heavy flavor have enhanced probability to be tagged. For this reason, the normalization of the background events arising from the simple application of the parametrization needs to be scaled. This normalization procedure is described in the next section.
%The scaling factor is obtained in a region, NN$_{QCD} < -0.1$,  defined in the next section.
%The goodness of the parametrization and the goodness of the resulting estimate in different kinematic regions is shown in Fig.\,\ref{fig:CR1} and \ref{fig:CR3} where we compare some of the kinematic properties of the distributions for tags in the data to the expected background in background events where $\met>70$~GeV, and where $\met >50$~GeV and we require an well-reconstructed lepton. 
The performance of the parametrizations and of the resulting estimate is shown in Fig.~\ref{fig:CR1} where we compare our background model to data in a QCD dominated region containing events with $\met>70$~GeV and  $\Delta \phi(\vec{\met}, \vec{j_2})<0.4$. We see that the tag rate parametrization produces a good modeling of the kinematic properties of the QCD events. When building the QCD model in samples containing sizeable contributions of non-QCD process such as $W$ + jets production, we apply the tag rate parametrization to our Monte Carlo simulation of non-QCD processes, and subtract the output from the QCD background estimate.  The normalization of the QCD background contribution is set as a scale factor derived in a control region, which multiplies the prediction obtained with the tag rate parametrization, as described in the next section.

The other backgrounds to single top quark production in this signature come from the production of a $W$ or a $Z$ boson in association with jets, top quark pair production through strong interactions, and pair production of heavy vector bosons. All these background processes are simulated using the {\sc pythia} Monte Carlo simulation program.

%All the $W$ and $Z$ boson samples produced in association with $b$- or $c$-quark, and events with 
%$W\rightarrow \tau \nu$ and $Z\rightarrow \tau \bar{\tau}$ are simulated with {\sc pythia}.
We normalized the $W$ and $Z$ boson + heavy flavor jets backgrounds
using the inclusive cross sections measured by CDF\,\cite{Abulencia:2005ix}. 
The measurements correspond to a factor of 1.4 with respect to
the {\sc pythia} LO predictions. 
%Since there are no cross section measurements for
%$W\rightarrow \tau \nu$ and $Z\rightarrow \tau \bar{\tau}$, we
%use the LO {\sc pythia} cross section corrected by a scale factor of 1.4.
In this way, $Z$/$W$ boson samples have the correct normalization with respect to their
inclusive production. However, the heavy flavor production simulated by
{\sc pythia} provides a possible source of systematic error. We assign a 40\%
uncertainty based on the total uncertainty of the $Z$ + heavy flavor jets cross section
measurement in CDF\,\cite{abulencia:032008}. The production of $W$ and $Z$ bosons in association
with light flavor jets is included in the multijet modeling.

Top quark pair production yields a non-negligible contribution to the background in the signal
region. Due to the large top quark mass and the leptonic decay of the $W$ boson originating from the top quark,
these events are energetic, have large $\met$, and have high jet multiplicity. As for the signal events, in this analysis simulated $t\overline{t}$ events were generated assuming a top quark mass of  175\,GeV/$c^2$. The cross section corresponding to that mass has been computed to be $6.7\pm 0.8$~pb\,\cite{cacciari-2004-0404}.
 
For the simulated diboson samples we use LO cross sections scaled by a factor corresponding 
to the ratio between the NLO and LO cross section prediction in {\sc mcfm}\,\cite{MCFM}. The boson decays are set to be inclusive.
An 11.5\% uncertainty on the {\sc mcfm} cross section is assigned to the diboson normalization\,\cite{DIBOSONS}.

\section{Event selection}
\label{sec:evsel}

\subsection{Preselection and topology requirements}

%To avoid trigger inefficiencies, we require jet and $\met$ cuts higher than the ones requried at trigger level: $\met \ge 50$~GeV, leading jet energy $E_{T,1} \ge 35$~GeV, and second leading jet $E_{T,2} \ge 25$~GeV. 

We define the signal region by selecting events in the kinematic region where the trigger is highly efficient: $\met > 50\,$GeV, $E^{j_1}_T > 35$\,GeV and $E^{j_2}_T > $25\,GeV, $\Delta R(\vec{j_1},\vec{j_2}) > 1$ and by requiring the number of jets to be no more than three, thus accepting events with extra radiation from the incoming or outgoing partons, or hadronically-decaying taus reconstructed as jets. In the case of three jet events, the jet with lowest $E_T$ is required to have $E^{j_3}_T \ge 15$~GeV.
Events with four or more jets with $\Et > 15$~GeV and $|\eta|<2.4$ are rejected to reduce backgrounds from QCD and $\ttbar$ production.
We veto events containing well-identified electrons or muons as identified in Ref.\,\cite{Aaltonen:2009jj} to ensure the independence of the analyzed samples. 
About 523\,000 events pass these preselection requirements, where the $s/b$ ratio for single top quark events after this selection is about 1/2\,800 assuming SM cross sections. At this stage of the analysis, the vast majority of the background events are QCD events where mismeasurement of the jet energies gives the very large $\met$. These events are characterized by having $\vec{\met}$ aligned in the axymuthal angle $\phi$ with one of the jets ($\vec{j_i}$) in the event, where the index $i$ runs over the jets in the event. We require $\Delta \phi (\vec{\met}, \vec{j_1}) > 1.5$ and $\Delta \phi (\vec{\met}, \vec{j_{2,3}}) > 0.4$ to reject such events. These cuts remove about an order of magnitude of QCD events, but still leave us with a $s/b$ of $\sim 1/340$, where the majority of the background is still composed of QCD multijet production. Finally, we require the presence of at least one jet identified as originating from a $b$ quark. We divide the sample in the three exclusive subsamples 1S, SJ and 2S defined in Sec.\,\ref{sec:SandBmodel}.
%characterized by events with one \secvtx-tagged jet (1S), one {\secvtx}- and one \jetprob-tagged jet (SJ), and two \secvtx-tagged jets (2S); 
This requirement brings the average $s/b$ ratio over the three subsamples to $\sim 1/50$, where the QCD background is still the dominant one.
Thus, we need to exploit additional properties of these events in order to further increase the purity of the sample.

\subsection{Neural network based event selection}
\label{sec:NNselection}

We introduce here a neural network approach to the event selection to recognize and separate QCD multijet events with mismeasured jets in which $\met$ is due to instrumental effects from events with $\met$ originating from neutrinos. In addition, the neural network is designed to reject events with mis-tagged light flavor jets. Using a neural network instead of a ``cut-cascade" approach to event selection allows the exploitation of the correlation between the many observables which provide discrimination between signal and backgrounds, and gives a single output thus simplifying the determination of the optimal cut.
The neural network model chosen is the Multi Layer Perceptron\,\cite{mlp} as implemented in the {\sc tmva} package\,\cite{TMVA}, found in {\sc root}\,\cite{root}.
\par
In this analysis the charged particle spectrometer is used in an innovative way to discriminate between events containing high energy neutrinos and QCD events. We introduce here the imbalance in the momentum flow in the transverse plane, and name it missing transverse momentum, or $\vec{\mpt}$, in analogy with the missing transverse energy, $\vec{\met}$. To compute $\vec{\mpt}$, we select charged particles with $0.5 < \Pt < 200\,$GeV/$c$ and $z$ position at the beamline compatible with the $z$ position of the primary vertex. The missing transverse momentum is then defined as $\vec{\mpt} = -\sum_{\text{tracks}} \vec{\Pt}$. In collisions producing high energy neutrinos, the magnitude of $\vec{\mpt}$ ($\mpt$) is proportional to the neutrino energy, while $\vec{\mpt}$ provides a good estimate of the neutrino direction. 
In QCD events with the $\met$ plus jets topology, where high energy neutrinos are rarely produced and the fluctuation of the charged-to-neutral ratio in jet fragmentation is the primary source of imbalance of the total transverse momentum, the magnitude of $\mpt$ is expected to be relatively low, and the vector $\vec{\mpt}$ to be often aligned in direction with the momentum of one of the most energetic jets.

As inputs into our neural network we use the following variables which describe the energy and momentum flow in the detector: the absolute amount of the missing transverse energy, $\met$; the absolute amount of the missing transverse momentum $\mpt$; the $\met$ significance defined as $\met$/$\sqrt{\sum \Et}$, where ${\sum \Et}$ is a scalar sum over the energy deposited in the calorimeter towers; the ratio of $\met$ to $\Ht$; the ratio of $\mht =  |-\sum_{\text{jets}} \vec{\Pt}|$ and $\met$; the invariant mass of $\vec{\met}$,  $\vec{j_1}$ and $\vec{j_2}$, $M(\vec{\met}, \vec{j_1}, \vec{j_2})$. 

We use the following angular variables: the azimuthal difference between $\vec{\mpt}$ and $\vec{\met}$, $\Delta \phi(\vec{\mpt},\vec{\met})$; the maximum of $\Delta \phi$ between any two jets $\vec{j_i}, \vec{j_k}$, $Max (\Delta \phi(\vec{j_i},\vec{j_k}))$; the maximum of the difference in $R$ between any two jets $\vec{j_i}, \vec{j_k}$, $Max(\Delta R(\vec{j_i},\vec{j_k}))$; the minimum of the difference in $\phi$ between $\vec{\met}$ and any jet $\vec{j_i}$, $Min(\Delta \phi(\vec{\met},\vec{j_i}))$; the minimum of the difference in $\phi$ between the   
$\vec{\mpt}$ and any jet $\vec{j_i}$, $Min(\Delta \phi(\vec{\mpt},\vec{j_i}))$; 
the difference in the azimuthal plane between the axis defined by the two most energetic jets in their rest frame, and the vector sum of the two jets in the lab frame,
$\phi^{*}$;
%the $\Delta \phi$ between $\vec{\mpt}$ and the direction of their boost; 
the event sphericity\,\cite{sphericity}.  

%S = 1.5 $\times (\lambda_2+\lambda_3$) where $\lambda_1 \ge \lambda_2 \ge \lambda_3$ are the eigenvalues of the sphericity tensor $M^{ab}= \sum _{jets} \vec{P_{j_i}} \vec{P_{j_k}}$, where $P_j^{a,b}$ are the vectors of the high $P_T$ jets in the event. 
We also use variables that discriminate between the fragmentation properties of heavy flavor quark jets and jets originating from light flavor quarks or gluons: by taking the charged particles with $p_T>0.5$\,GeV/$c$ and $|d_0/\sigma_{d_0}|>2.5$ and contained inside the jet cone, we build the variable $\sum \Pt^{chgd}/\Pt^{j}$ for the $E_T$ leading and second leading jets.
%where $\Pt^{chgd}$ is the transverse momentum of a charged particle contained in the jet cone, and significantly displaced from the collision point. 
The 15 variables used as inputs to the neural network are summarized in Table\,\ref{tab:nnvarQCDNN}. 
%
%\begin{table}[hbtp]
%\begin{center}
%\caption{Input variables to the neural network devised to suppress the QCD background, and the background coming from production of light flavor jets.}
%\begin{tabular}{l}
%\hline\hline
%Variable 
%\hline
%$\met$ \\
%\hline
%$\mpt$ \\
%\hline
%$\met$/$\sqrt{\sum E_T}$ \\
%\hline
%$\met$/$H_T$ \\
%\hline
%$\mht$ / $\met$ \\
%Invariant mass of $\vec{\met}$ , $\vec{j_1}$ and $\vec{j_2}$, 
%$M(\vec{\met},\vec{j_1},\vec{j_2})$ \\
%\hline
%$\Delta \phi(\vec{\met},\vec{\mpt})$ \\
%\hline
%$Max (\Delta \phi(\vec{j_i},\vec{j_k}))$ \\
%\hline
%$Max (\Delta R(\vec{j_i},\vec{j_k}))$ \\
%\hline
%$Min (\Delta \phi(\vec{\met},\vec{j_i}))$ \\
%\hline
%$Min (\Delta \phi(\vec{\mpt},\vec{j_i}))$ \\
%\hline
%$\Delta \phi(\vec{j_1}, \vec{j_2})$ in the 2-jet rest frame \\
%$\phi^{*}$ \\
%\hline
%Sphericity\\
%\hline
%$\sum \Pt^{chgd}/\Pt^{j^1}$\\
%\hline
%$\sum \Pt^{chgd}/\Pt^{j^2}$\\
%\hline\hline
%\end{tabular}
%\label{tab:nnvarQCDNN}
%\end{center}
%\end{table}

\begin{table*}[hbtp]
\begin{center}
\caption{Input variables to the neural network devised to suppress the multijet background.}
\begin{tabular}{ll}
\hline\hline
Variable  & Description \\
\hline
$\met$ & Absolute amount of the missing transverse energy\\
%\hline
$\mpt$ & Absolute amount of the missing transverse momentum\\
%\hline
$\met$/$\sqrt{\sum E_T}$ & Missing $E_T$ significance \\
%\hline
$\met$/$H_T$  & Ratio of $\met$ to $\Ht$ \\
%\hline
$\mht$ / $\met$ & Ratio of $\mht$ to $\met$ \\
%\hline
$M(\vec{\met},\vec{j_1},\vec{j_2})$ & Invariant mass of $\met$, $\vec{j_i}$ and $\vec{j_2}$ \\
%\hline
$\Delta \phi(\vec{\met},\vec{\mpt})$ & Azymuthal difference between $\met$ and $\mpt$ \\
%\hline
$Max (\Delta \phi(\vec{j_i},\vec{j_k}))$ & Maximum of $\Delta \phi$ between any two jets $\vec{j_i}, \vec{j_k}$\\
%\hline
$Max (\Delta R(\vec{j_i},\vec{j_k}))$ & Maximum of $\Delta R$ between any two jets $\vec{j_i}, \vec{j_k}$\\
%\hline
$Min (\Delta \phi(\vec{\met},\vec{j_i}))$ & Minimum of $\Delta \phi$ between $\vec{\met}$ and any jet $\vec{j_i}$ \\
%\hline
$Min (\Delta \phi(\vec{\mpt},\vec{j_i}))$ & Minimum of $\Delta \phi$ between $\vec{\mpt}$ and any jet $\vec{j_i}$ \\
%\hline
%$\Delta \phi(\vec{j_1}, \vec{j_2})$ in the 2-jet rest frame \\
$\phi^{*}$ & $\Delta \phi$ between the ($\vec{j^1},\vec{j^2}$) axis in their rest frame, and their vector sum in the lab frame \\
%\hline
Sphericity & $S = \frac{3}{2} (\lambda_2 + \lambda_3)$\,\cite{sphericity} \\
%\hline
$\sum \Pt^{chgd}/\Pt^{j_1}$ & Fraction of $\Pt^{j_1}$ carried by charged particles displaced from the primary vertex \\
%\hline
$\sum \Pt^{chgd}/\Pt^{j_2}$ &  Fraction of $\Pt^{j_2}$ carried by charged particles displaced from the primary vertex  \\
\hline\hline
\end{tabular}
\label{tab:nnvarQCDNN}
\end{center}
\end{table*}

\begin{figure*}[htbp]
\centering
\subfigure
%[$\met$]
{\includegraphics[width= .32 \linewidth]{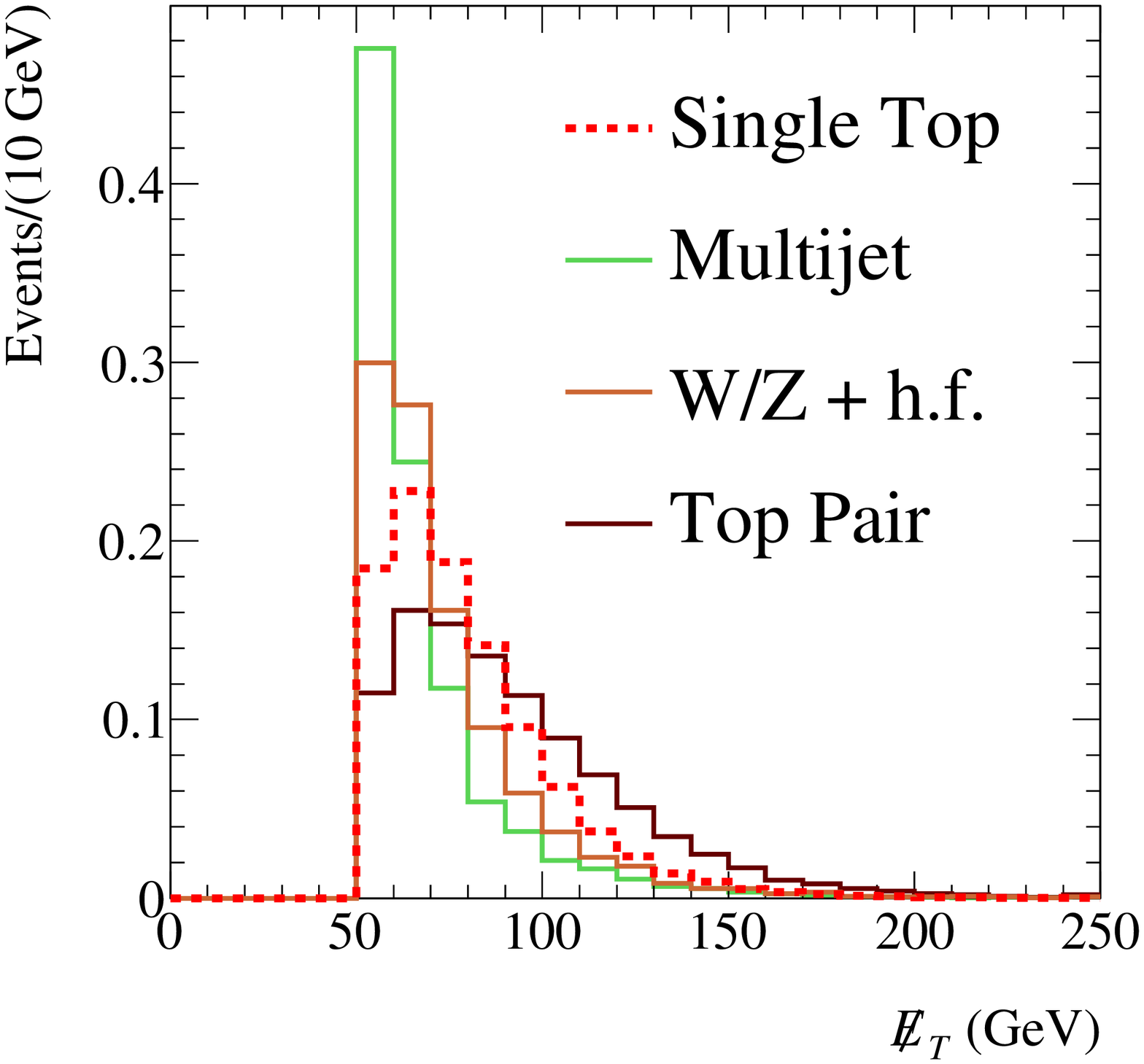}}
\subfigure
%[$\mpt$]
{\includegraphics[width= .32 \linewidth]{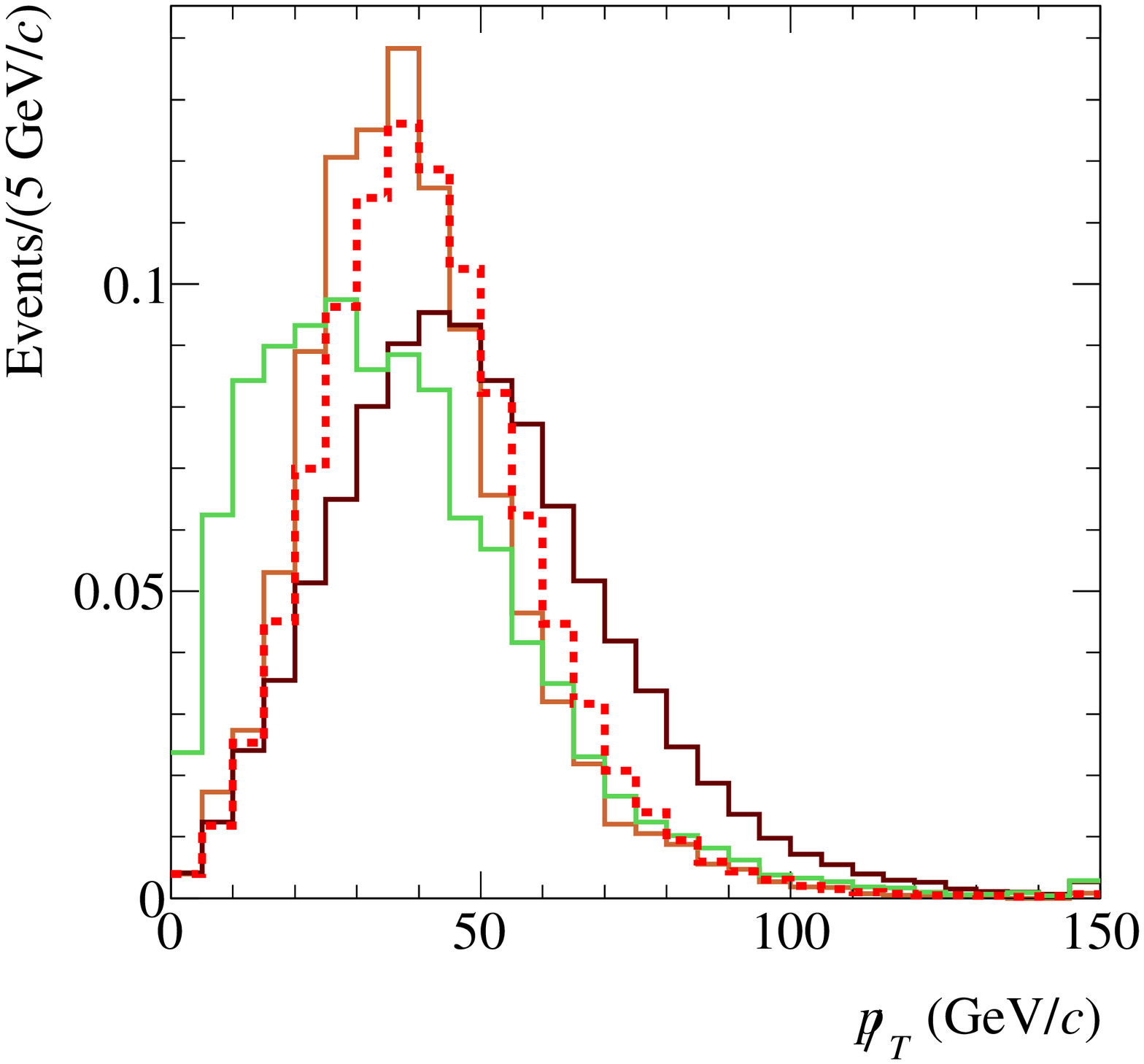}}
\subfigure
%[$\met/\sqrt{\sum \Et}$]
{\includegraphics[width= .32 \linewidth]{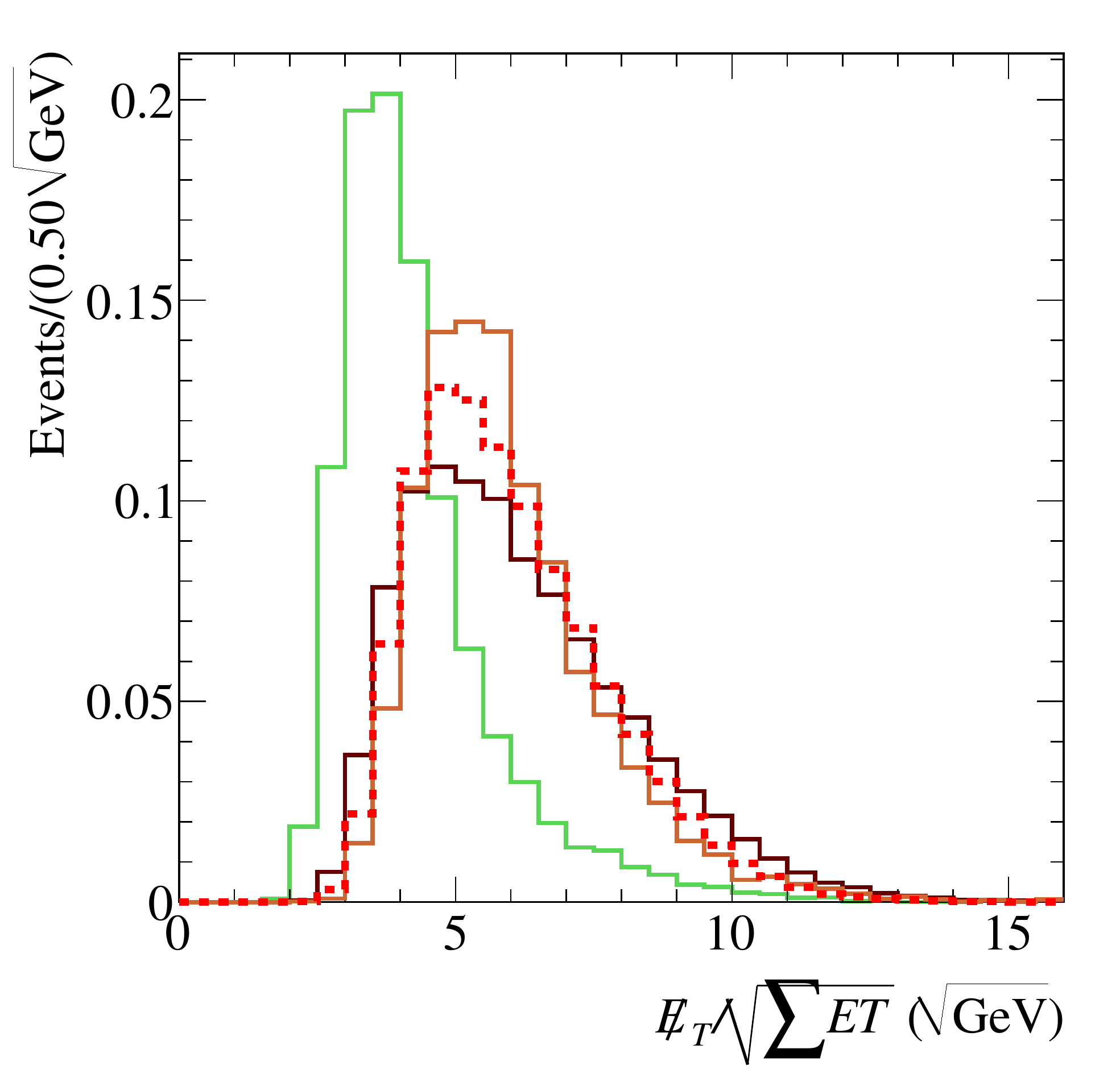}}
\subfigure
%[$\met/\Ht$]
{\includegraphics[width= .32 \linewidth]{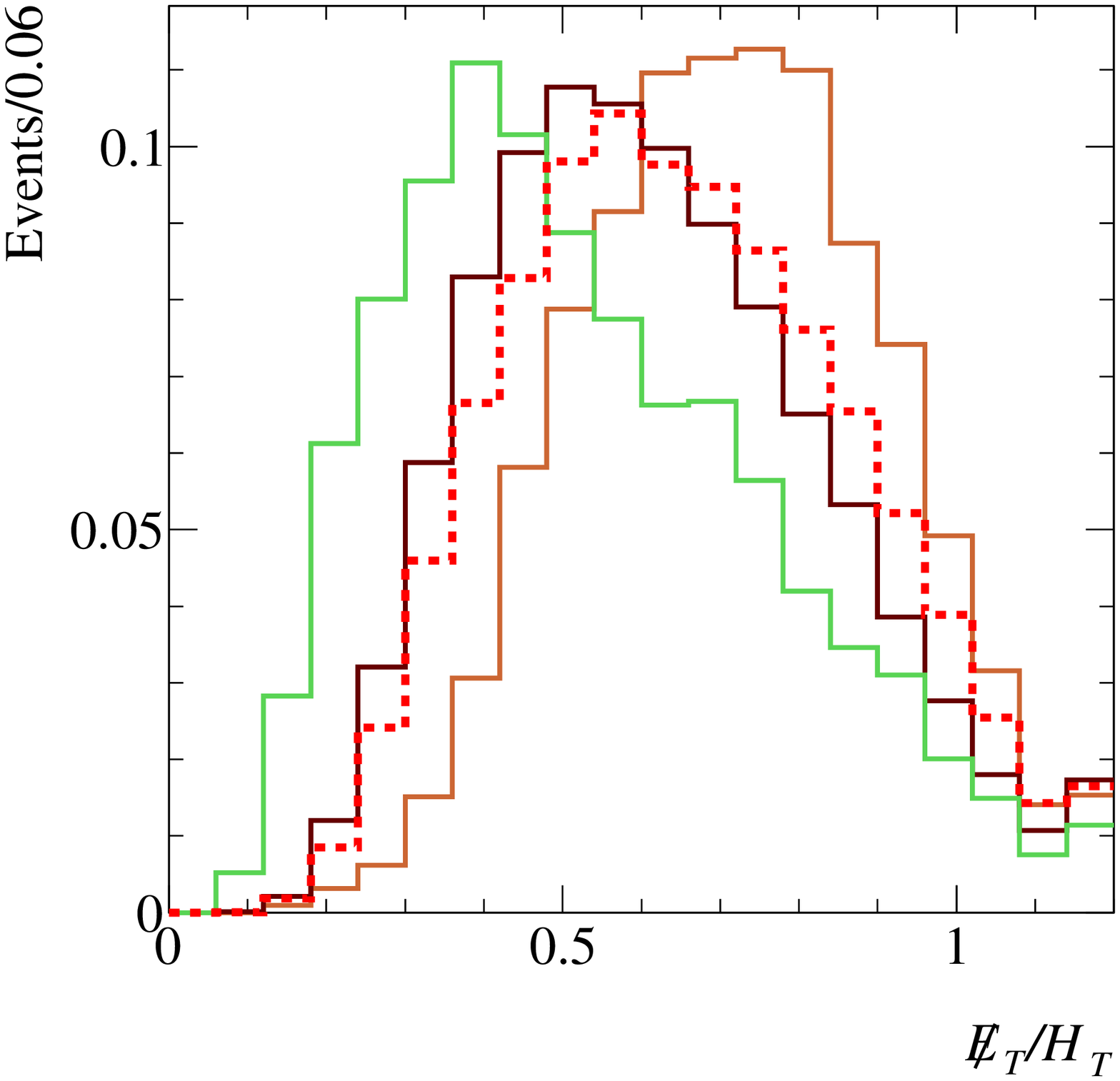}}
\subfigure
%[$\mht/\met$]
{\includegraphics[width= .32 \linewidth]{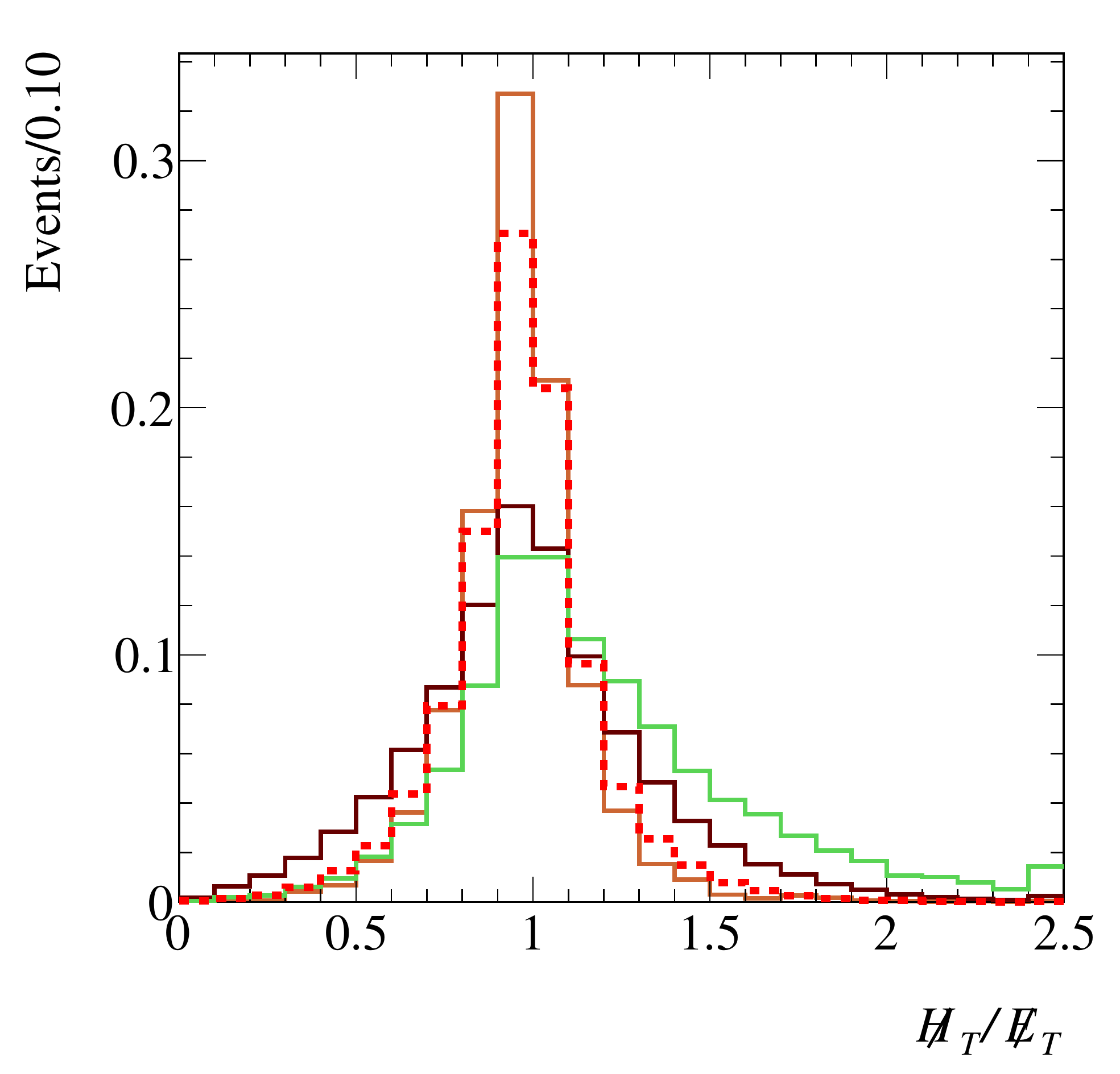}}
\subfigure
%[Invariant mass of $\met, jet_1$ and $jet_2$]
{\includegraphics[width= .32 \linewidth]{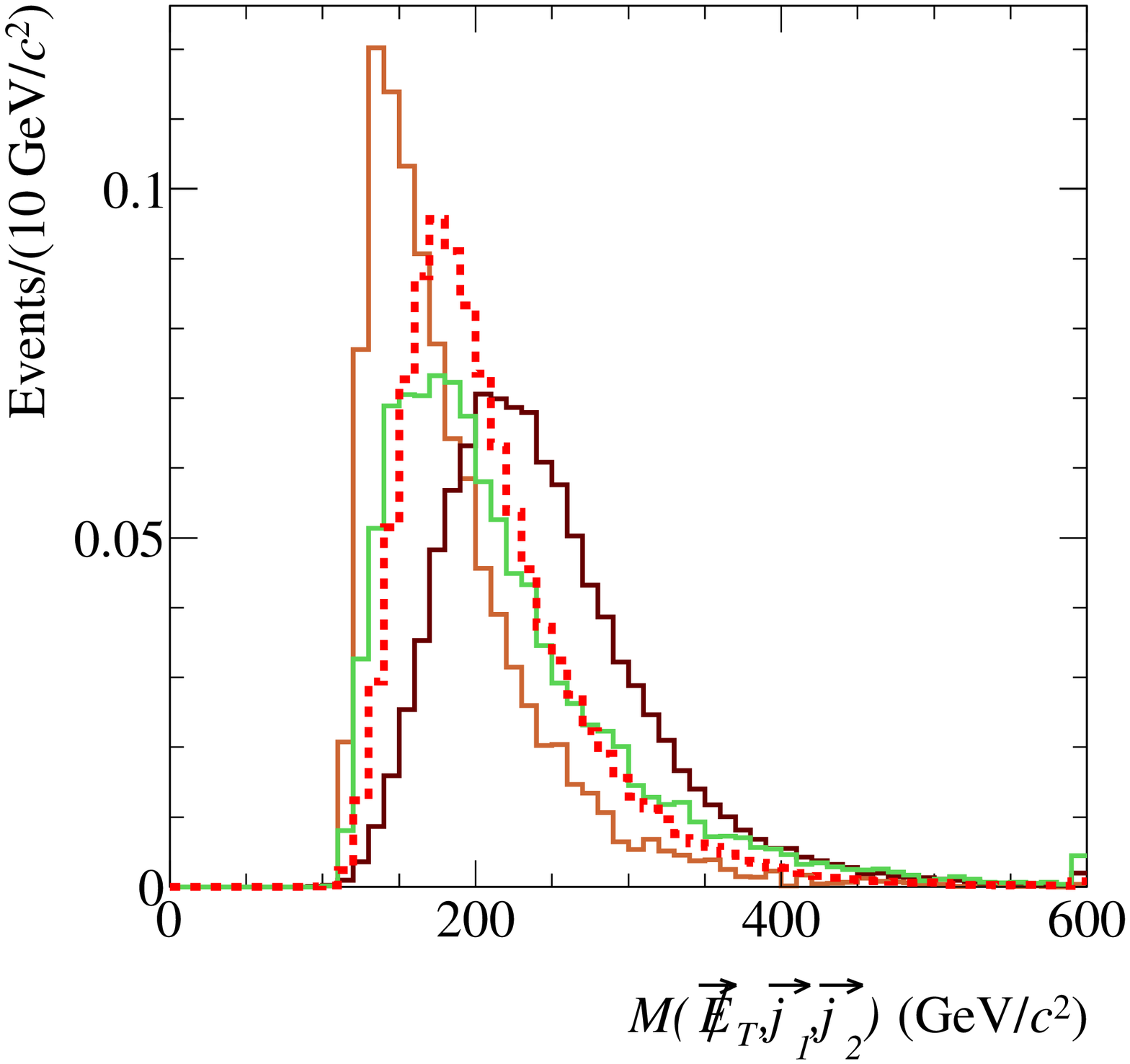}}
\subfigure
%[$\Delta \phi(\met,\mpt)$]
{\includegraphics[width= .32 \linewidth]{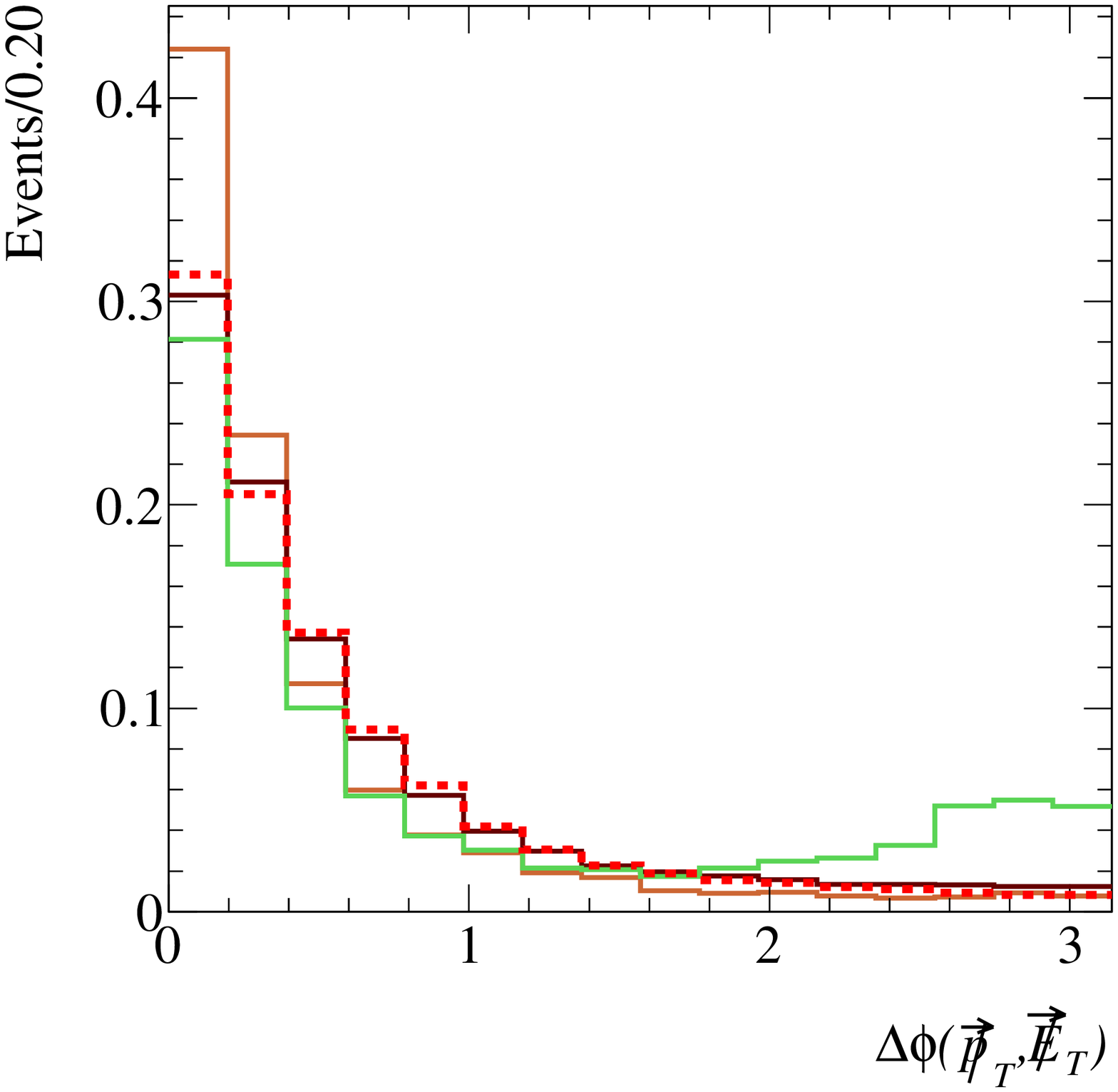}}
\subfigure
%[Max($\Delta \phi(jet_i,jet_j)$]
{\includegraphics[width= .32 \linewidth]{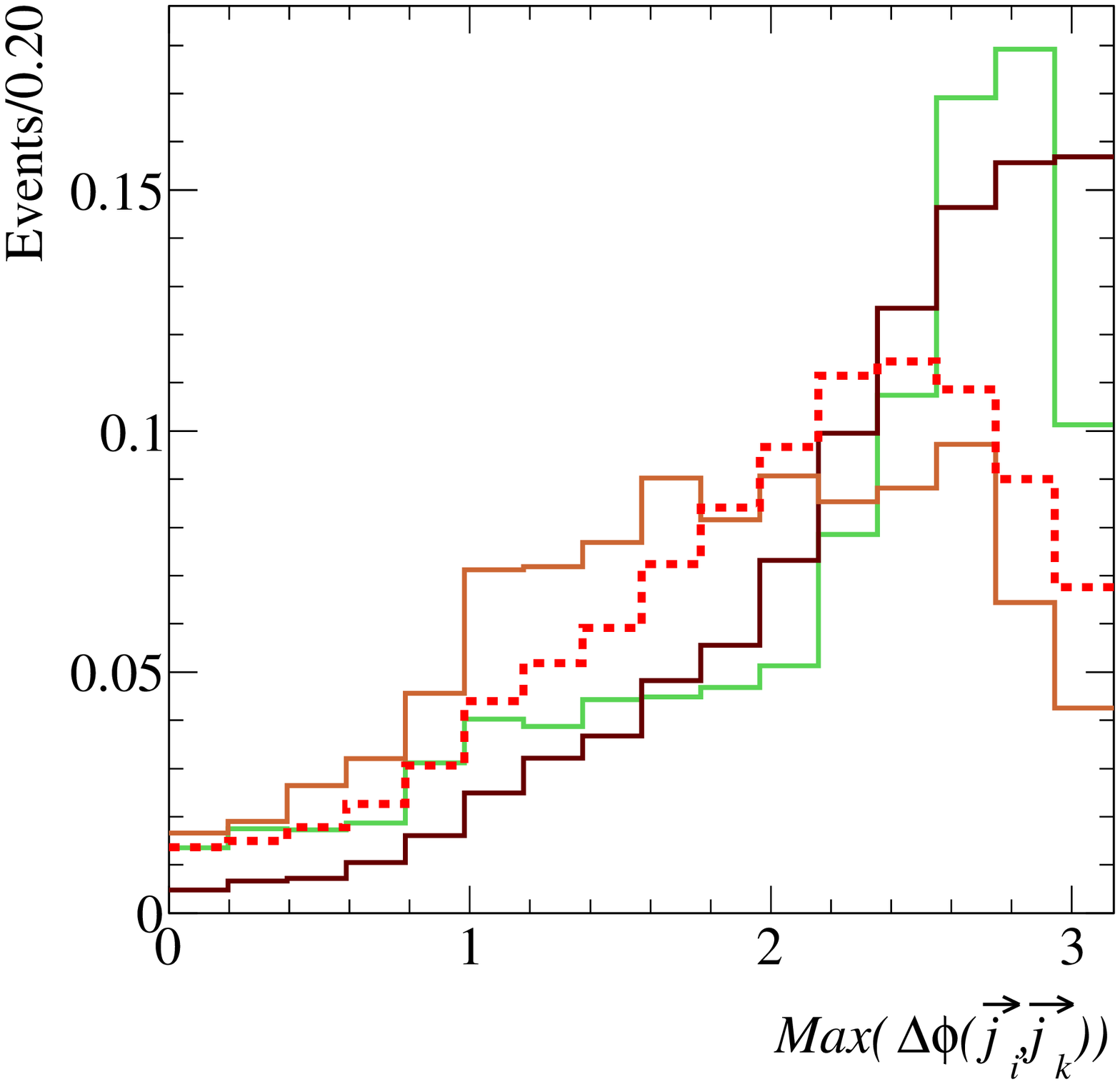}}
\subfigure
%[Max($\Delta R(jet_i,jet_j)$]
{\includegraphics[width= .32 \linewidth]{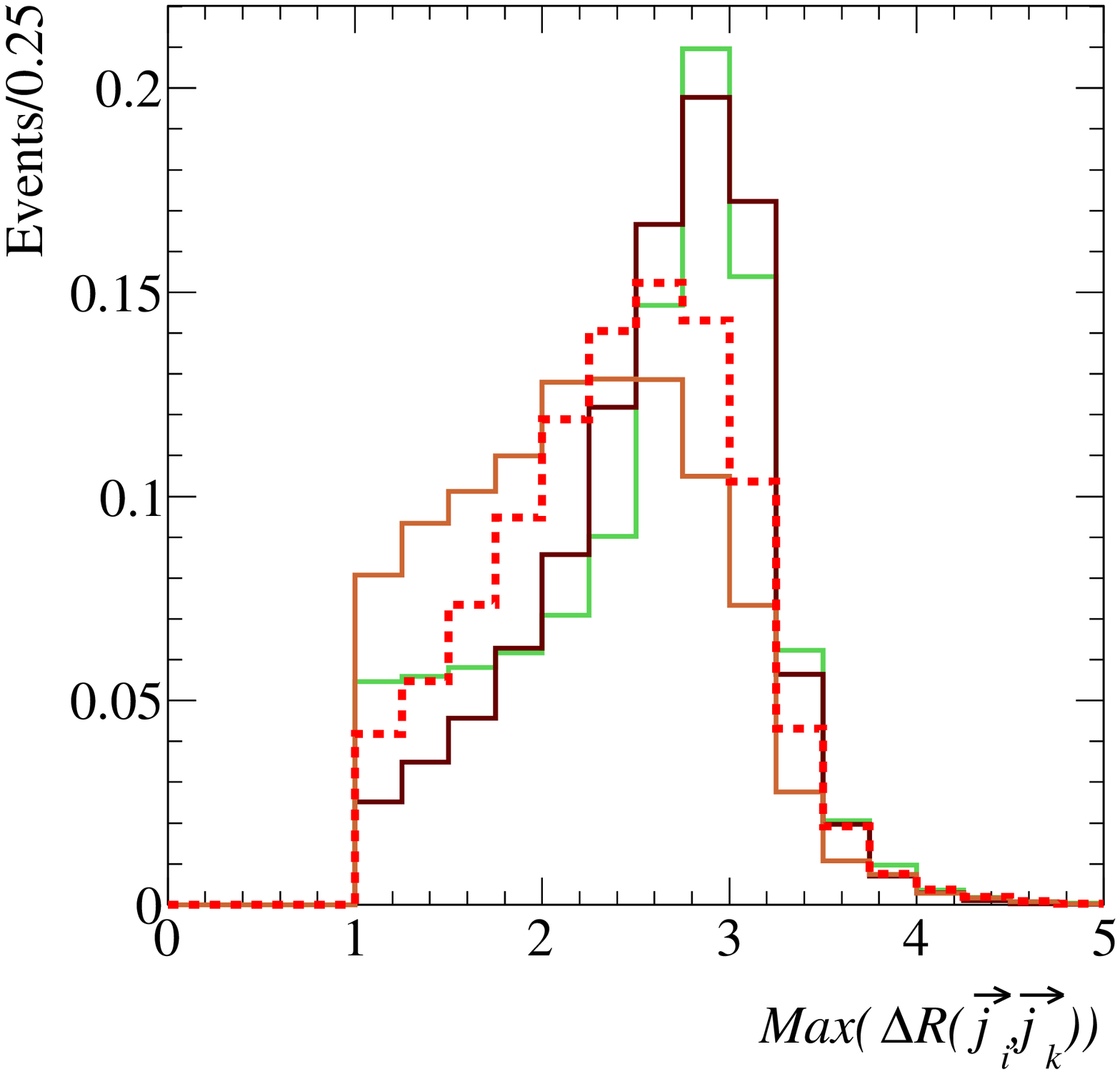}}
\subfigure
%[Min($\Delta \phi (\met,jet_i))$]
{\includegraphics[width= .32 \linewidth]{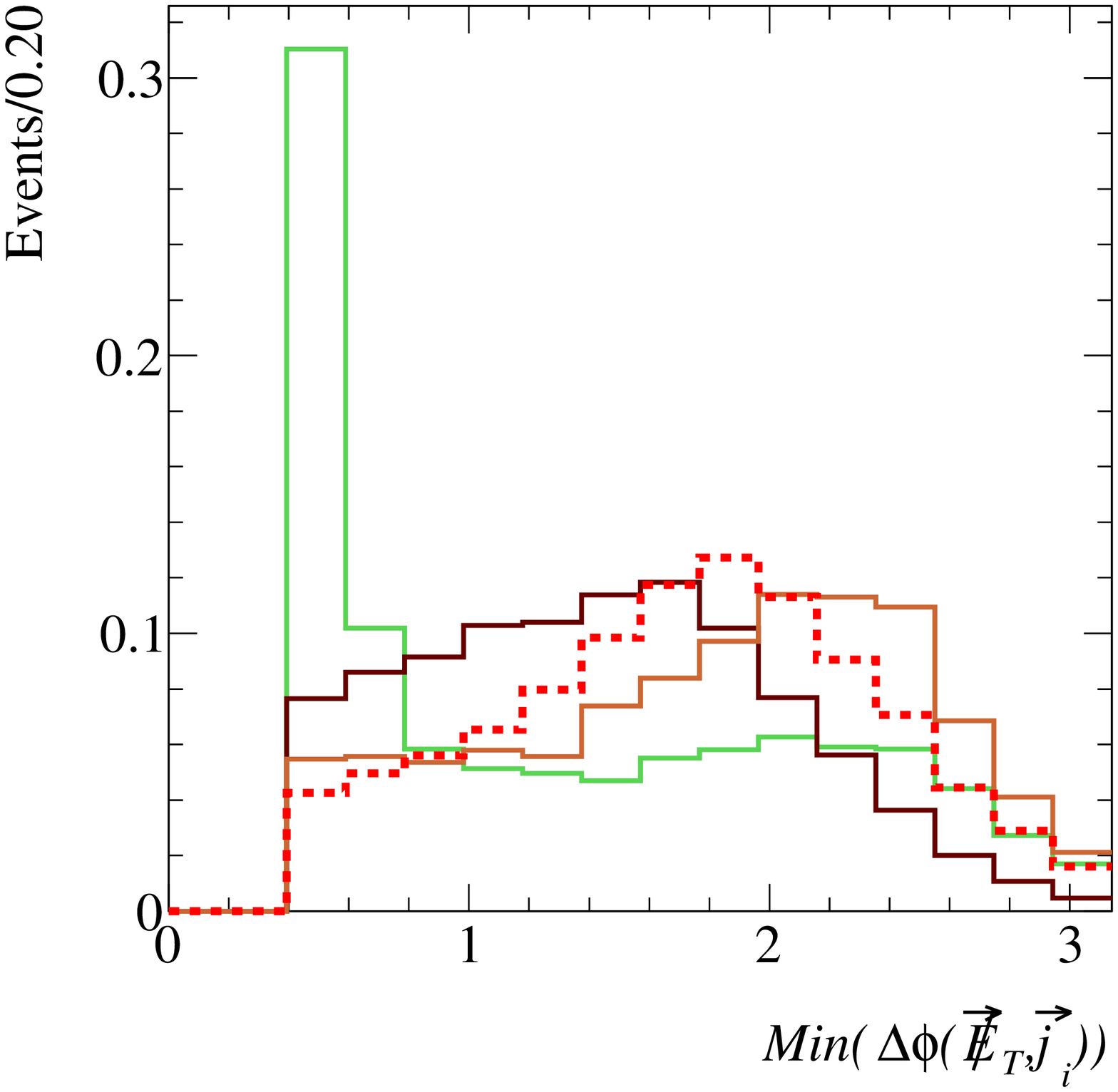}}
\subfigure
%[Min($\Delta \phi (\met,jet_i))$]
{\includegraphics[width= .32 \linewidth]{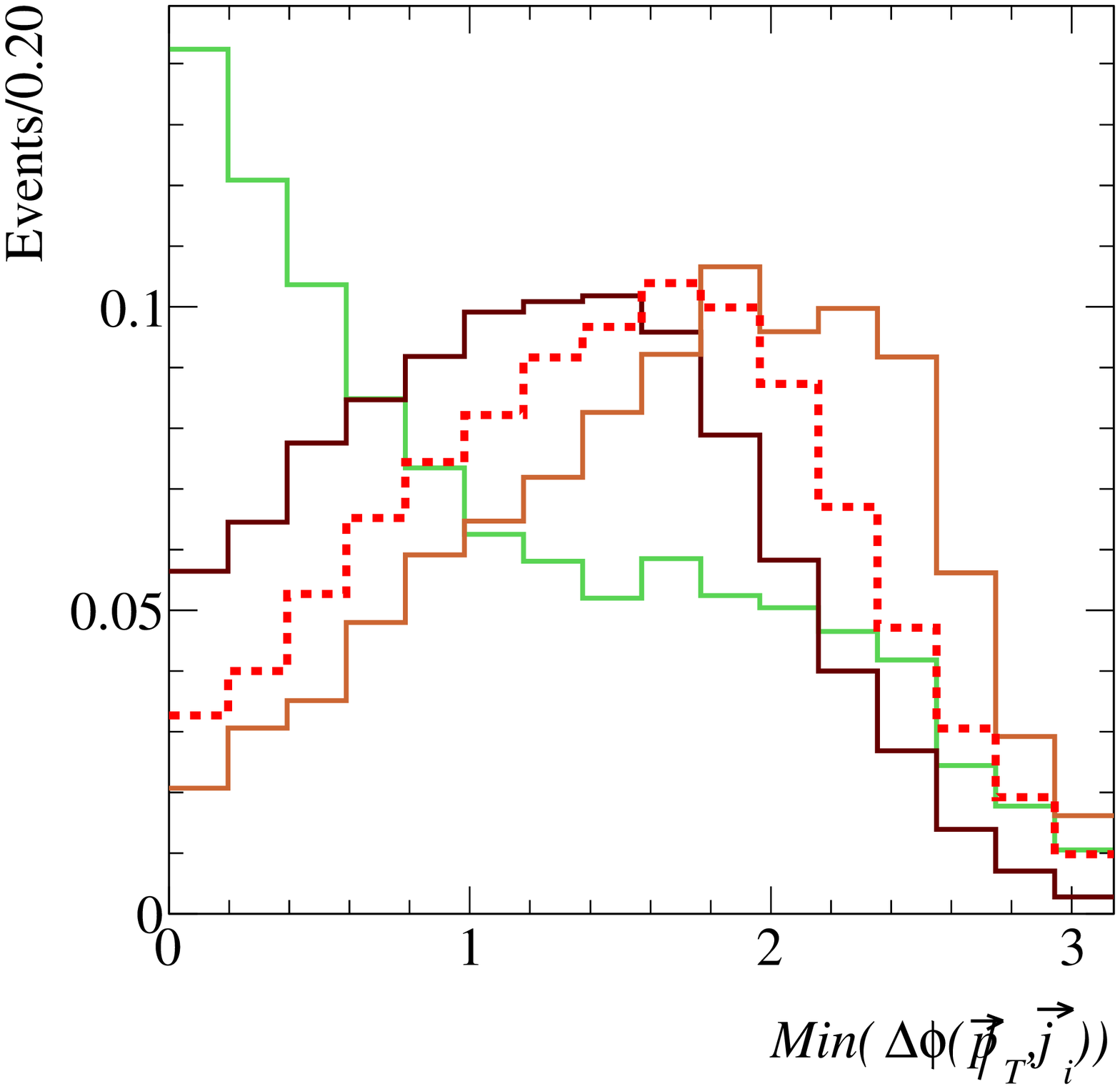}}
\subfigure
%[$\Delta \phi(jet_1,jet_2)$ with respect to the dijet system direction]
{\includegraphics[width= .32 \linewidth]{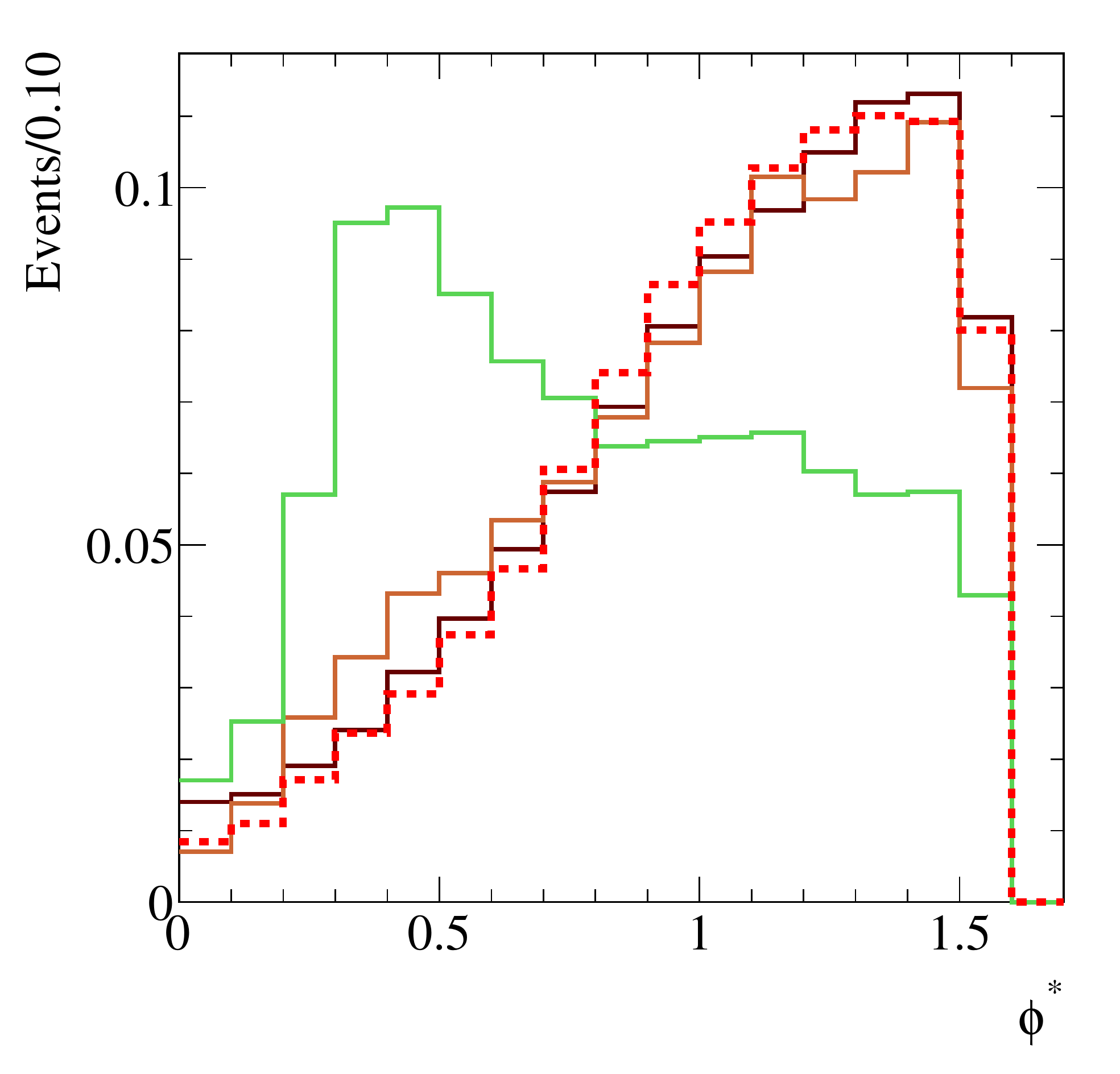}}
\caption{Kinematic distributions for signal and background events passing the event preselection. The three subsamples are summed together in their respective proportions. All histograms are normalized to unit area.} 
\label{fig:kinsel1}
\end{figure*}

\begin{figure*}[htbp]
\centering
\subfigure%[Sphericity]
{\includegraphics[width= .32 \linewidth]{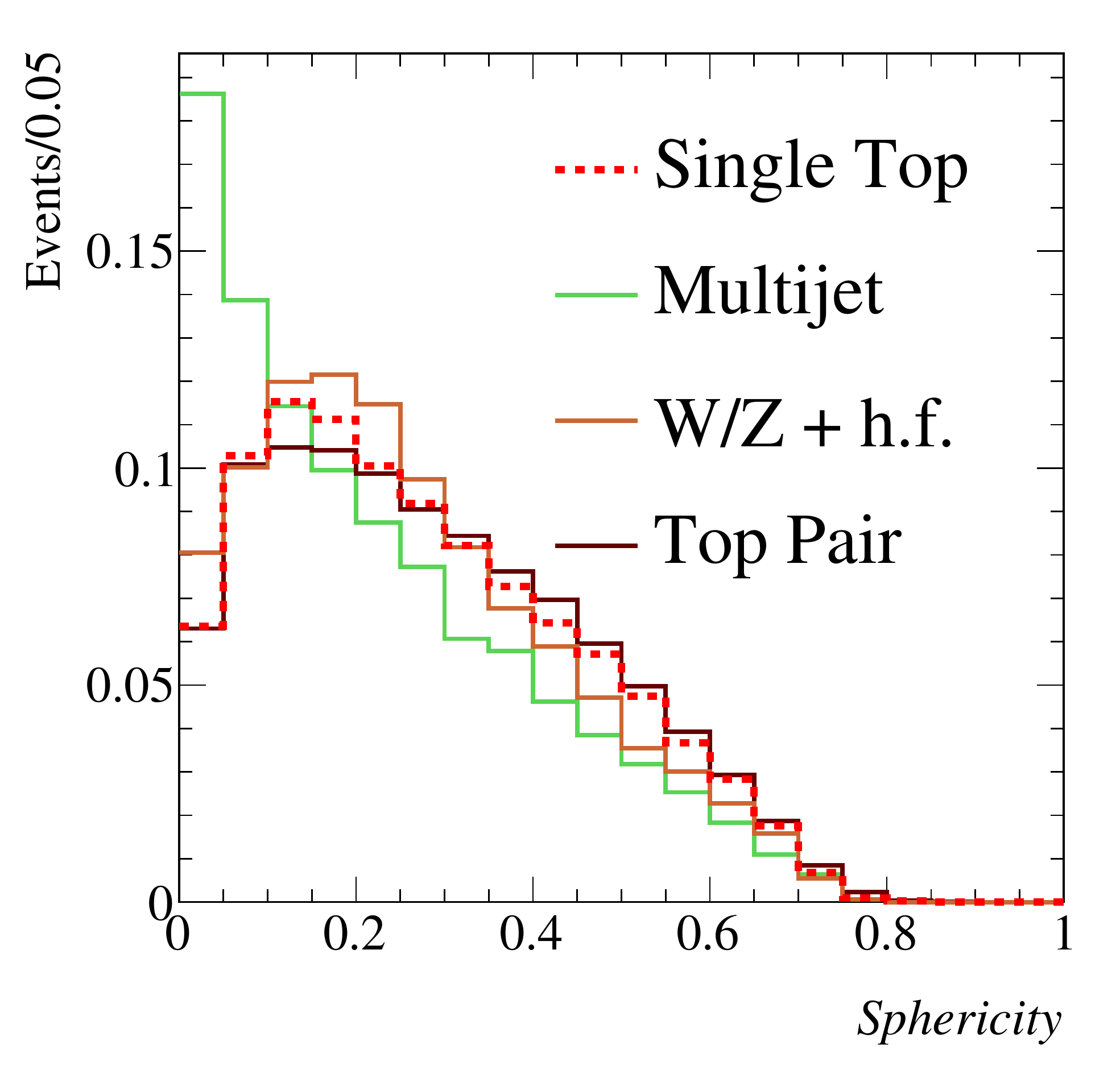}}
\subfigure
%[$\sum \Pt^{chgd}/\Pt^{jet_1}$]
{\includegraphics[width= .32 \linewidth]{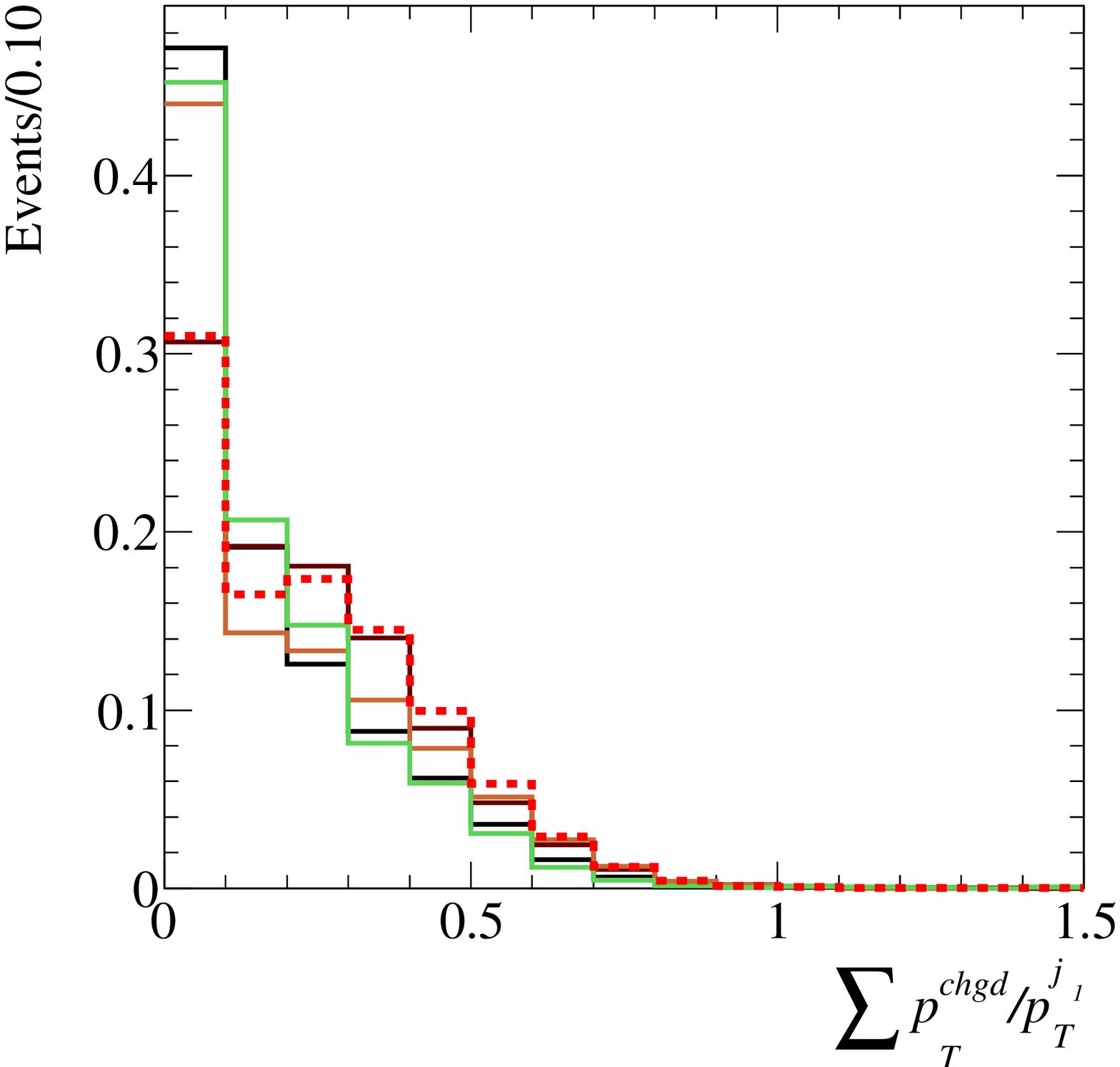}}
\subfigure
%[$\sum \Pt^{chgd}/\Pt^{jet_2}$]
{\includegraphics[width= .32 \linewidth]{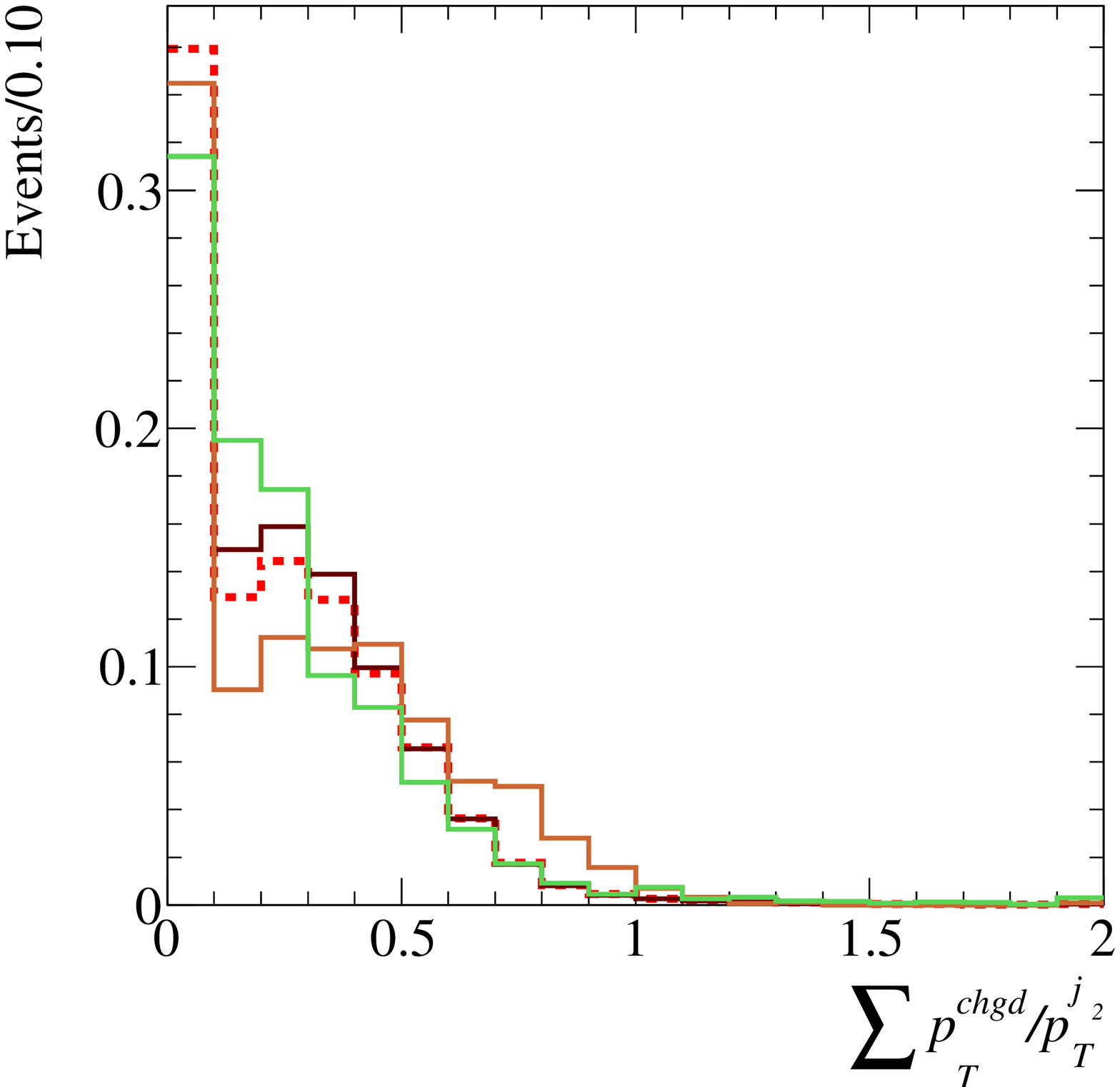}}
\caption{Kinematic distributions for signal and background events passing the event preselection. The three subsamples are summed together in their respective proportions. All histograms are normalized to unit area.} 
\label{fig:kinsel2}
\end{figure*}

 Comparisons of the kinematic distributions for background and signal events for the 15 variables are shown in Figs.\,\ref{fig:kinsel1} and\,\ref{fig:kinsel2}, where for simplicity the three subsamples 1S, 2S and SJ are combined. 

\begin{figure*}[htbp]
  \centering
%  \subfigure[]{\includegraphics[width=0.40\textwidth]{./newfinalPlots8/PreSelection_QCDNN_1Tag_Validation}}
%  \subfigure[]{\includegraphics[width=0.40\textwidth]{./newfinalPlots8/PreSelection_QCDNN_TT_Validation}}
%  \subfigure[]{\includegraphics[width=0.40\textwidth]{./newfinalPlots8/PreSelection_QCDNN_TJ_Validation}}\\
\subfigure[]{\includegraphics[width=8cm]{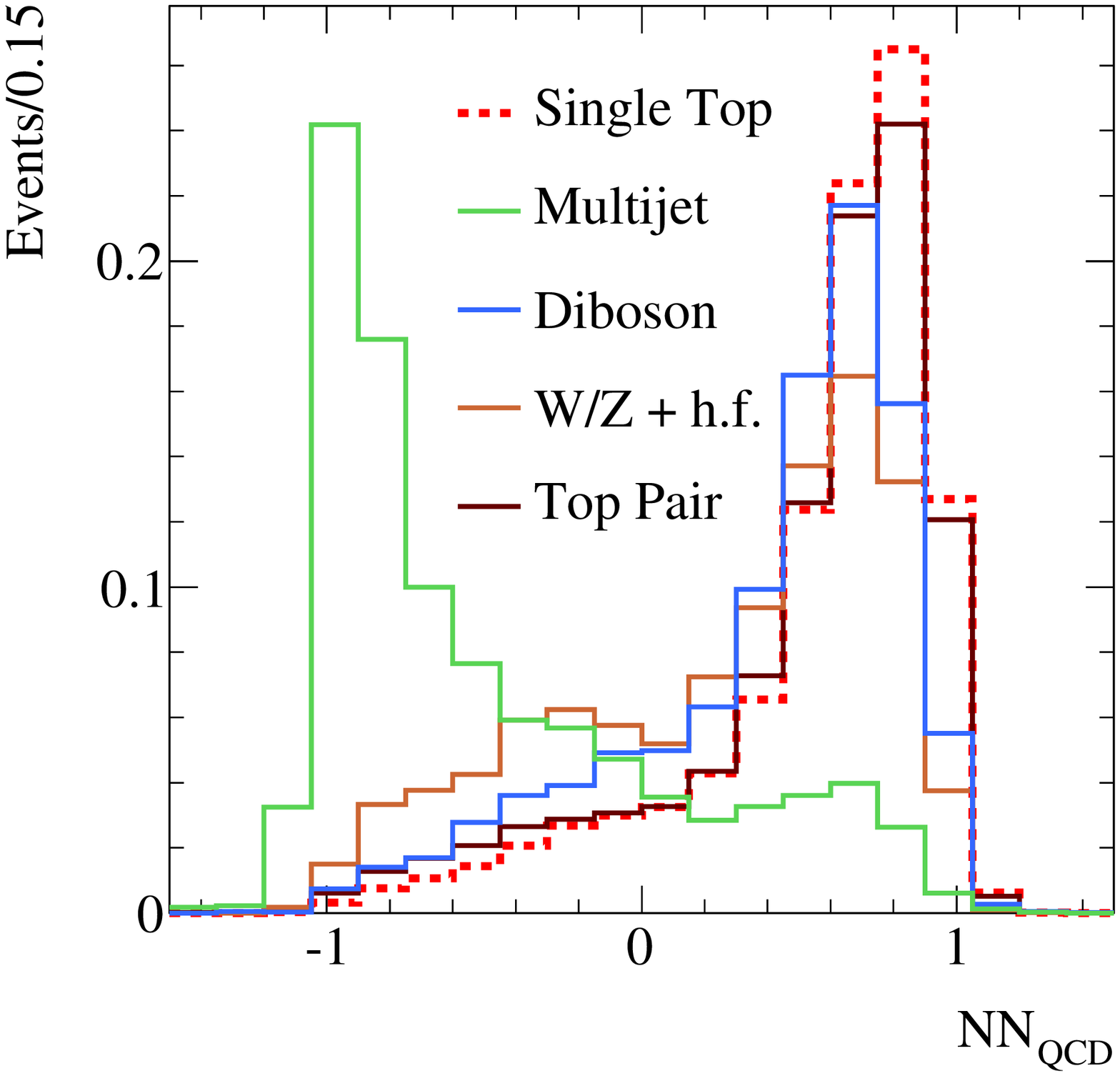}}
\subfigure[]{\includegraphics[width=8cm]{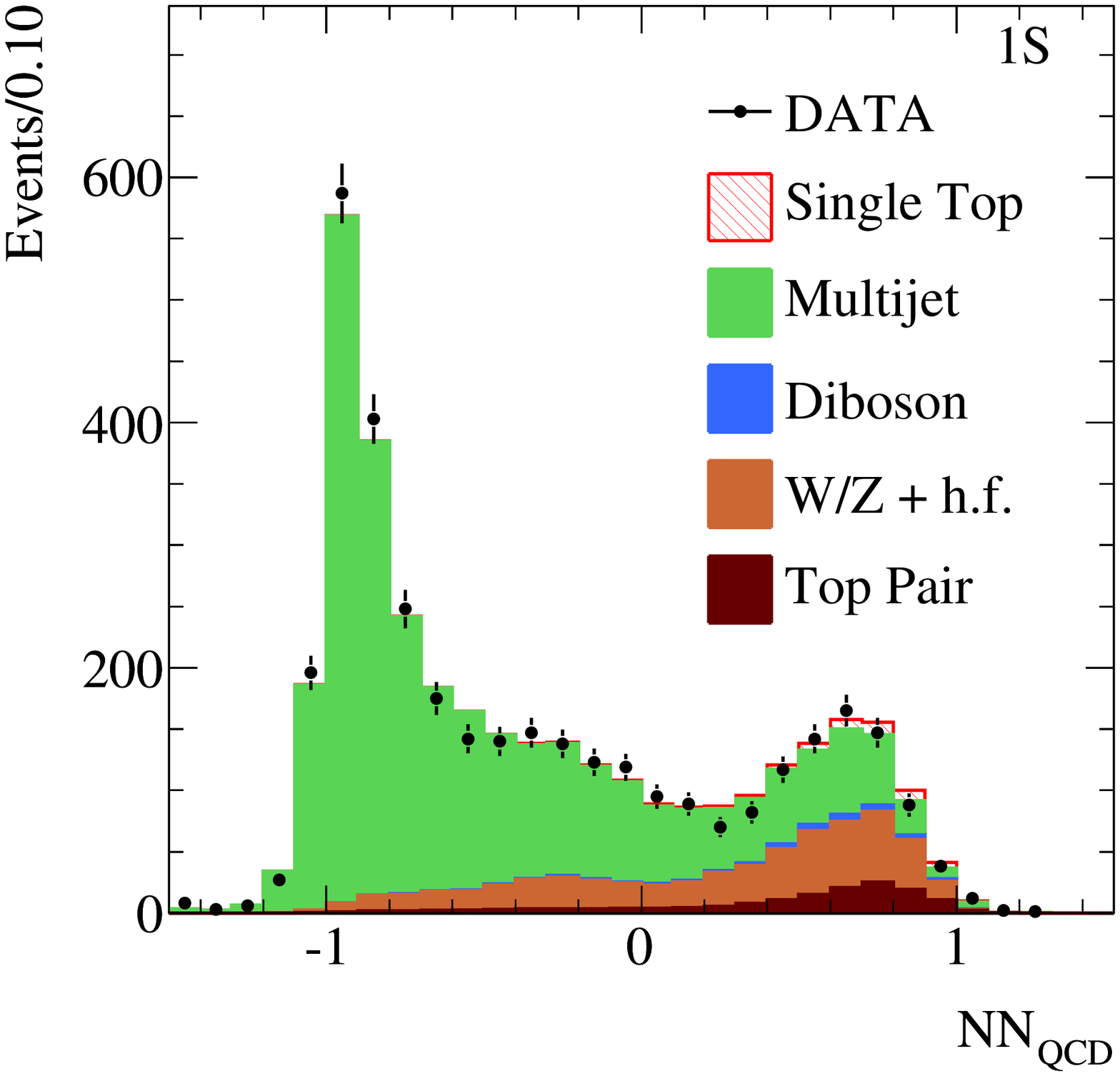}}
\subfigure[]{\includegraphics[width=8cm]{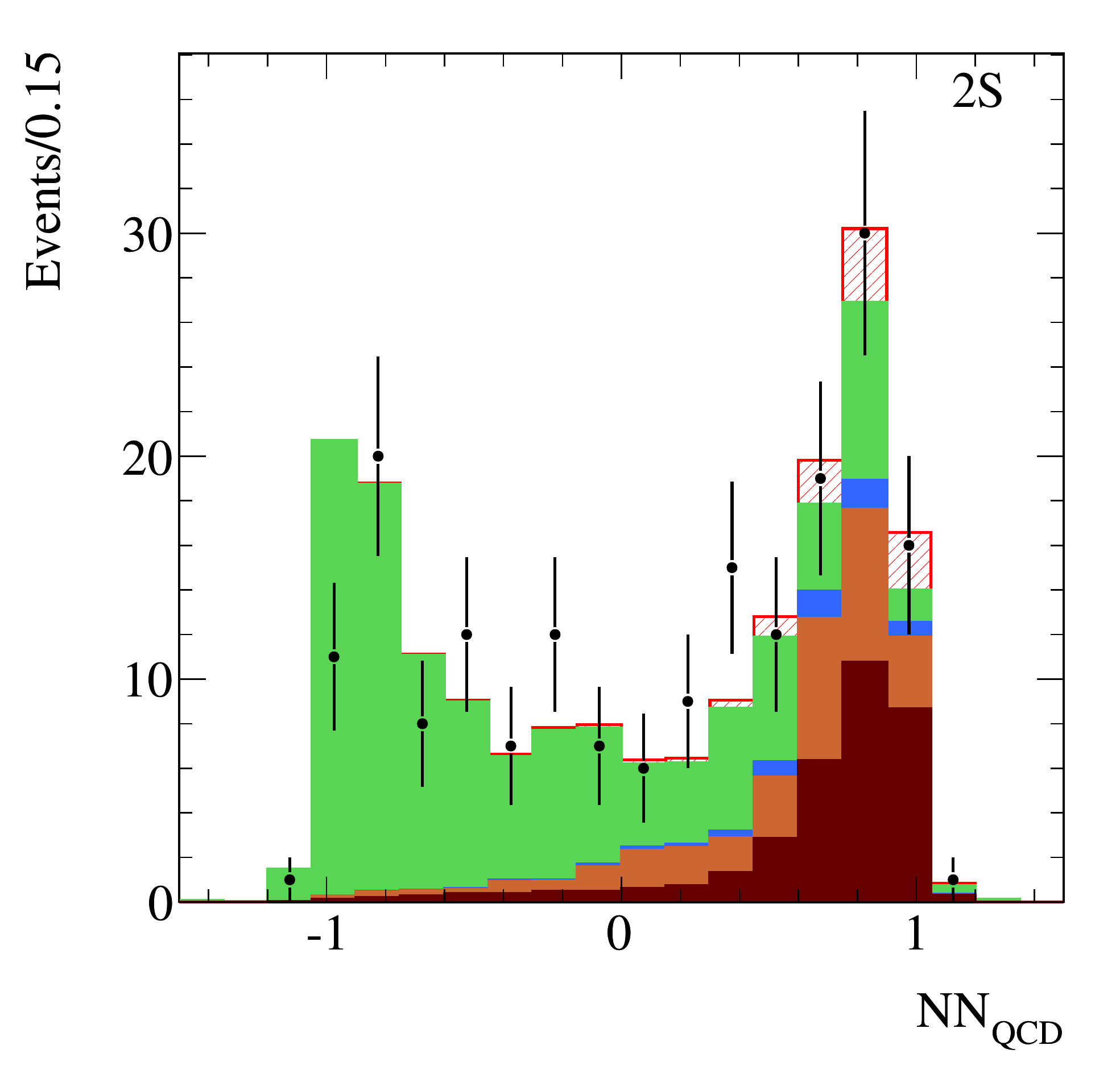}}
\subfigure[]{\includegraphics[width=8cm]{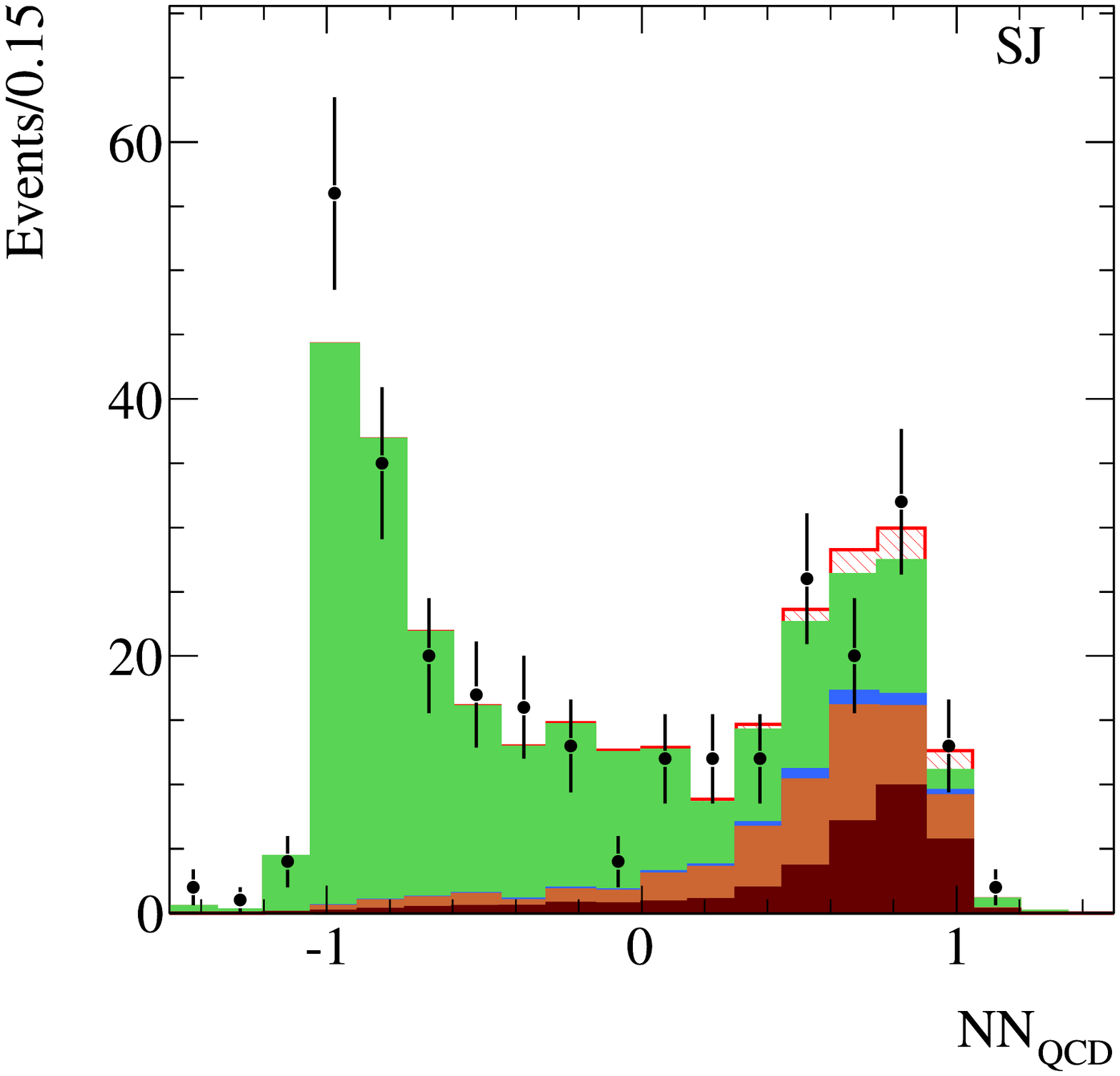}}
  \caption{Distributions of the NN$_{\rm QCD}$ output for events passing the event pre-selection. (a) shows the distribution for the signal and the backgrounds normalized to unity. The remaining three plots show the same distribution for the three subsamples, 1S (b) 2S (c) and SJ (d), where the background and signal predictions are stacked according to predictions, and compared to data events. As can be seen from the plots, the kinematics of the QCD background events are very different from the signal and the other backgrounds. These events are removed using the cut NN$_{\rm QCD} < -0.1$. The remaining events are used to scan for the presence of the single top quark signal. }
  \label{fig:outQCDNN}
\end{figure*}

The QCD background kinematics do not vary significantly with the heavy flavor content so only one neural network is used for the three $b$-tagged subsamples. The single top quark signal used for the training is a mixture of Monte Carlo simulated $s$-channel events (50\%) and $t$-channel events (50\%), which corresponds to the predicted signal composition after preselection.  For the background, we use the multijet background model described in Sec.\,\ref{sec:SandBmodel}. 
%The neuron activation function is the hyperbolic tangent. %We use a LearningRate of 0.01 and a TestRate of 5. 
%The other parameters are set to the default values in  {\sc tmva}; the number of epochs for the training is set to 500. 
All samples are split into two subsamples: one for training the neural network, and one for making predictions of the neural network output and for testing for overtraining. The distributions for training and testing samples are in good agreement. Both samples contain 21\,000 signal and 14\,000 multijet background events.
Among the configurations investigated, the one which performs best uses all 15 variables defined above as inputs, two hidden layers with 30 and 15 nodes respectively, and one output node. Figure\,\ref{fig:outQCDNN} shows the distribution of the value of the output node, NN$_{\rm QCD}$.

%Table~\ref{tab:Events_all_SR} lists the expected and observed event yields in signal region.

%\begin{table}[!ht]
%  \begin{center}
%    \begin{tabular}{|l|c|c|c|}
%\hline
%      		& Single top      & Data (L=2.1fb$^{-1}$)             \\
%      \hline
%% Scaled
%Trigger driven cuts & 331.7   & 523207 \\ 
%Lepton veto                      & 181.3 & 500813    \\
%\met not aligned to jets   &  164.6 &  56301\\
%b-tag requirement           & 75.1   & 4010 \\
%$N_{QCD}$ cut                &  64 & 1411 \\\hline
%\hline
%    \end{tabular}
%    \caption{Event selection}
%    \label{tab:EventsSel}
%  \end{center}
%\end{table}

% EVENTS TABLE

%\begin{table}[!ht]
%  \begin{center}
%    \begin{tabular}{|l|c|c|l|}
%\hline
%      		& Single top      & Data (L=2.1fb$^{-1}$)     &	$s/b$        \\
%      \hline
%% Scaled

%% WITH DETAIL

%%Trigger driven cuts & 331.7   & 523207  & $\sim$ 1/1575 \\ 
%%Lepton veto                      & 181.3 & 500813  & $\sim$  1/2767 \\
%%\met not aligned to jets   &  164.6 &  56301& $\sim$ 1/343 \\
%%b-tag requirement           & 75.1   & 4010 & $\sim$ 1/53 \\
%%NN$_{\rm QCD}$ cut                &  64 & 1411 & $\sim$ 1/22 \\\hline

%Trigger driven cuts & 331.7   & 523207  & $\sim$ 1/1600 \\ 
%Lepton veto                      & 181.3 & 500813  & $\sim$  1/2800 \\
%\met not aligned to jets   &  164.6 &  56301& $\sim$ 1/340 \\
%b-tag requirement           & 75.1   & 4010 & $\sim$ 1/50 \\
%NN$_{\rm QCD}$ cut                &  64 & 1411 & $\sim$ 1/20 \\\hline

\begin{figure*}[htbp]
  \centering
\subfigure
{\includegraphics[width=8cm]{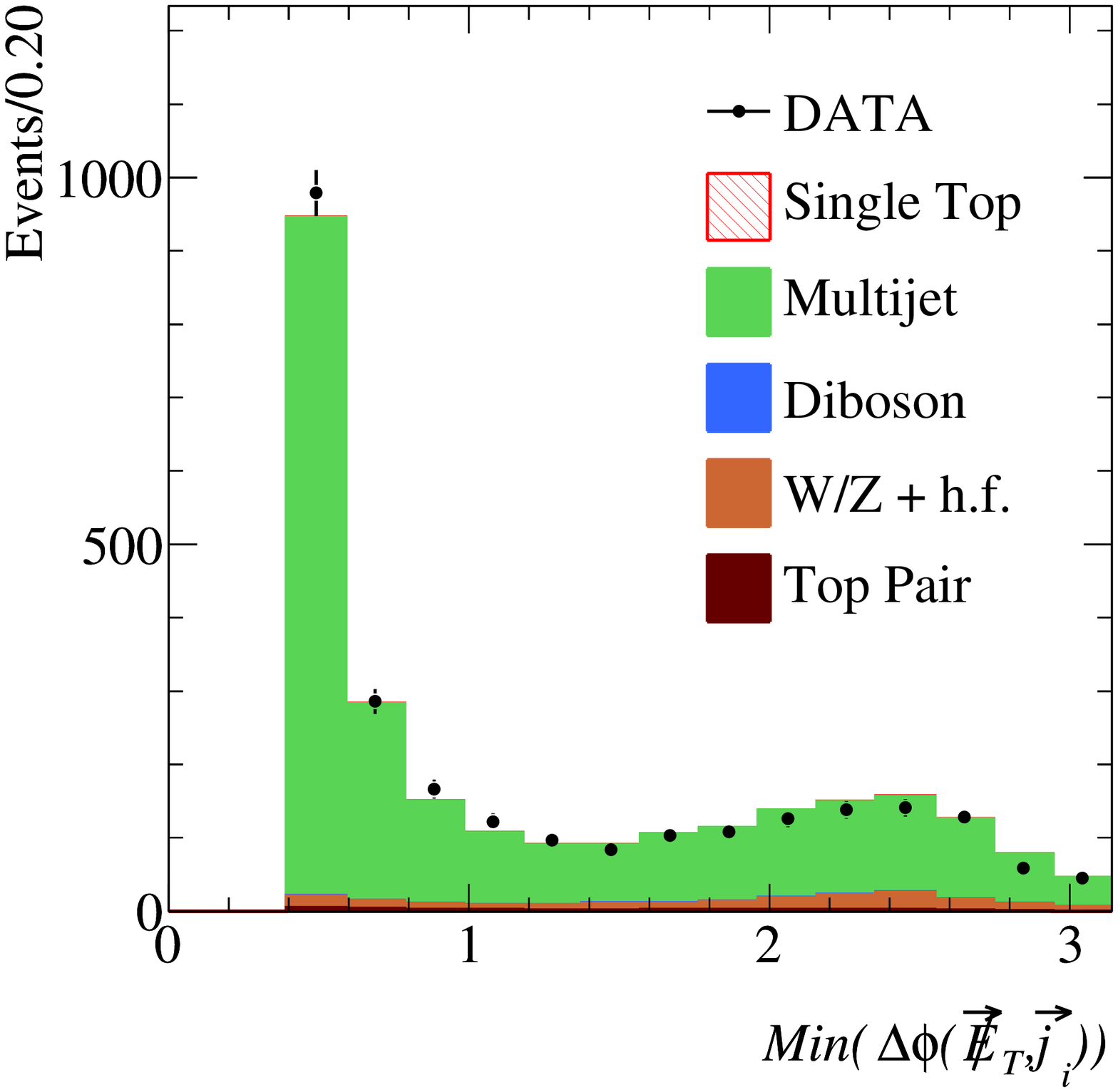}}
\subfigure
{\includegraphics[width=8cm]{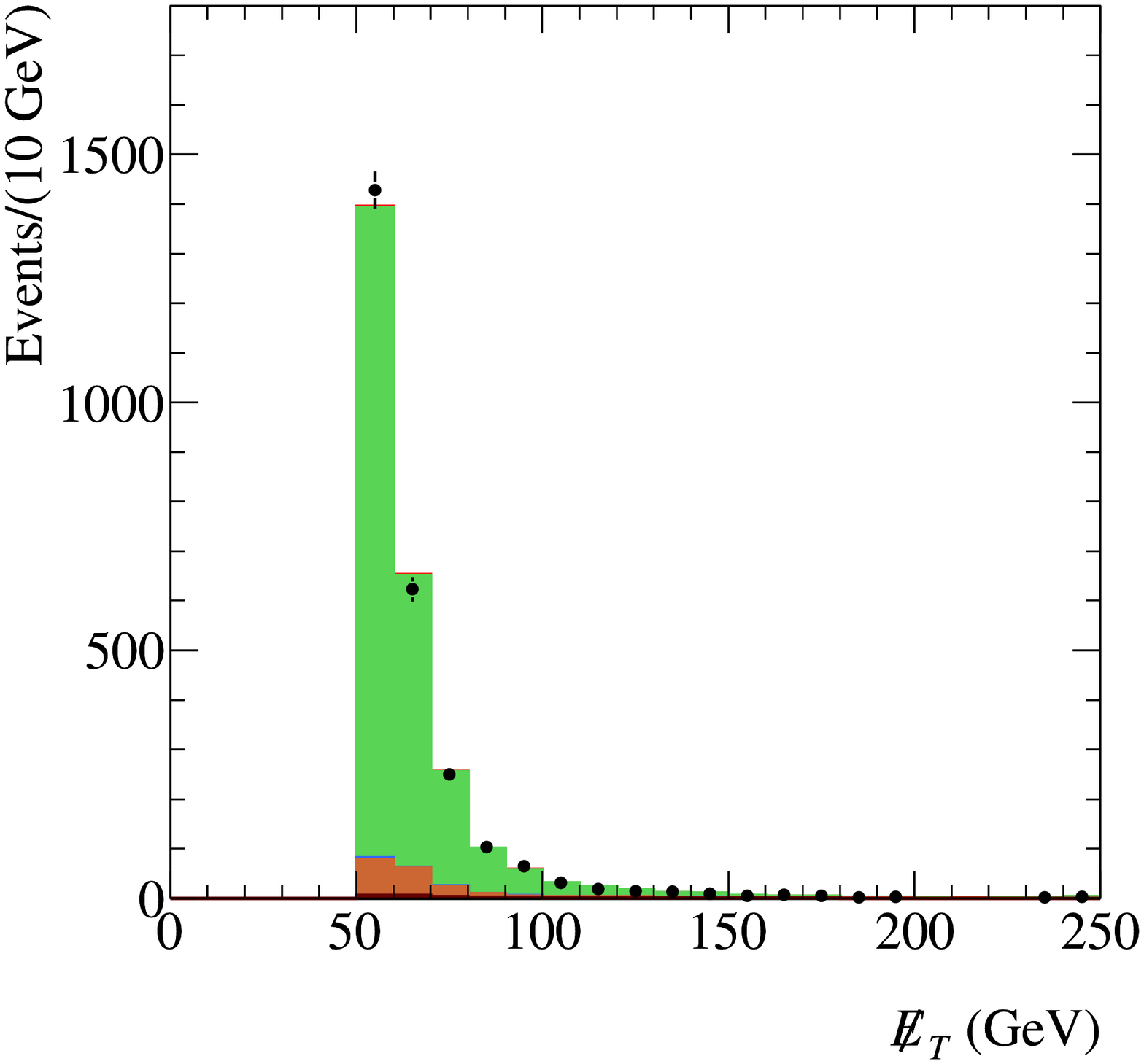}}
\subfigure
%[$\Delta\phi(\mpt,\met)$]
{\includegraphics[width=8cm]{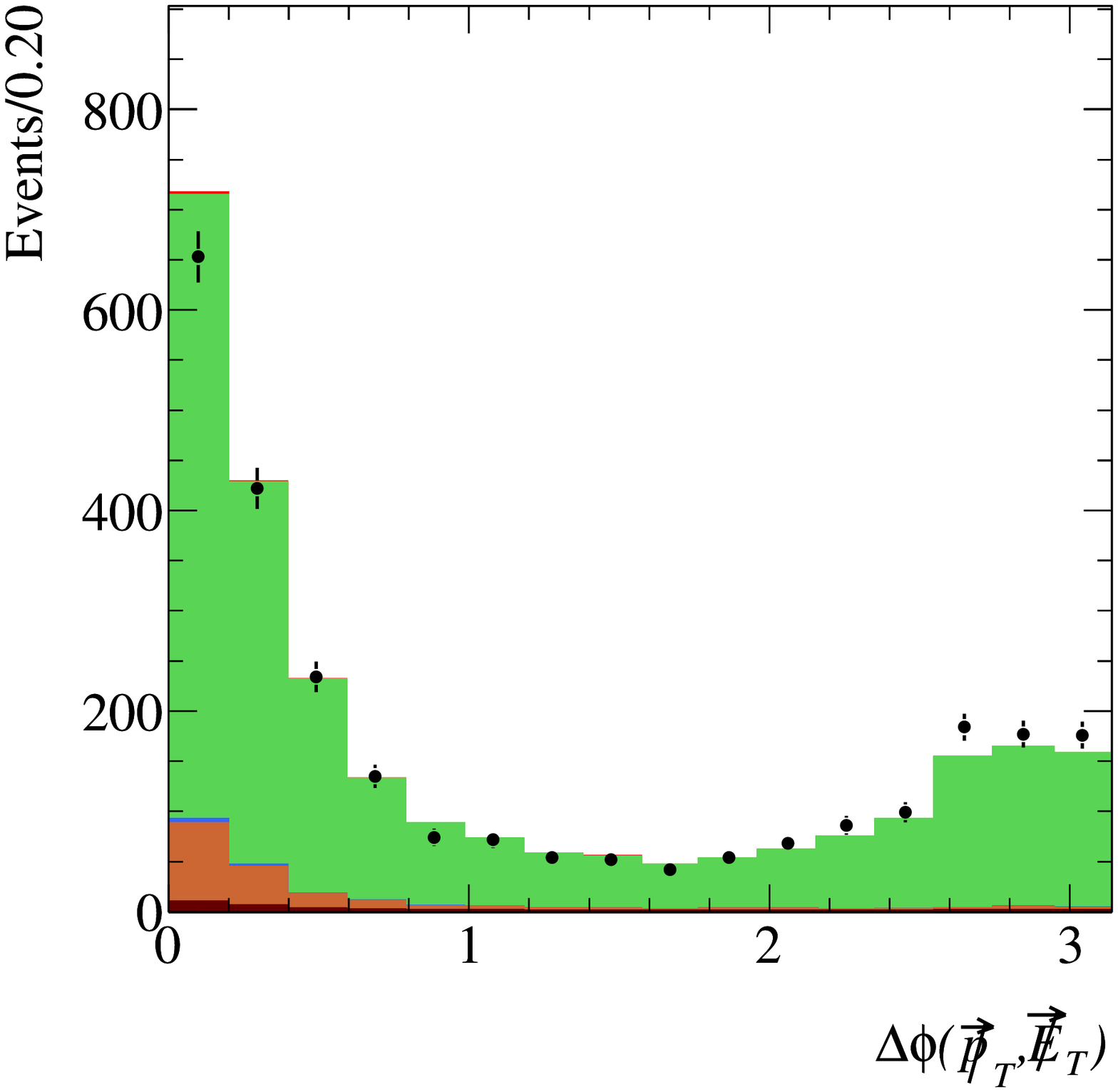}}
\subfigure
%[Invariant mass of $\met$ , $jet_1$ and $jet_2$]
{\includegraphics[width=8cm]{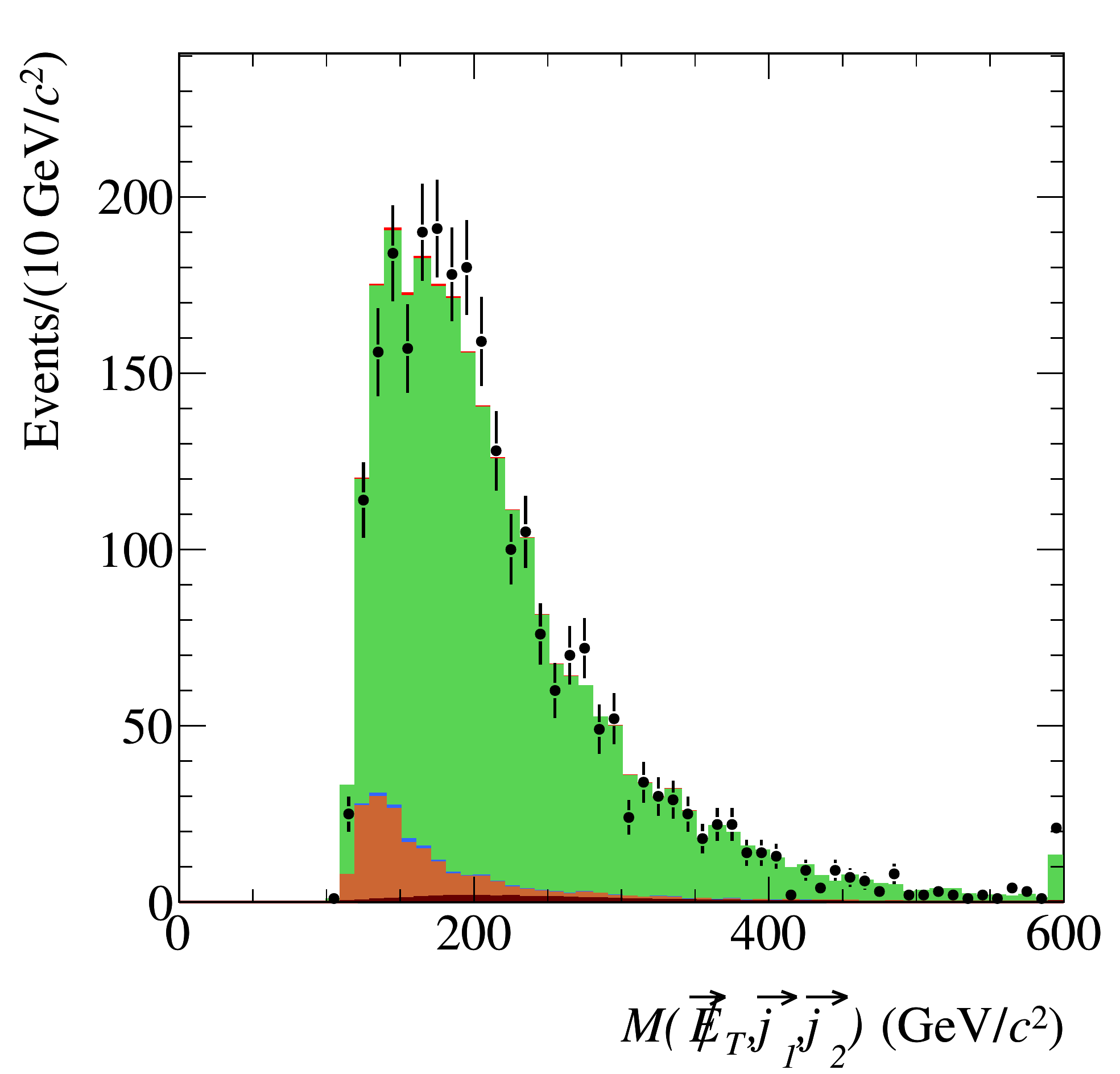}}
%[Min($\Delta\phi(\met,jet_i)$)]
  \caption{Comparison of the QCD background modeling to data in the NN$_{\rm QCD} < -0.1$ control region. The three subsamples are summed together in their respective proportions. Good agreement is visible between data and the background model.}
     \label{fig:CR3}
\end{figure*}

%
%\hline
%    \end{tabular}
%    \caption{Event selection}
%    \label{tab:EventsSel}
%  \end{center}
%\end{table}
The signal region is defined as the sample of events surviving the cut on NN$_{\rm QCD}$ devised to maximize the background rejection while retaining high signal efficiency. By requiring NN$_{\rm QCD}>$\,-0.1, we reduce the multijet contribution by 77\%, while keeping 91\% of the signal. The overall backgrounds are reduced by 65\% thus bringing the $s/b$ ratio from 1/50 to 1/20. The signal significance $s/\sqrt{s+b}$ is increased by 50\%, from 1.2 to 1.7. 
%The value of the NN$_{\rm QCD}$ is that which maximizes the significance ($S/\sqrt{S+B}$) of the remaining sample.
%All variables provide a description of events with $\met$ coming from mismeasured jets or of events with mistags and help improving the NN performance.  On the other hand, the use of additional variables negligibly affects performance.
We look at the background-dominated region with NN$_{\rm QCD} < -0.1$ to verify that the background model properly describes the data. Figure\,\ref{fig:CR3} shows distributions of data events superimposed to the sum of the expected backgrounds; the background model properly describes the data within uncertainties.
Moreover, data events in the same region are used to compute the normalization for multijet production, by comparing the prediction given by weighting the events for the tag rate parametrization, and the observed number of data events. We find that the predictions must be multiplied by scale factors that depend on the combination of $b$-tagging algorithms used. The values extracted are  $1.08\pm0.05$  for events in the 1S subsample, $0.79\pm0.10$ for 2S events, and $0.76\pm0.07$ for SJ events.
Table\,\ref{tab:Events_all_SR} shows the contribution of signal and background events in the signal sample, divided into three subsamples under study.
\begin{table}[!ht]
  \begin{center}
    \caption{Number of predicted and observed events in the signal region defined by requiring NN$_{\rm QCD} > -0.1$ in the subsample with exactly one \secvtx-tagged jet (1S), two \secvtx-tagged jets (2S) and one \secvtx- and one \jetprob-tagged jet (SJ). The notation ``h.f." stands for heavy flavor jets. The uncertainty in the predicted number of events is due to the theoretical cross section uncertainty and to the uncertainty on signal and background modeling.}
    \begin{tabular}{lccc}
    %\multicolumn{4}{l}{\small\textbf{CDF Run II Preliminary, 2.1 fb$^{-1}$}\normalsize}\\
      \hline\hline
      Process		&1b-tag  (1S)    & 2 $b$-tags(2S)  & 2 $b$-tags(SJ)               \\
      \hline
% Scaled
$s$-channel &15.7$\pm$2.0&7.6$\pm$0.9&6.3$\pm$0.8\\
$t$-channel &31.2$\pm$4.9&1.7$\pm$0.2&1.6$\pm$0.2\\ 
$t \bar t$ &125$\pm$23&30.3$\pm$5.8&29.2$\pm$5.7\\
$WW/WZ/ZZ$ &33.0$\pm$6.5&4.9$\pm$0.6&4.2$\pm$0.6\\
W + h.f.&269$\pm$113&12.7$\pm$7.5&22.7$\pm$13.7\\
Z + h.f&105$\pm$53&11.8$\pm$5.8&11.8$\pm$6.0\\
multijet&592$\pm$27&28.9$\pm$3.8&58.5$\pm$5.8\\ \hline
%\vskip 0.5
%Signal&46.8$\pm$5.2&9.3$\pm$1.0&7.9$\pm$0.8\\
%Background&1125$\pm$169&89$\pm$15&126$\pm$21\\
Total&1172$\pm$169&98$\pm$15&134$\pm$21\\ \hline
Observed &1167&113&131\\
\hline\hline
    \end{tabular}
%    \caption{Number of predicted and observed events in the signal region in the subsample with exactly one \secvtx-tagged jet (1S), two \secvtx-tagged jets (2S) and one \secvtx- and one \jetprob-tagged jet (SJ). The notation ``h.f." stands for heavy flavor jets. The uncertainty in the predicted number of events is due to the theoretical cross section uncertainty, and to the uncertainty on signal and background modeling.}
    \label{tab:Events_all_SR}
  \end{center}
\end{table}
After requiring NN$_{\rm QCD} > -0.1$, the dominant backgrounds are from multijet production, $W/Z$ + heavy flavor jets events, and \ttbar\, events. 
The multijet contribution after the NN$_{\rm QCD}>-0.1$ cut now primarily consists of events with true missing energy coming from a W or Z boson, 
accompanied by light-flavor jets misidentified as $b$-jets.
%We studied the properties of those events to develop a NN with the goal of discriminating the surviving backgrounds %from the interesting signal. 

%It should be noted at this stage that the ``QCD multijet" background in the signal region consists more of events with mis-tagged W/Z + heavy flavor jets rather than QCD production of jets.

We use our simulated single top quark data sample to investigate which decay modes of the W bosons from top quark
decays survive the event selection including the lepton veto. We find that the largest category of single top quark decays is $W \to \tau \nu$ ($\sim 50\%$ of decays), followed by $W \to \mu\nu$ ($\sim 30\%$) and  $W \to e \nu$ ($\sim 20\%$). The presence of hadronic $W$ decays is suppressed by the requirement of large $\met$, and by the NN$_{\rm QCD} > - 0.1$ cut, so that the fraction of all-hadronic single top decays is found to be negligible.

\section{Discriminating the signal from remaining backgrounds}
\label{sec:NNdisc}

In the previous section we described an event selection which enhances the signal purity of the sample by suppressing the presence of backgrounds that do not produce real neutrinos. At this stage of the analysis, the $s/b$ is about 1/20, where the main background processes all produce neutrinos. 
Unfortunately, all of the surviving backgrounds have topology and kinematics which are very similar to single top quark events. In addition, the systematic uncertainty on the background prediction is approximately 2 to 4 times the size of the signal we seek, depending on the subsample.  \par
Further discrimination of the signal from the background is required.  In order to increase the statistical power of the analysis, and to minimize the effect of the background systematic uncertainties, we study the signal sample to take advantage of the small residual differences between the signal and backgrounds.
Another neural network will be used for this purpose, where events which appear to be more signal-like are used to test for the presence of single top quark production and to measure the cross section, and events which appear to be more background-like are used to constrain the uncertain background rates. 

\begin{figure}
  \centering
{\includegraphics[width=8cm]{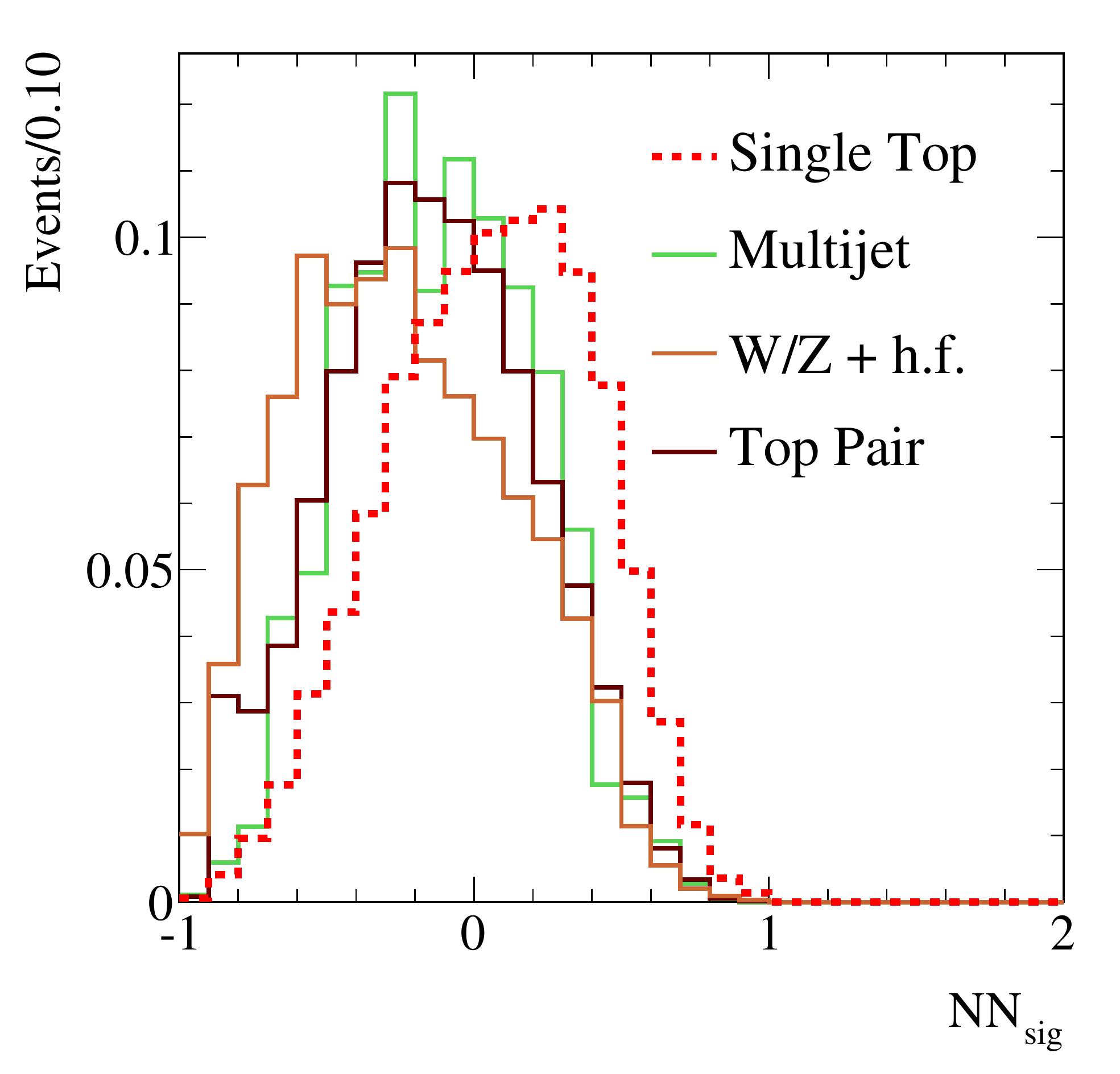}}
    \caption{Distribution of $NN_{\rm sig}$ for the signal and main background processes in the signal region. The three subsamples are summed together in their respective proportions. All histograms are normalized to unity. A small but important residual discrimination of the signal from the backgrounds is obtained.}
     \label{fig:outFinalMLPshape}
\end{figure}

\begin{figure*}
  \centering
\subfigure[]
%[Events with exactly one \secvtx-tagged jet]
{\includegraphics[width=8cm]{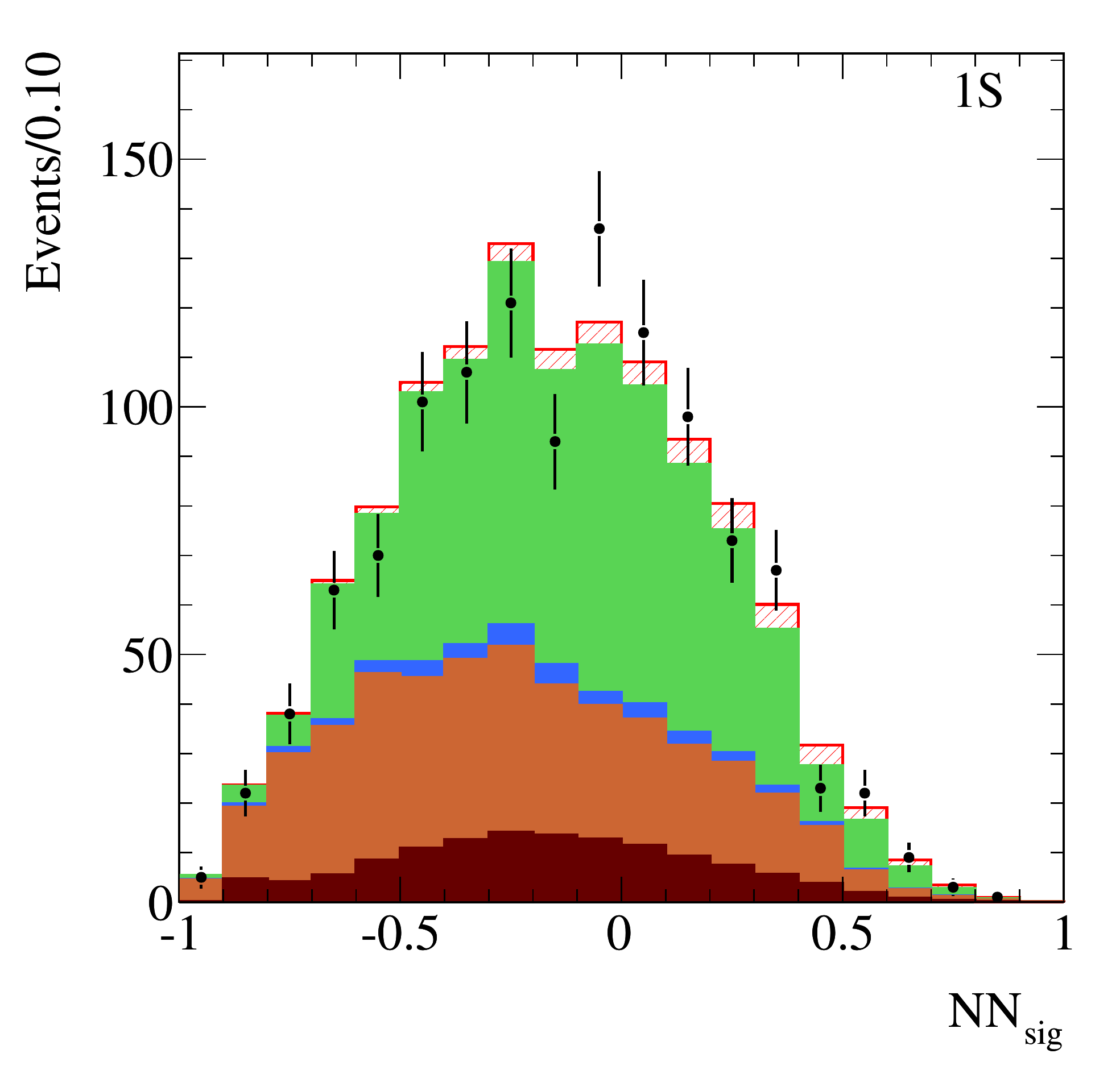}}
\subfigure[]
%[Events with two {\secvtx}-tagged jets]
{\includegraphics[width=8cm]{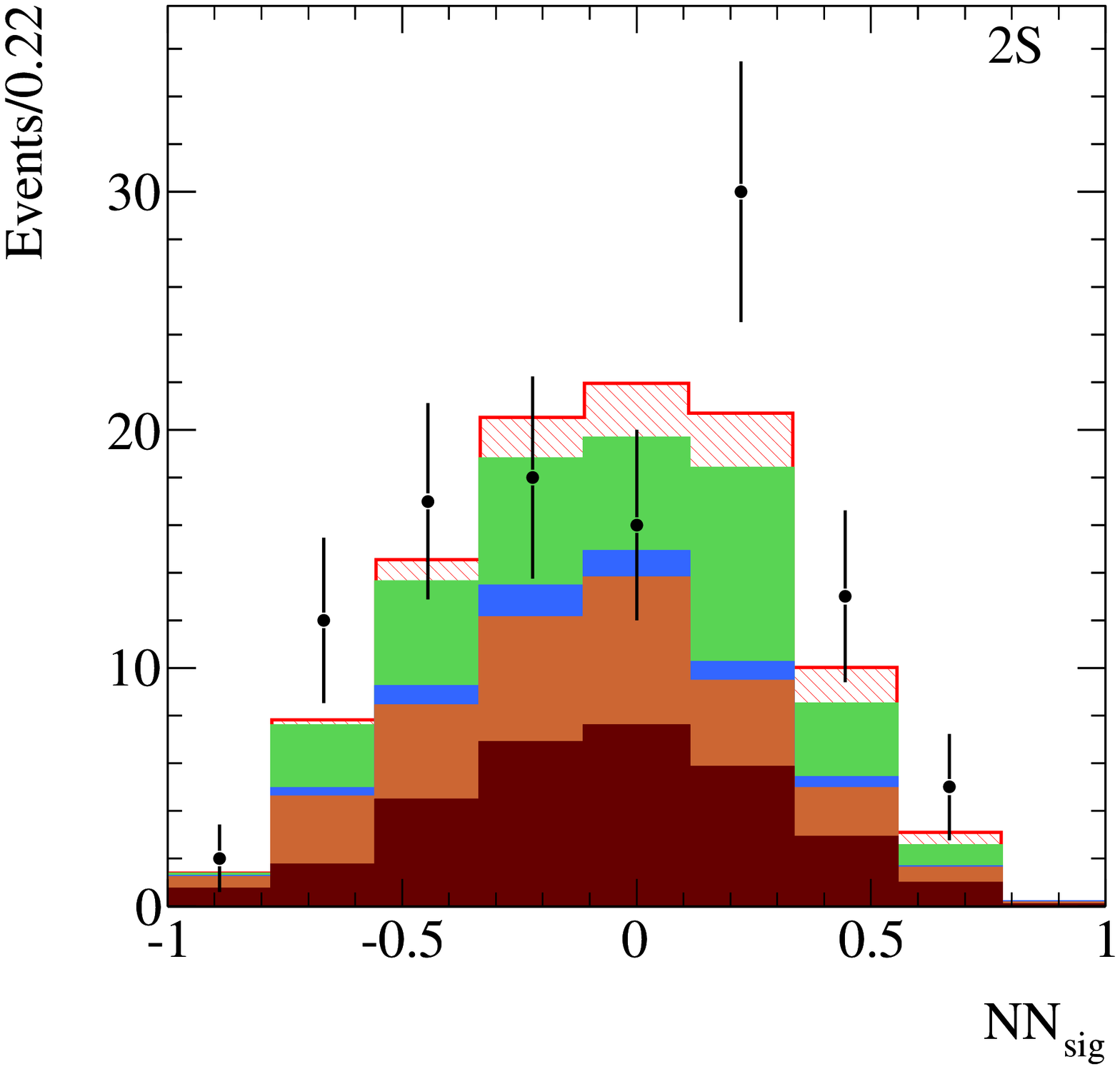}}
\subfigure[]
%[Events with one {\secvtx}-tagged and one \jetprob-tagged jet]
{\includegraphics[width=8cm]{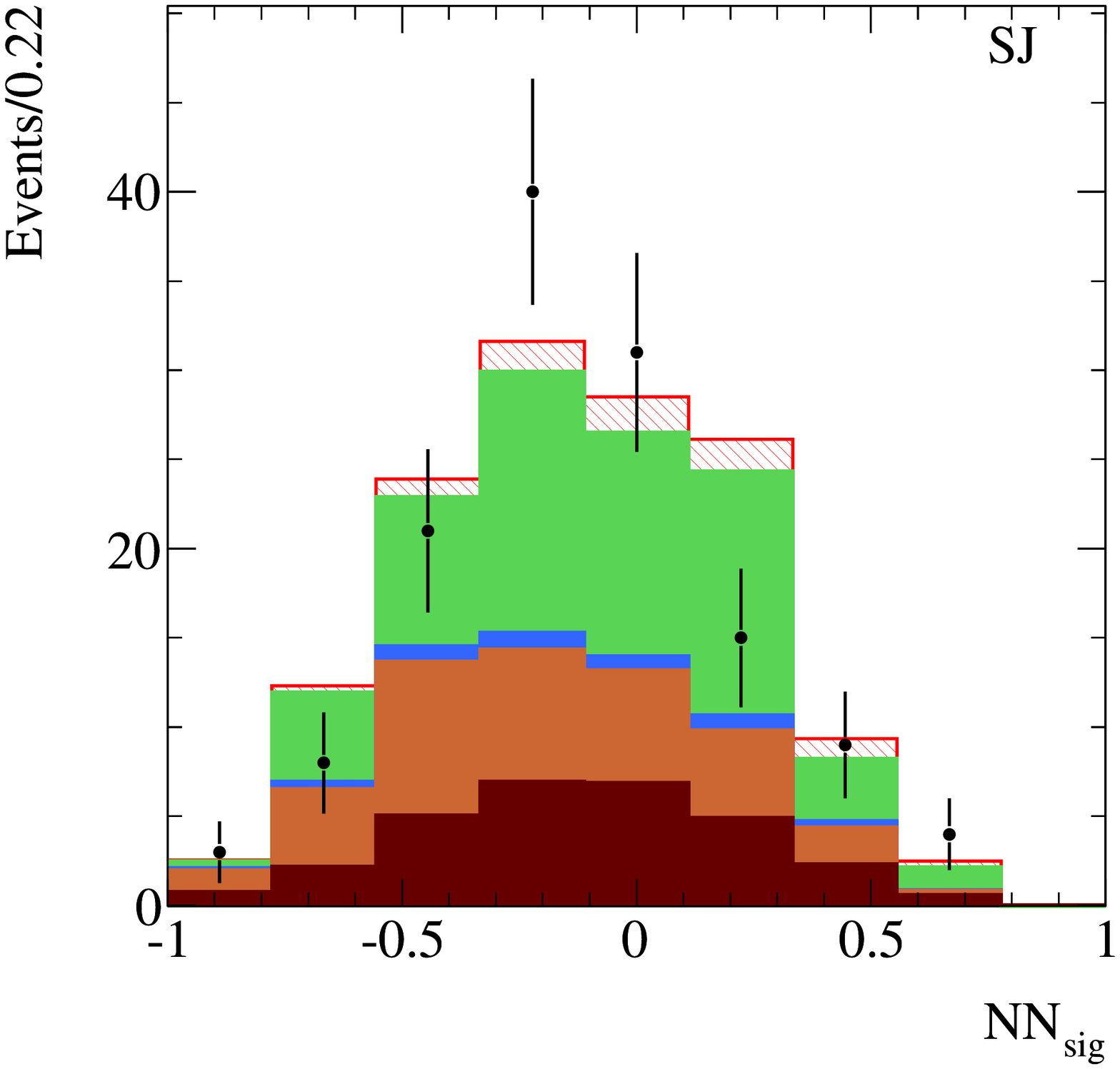}}
\subfigure
{\includegraphics[width=6cm]{newfinalPlots8/FullLegend\gray}}
    \caption{NN$_{\rm sig}$ discriminant output distributions in signal region, for the 1S subsample\,(a), 2S\,(b), SJ\,(c).}
     \label{fig:outFinalMLP}
\end{figure*}

We use the following variables to discriminate between signal and background processes: the invariant mass of $\vec{j_2}$ and $\vec{\met}$, $M(\vec{j_2},\vec{\met})$;
%the second jet in the $W \to \tau \nu$ events is often a $\tau$, thus this variable is the reconstructed $W$ transverse mass for this background events; 
$H_T$; $Min(\Delta \phi(\vec{\met},\vec{j_1}))$; $\sum \Pt^{chgd}/\Pt^{j_1}$; $\sum \Pt^{chgd}/\Pt^{j_2}$; $\met$; $\mpt$; 
%the difference in the azimuthal plane between the axis defined by the two most energetic jets in their rest frame, and the vector sum of the two jets in the lab frame,
$\phi^{*}$;
%the $\Delta \phi$ between the direction of the two $E_T$ leading jets in the two-jet rest frame and the direction of the boost; 
$\met/H_T$; $M(\vec{\met},\vec{j_1},\vec{j_2})$; the invariant mass of all jets, $M(alljets)$. 
\begin{table*}[hbtp]
\begin{center}
\caption{Input variables to the neural network aimed at discriminating single top quark production from the backgrounds remaining after the NN$_{\rm QCD} > -0.1$ requirement.}
\begin{tabular}{ll}
\hline\hline
Variable & Description \\
\hline
$M(\vec{\met},\vec{j_2})$ & Invariant mass of $\met$ and $\vec{j_2}$ \\
%\hline
%Scalar sum of jet $\Pt$'s, 
$\Ht$ & Scalar sum of the jet energies \\
%\hline
$Min(\Delta \phi(\vec{\met},\vec{j_i}))$ & Minimum of $\Delta \phi$ between $\vec{\met}$ and any jet $\vec{j_i}$ \\
%\hline
$\sum \Pt^{chgd}/\Pt^{j_1}$ & Fraction of $\Pt^{j_1}$ carried by charged particles displaced from the primary vertex\\
%\hline
$\sum \Pt^{chgd}/\Pt^{j_2}$ & Fraction of $\Pt^{j_2}$ carried by charged particles displaced from the primary vertex\\
%\hline
%Magnitude of $\vec{\met}$ \\
$\met$ & Missing transverse energy \\
%\hline
%Magnitude of $\vec{\mpt}$ \\
$\mpt$ & Missing transverse momentum \\
%\hline
$\phi^{*}$ & $\Delta \phi$ between the ($\vec{j^1},\vec{j^2}$) axis in their rest frame, and their vector sum in the lab frame \\
%$\Delta \phi(\vec{j_1},\vec{j_2})$ in the 2-jet rest frame \\
%\hline
$\met$/$H_T$ & Ratio between $\met$ and $H_T$ \\
%\hline
$M(\vec{\met},\vec{j_1},\vec{j_2})$ & Invariant mass of $\met$, $\vec{j_i}$ and $\vec{j_2}$ \\
%\hline
$M(all jets)$ & Invariant mass of all jets in the event\\
\hline\hline
\end{tabular}
\end{center}
\label{tab:nnvar}
\end{table*}
All the above observables are used as inputs to a multi-layer-perceptron neural network trained to distinguish the signal from backgrounds in the sample with NN$_{\rm QCD} > -0.1$.  We use the simulated single top quark $s$- and $t$-channel samples in their expected proportions to build the signal sample (approximately 50$\%$-50$\%$). For training purposes, we select the background processes which account for more than 5\% of the total background: multijet, $W\to \tau \nu$ plus heavy flavor jets, $Z\to \nu \nu$ plus heavy flavor jets, and $\ttbar$ production. Both training and test samples contain 39\,000 signal and 42\,000 background events. The network architecture consists of an input layer with 11 nodes corresponding to the input variables shown in Table\,\ref{tab:nnvar}, plus one bias node; one hidden layer with 22 nodes and one hidden layer with 11 nodes, and an output layer with one output node, which we label NN$_{\rm sig}$. We compare the NN$_{\rm sig}$ output distribution between the training and testing samples and find good agreement.
\begin{figure*}[htbp]
\centering
\subfigure
%[Invariant mass of $\met$ and 2nd jet]
{\includegraphics[width=.32\linewidth]{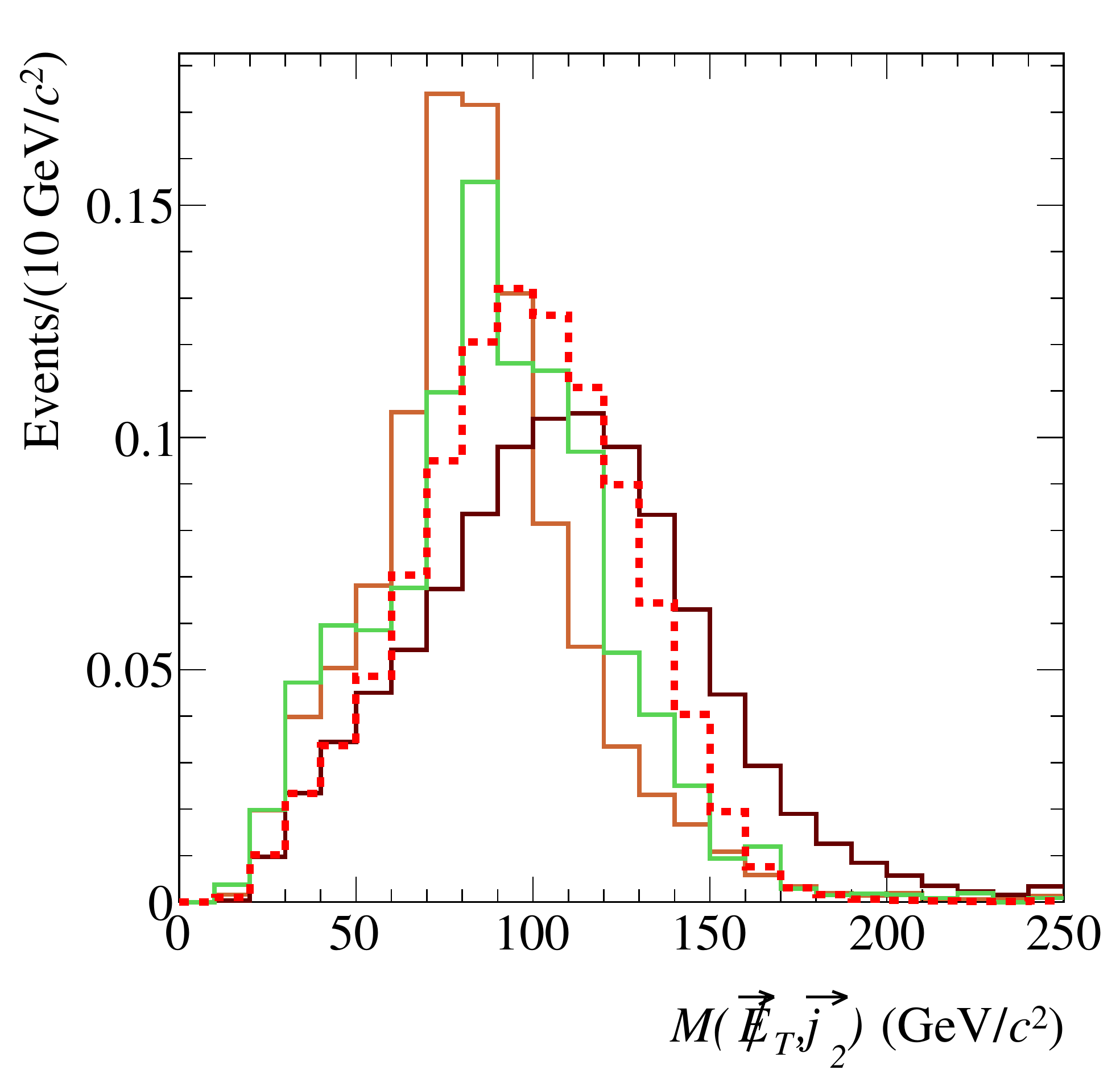}}
\subfigure
%[Scalar sum of jet $\Pt$'s, $\f$]
{\includegraphics[width=.32\linewidth]{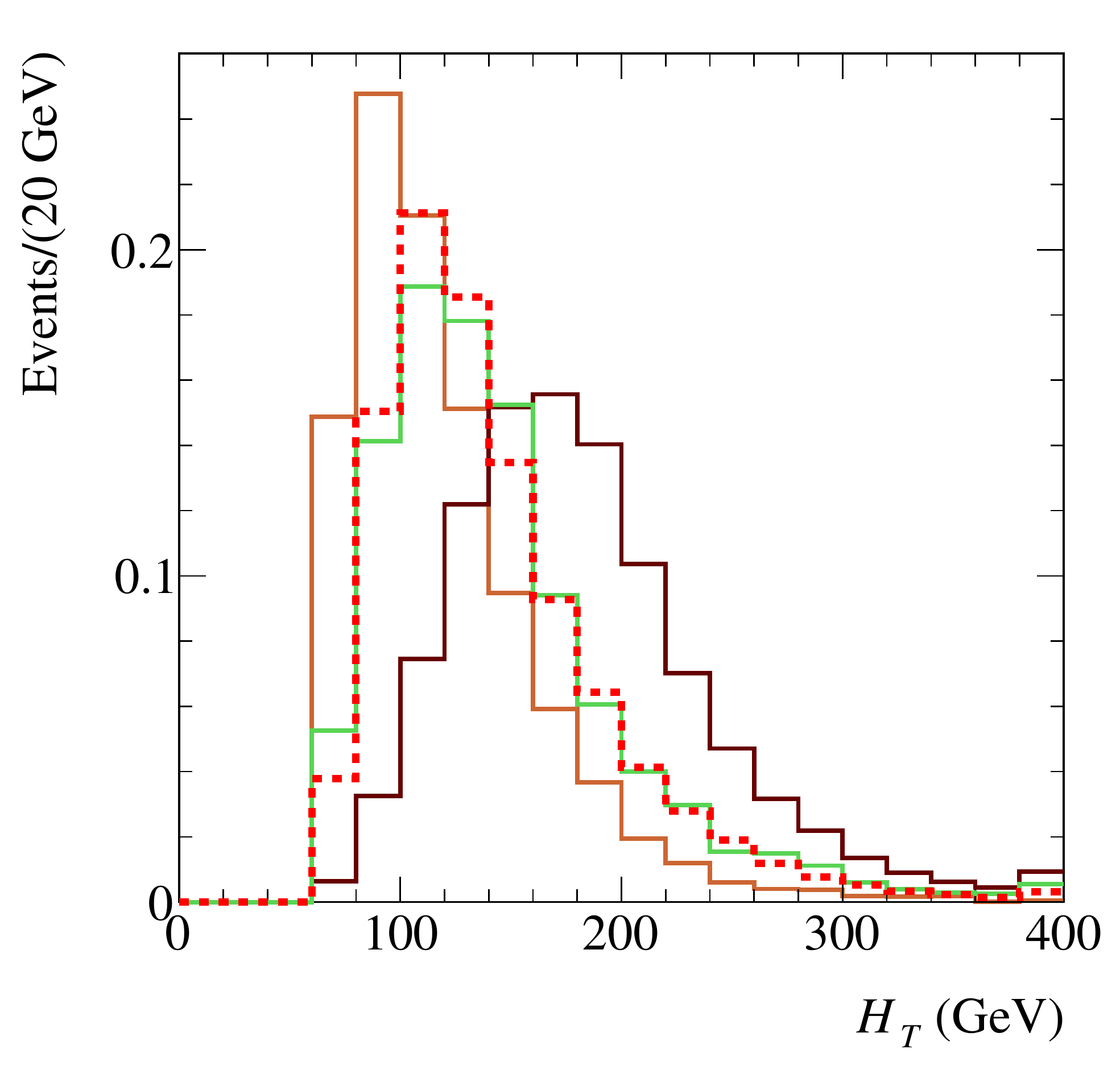}}
\subfigure
%[Min($\Delta \phi(\met,jet_i)$)]
{\includegraphics[width=.32\linewidth]{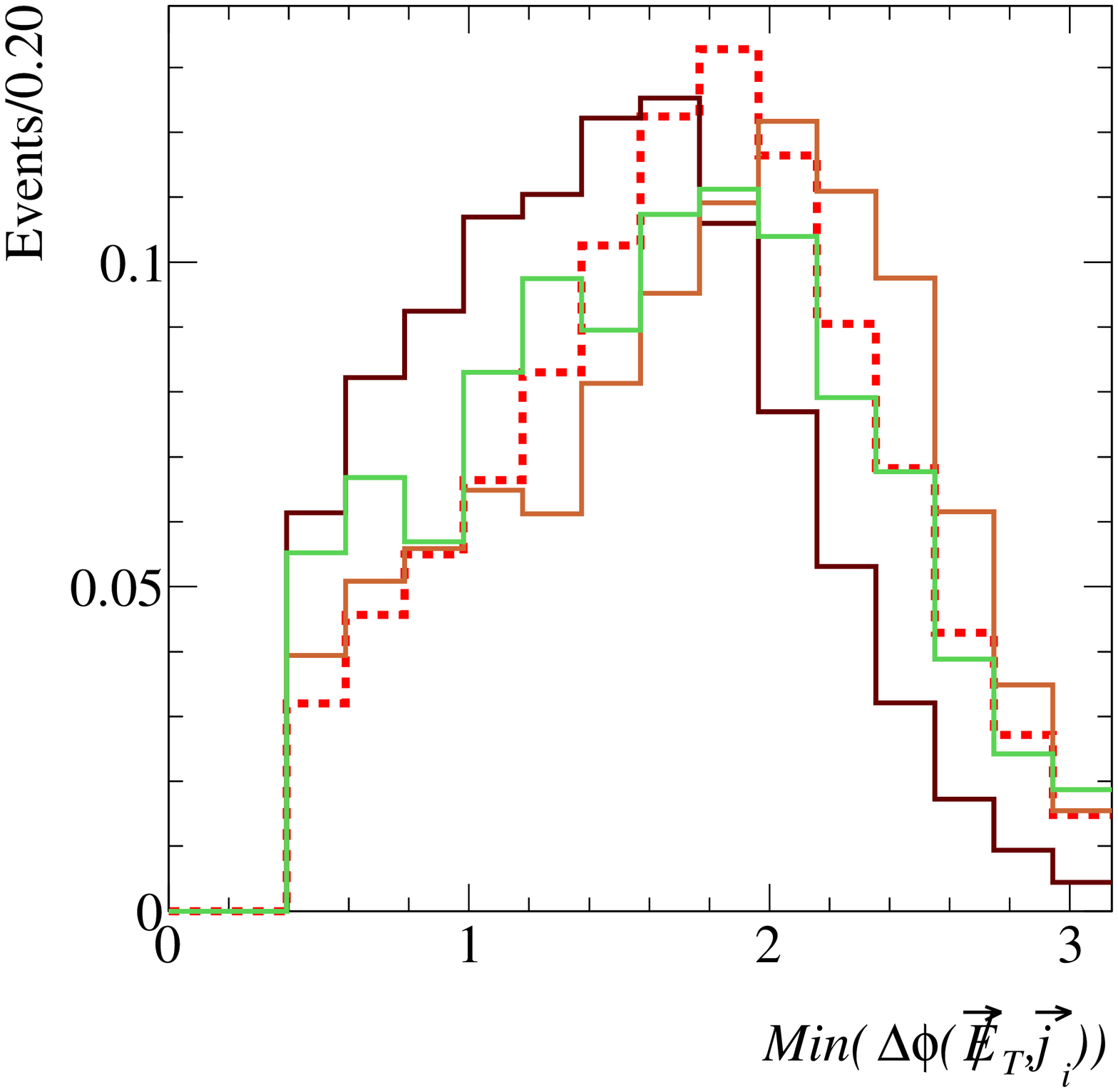}}
\subfigure
%[$\sum \Pt^{chgd}/\Pt^{jet}$ for the leading jet]
{\includegraphics[width=.32\linewidth]{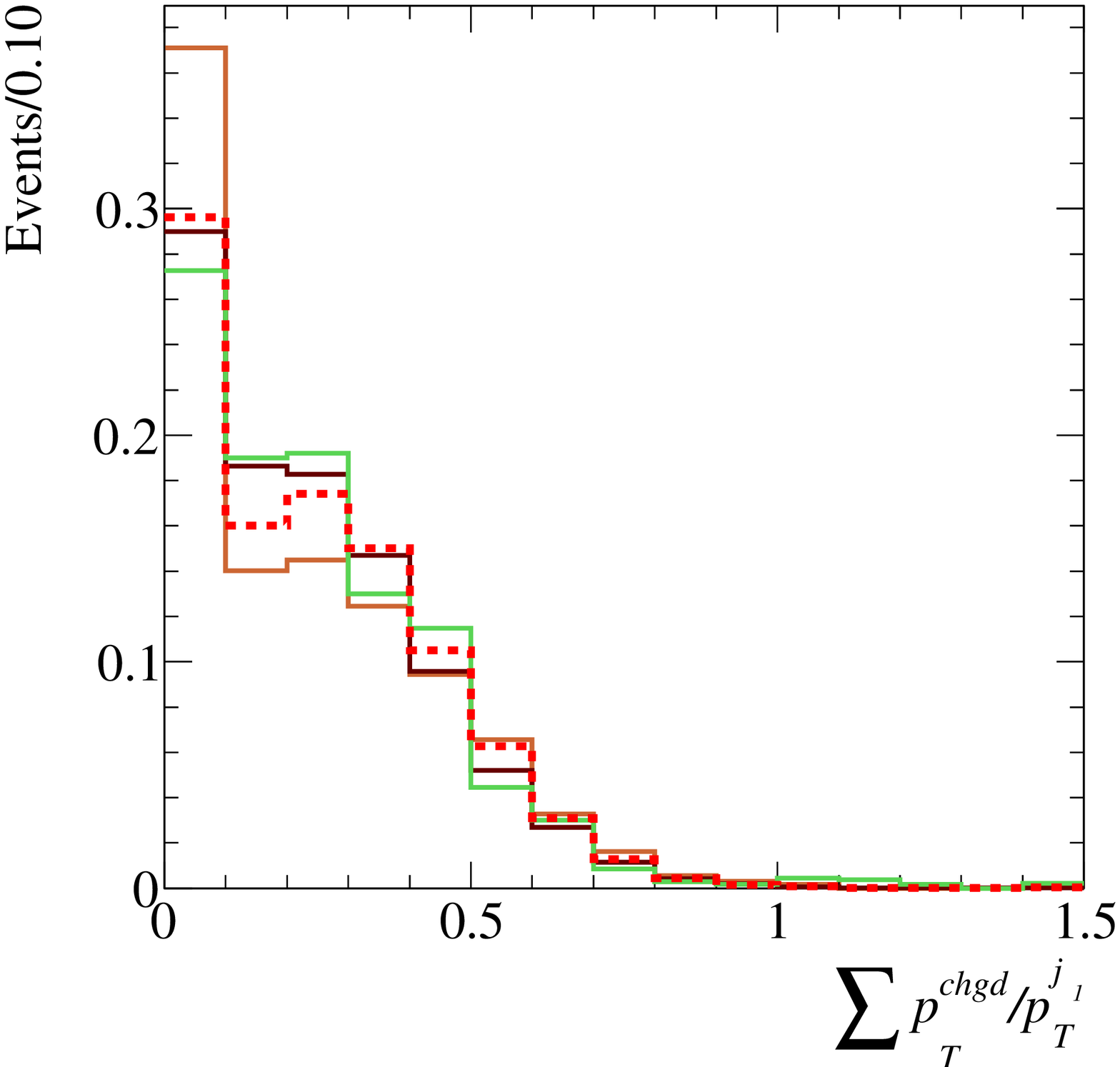}}
\subfigure
%[$\sum \Pt^{chgd}/\Pt^{jet}$ for the leading jet]
{\includegraphics[width=.32\linewidth]{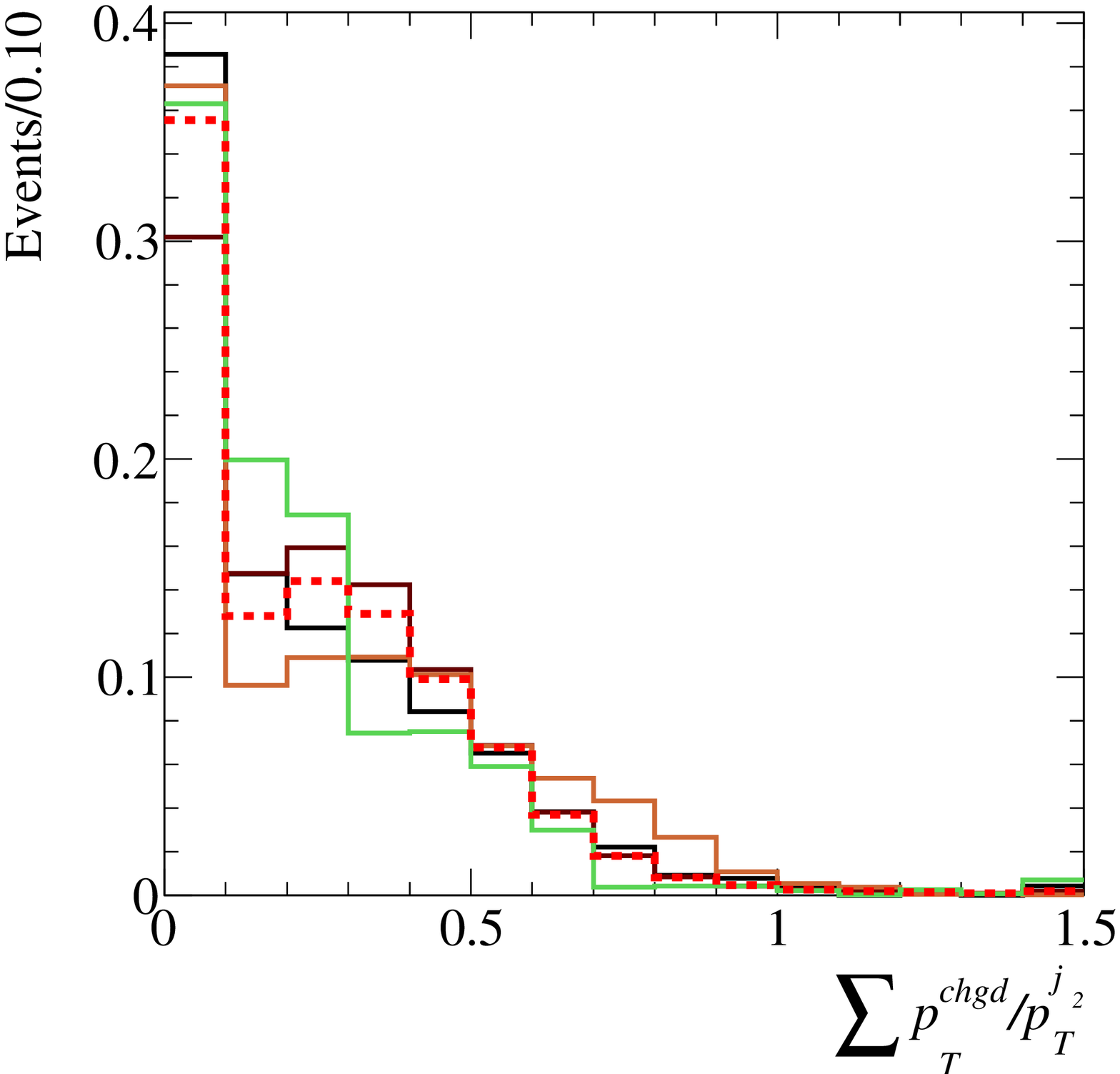}}
\subfigure
%[$\met$]
{\includegraphics[width=.32\linewidth]{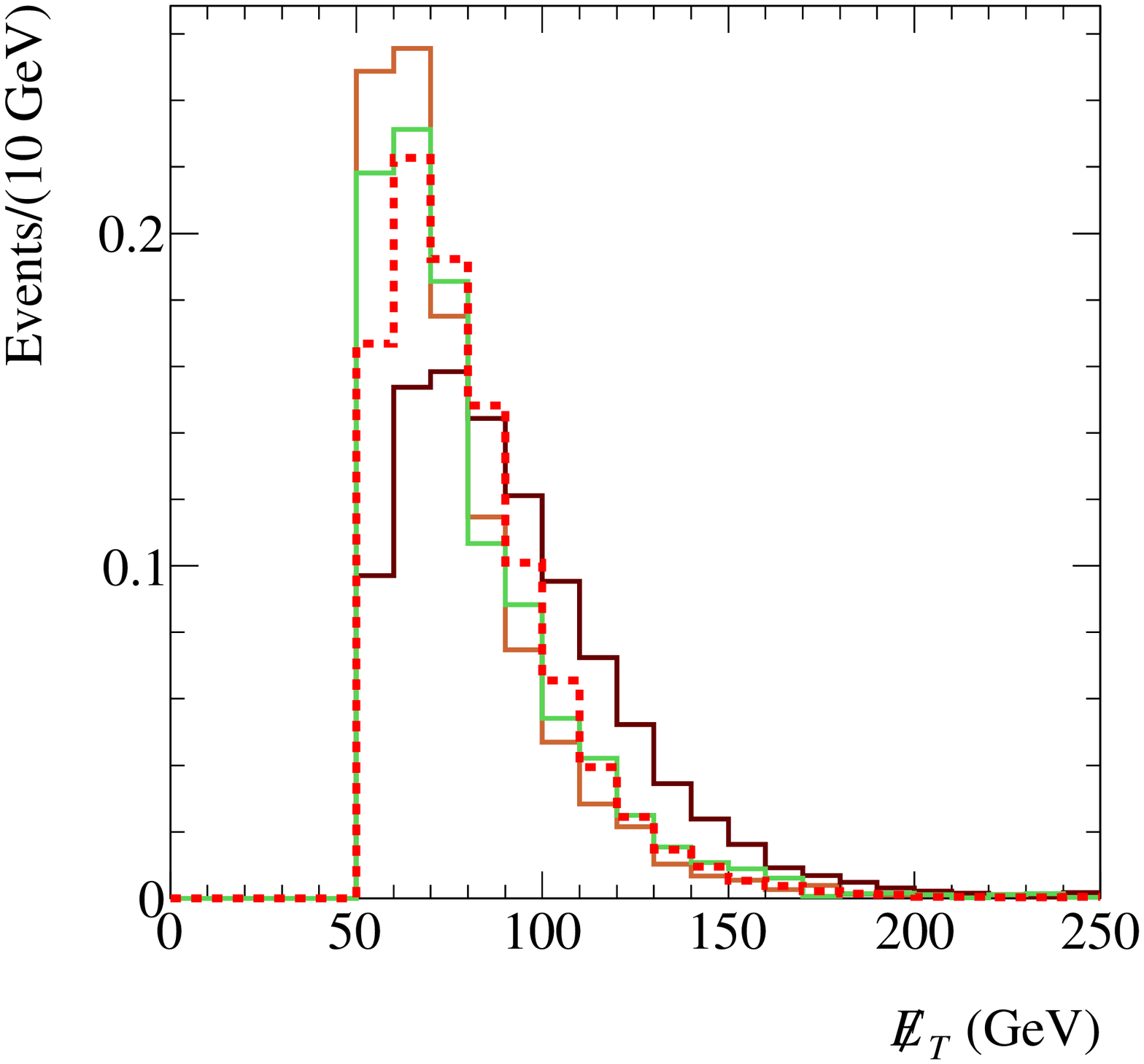}}
\subfigure
%[$\mpt$]
{\includegraphics[width=.32\linewidth]{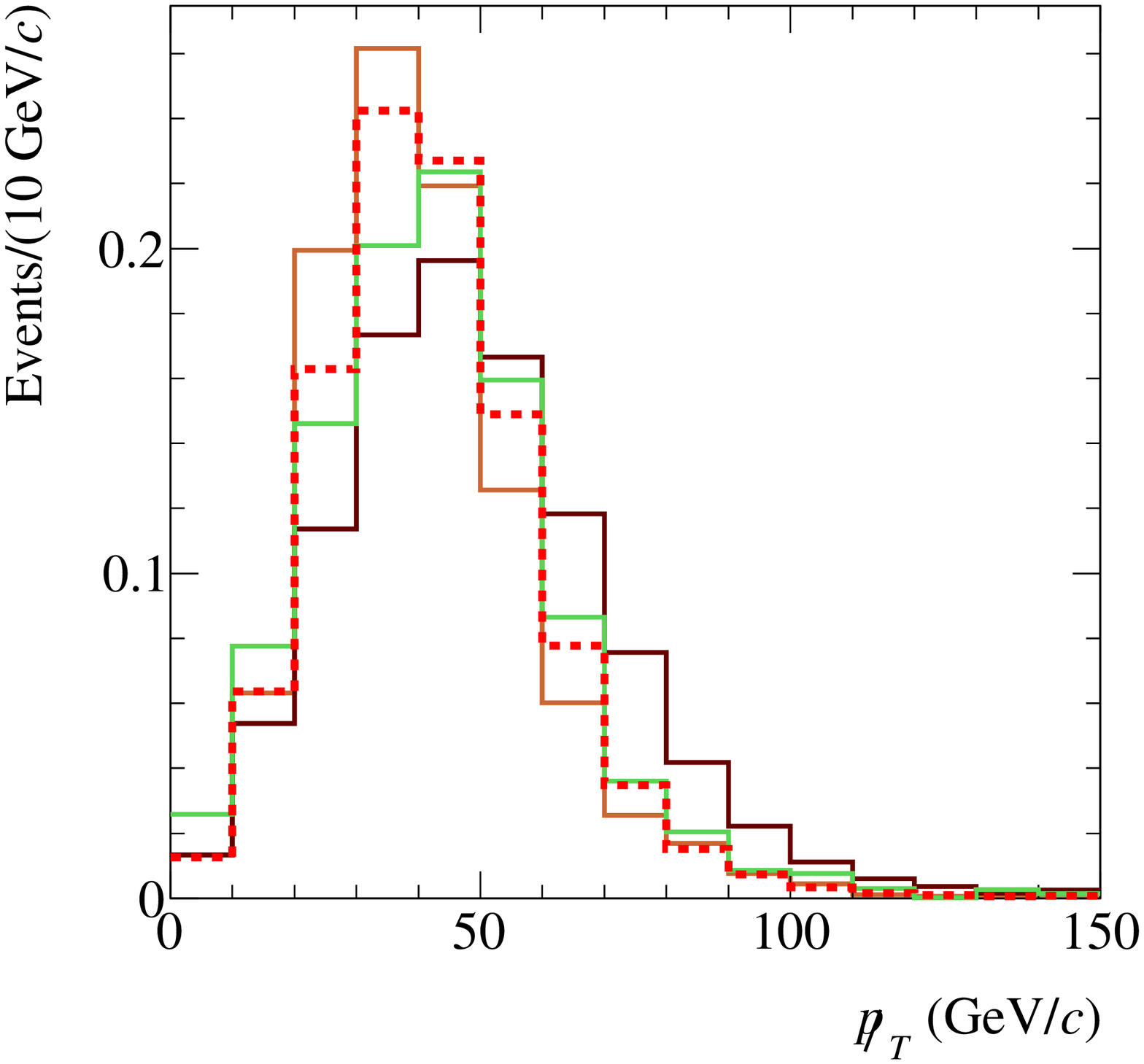}}
\subfigure
%[$\Delta \phi(jet_1,jet_2)$ with respect to the dijet system direction]
{\includegraphics[width=.32\linewidth]{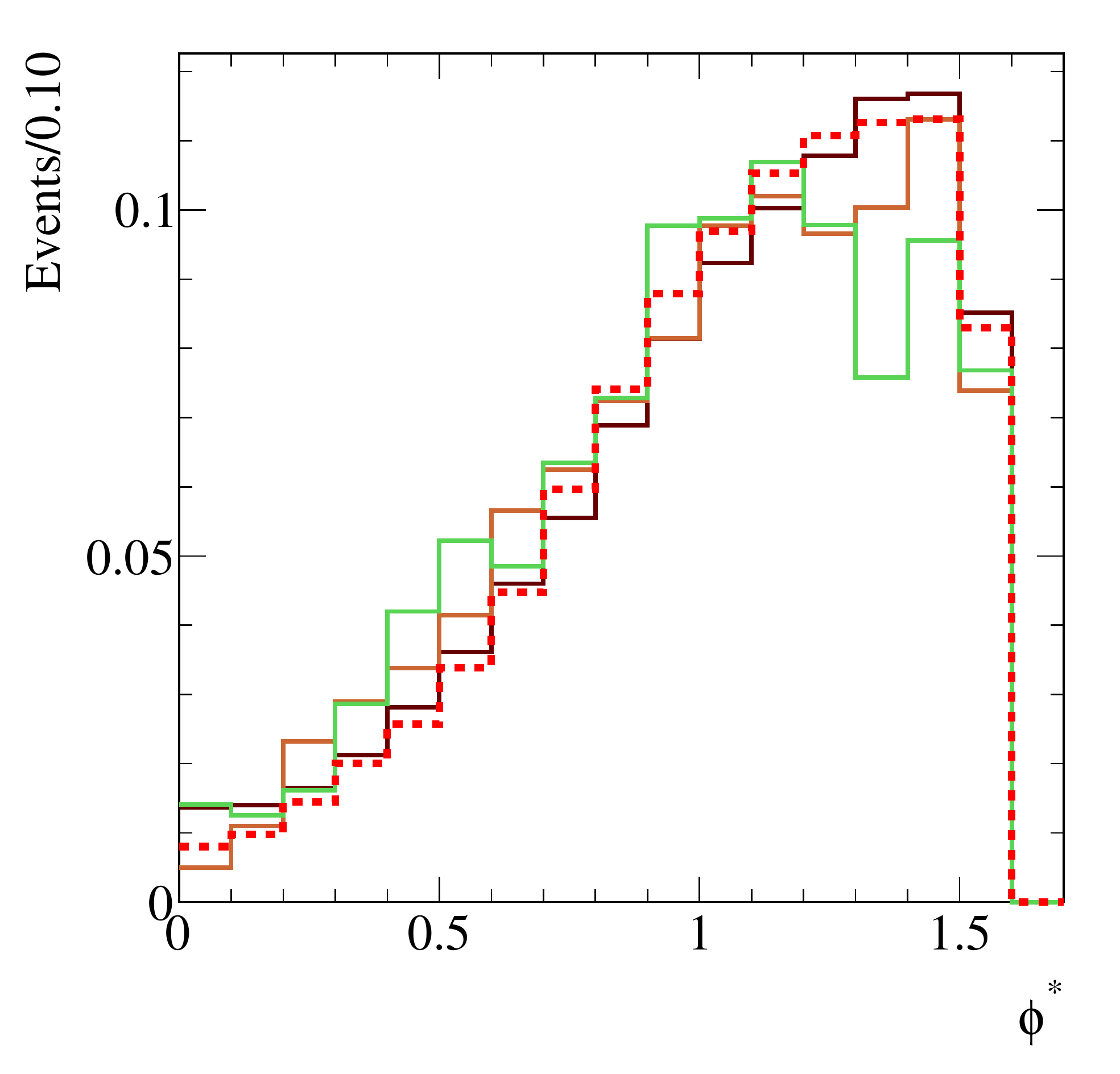}}
\subfigure
%[$\mht/\met$]
{\includegraphics[width=.32\linewidth]{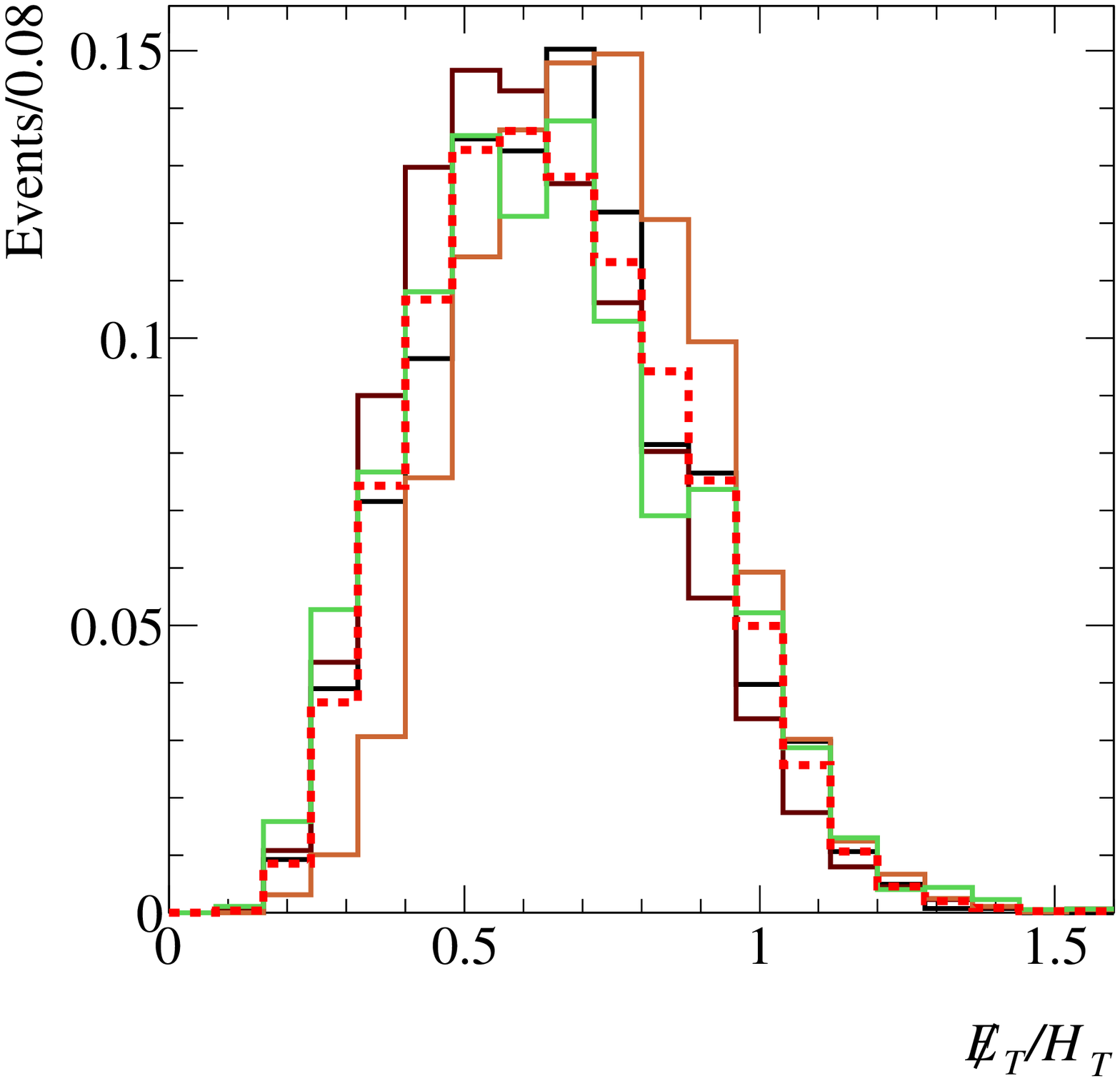}}
\subfigure
%[Invariant mass of $\met$ , $jet_1$ and $jet_2$]
{\includegraphics[width=.32\linewidth]{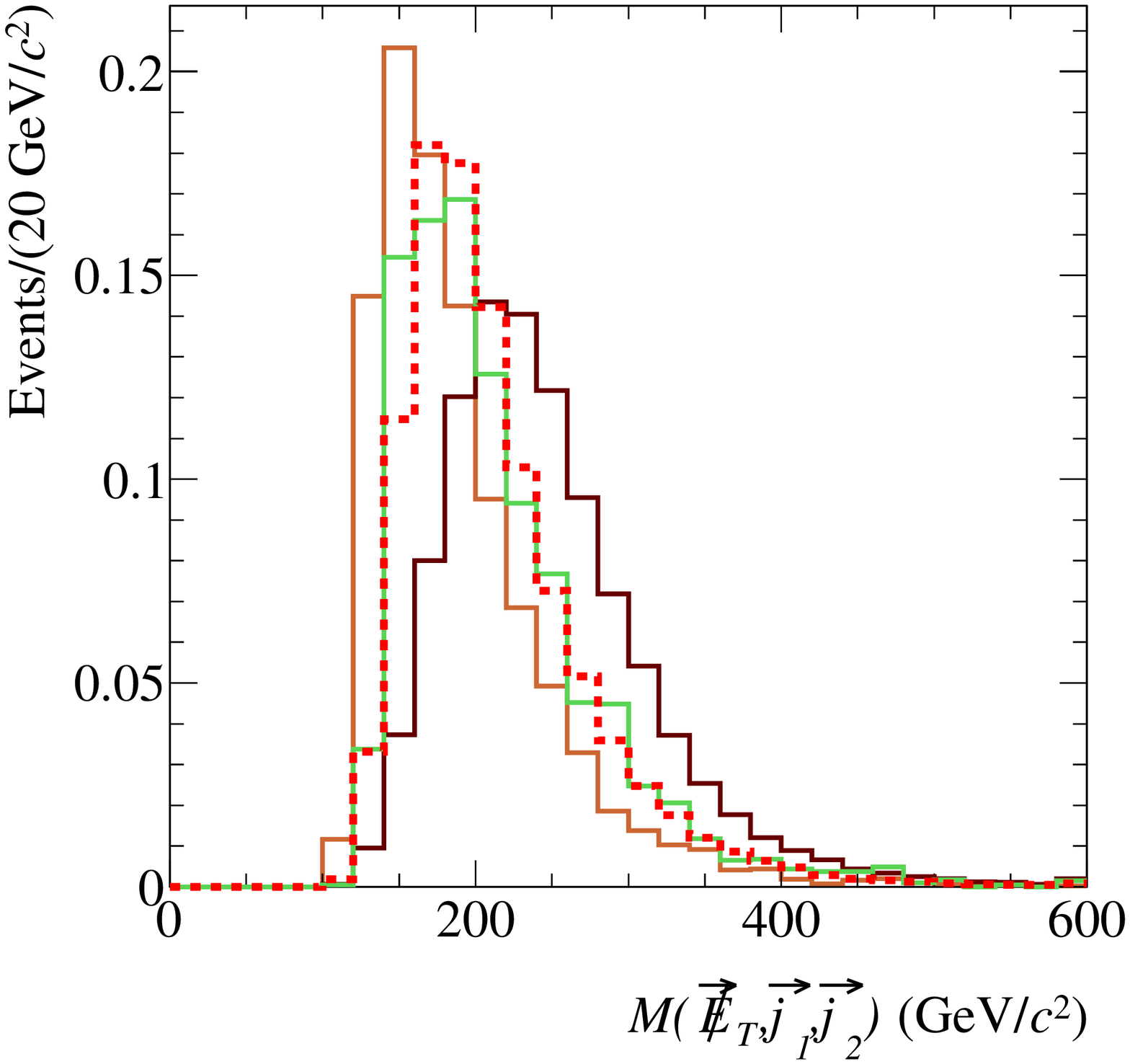}}
\subfigure
%[Invariant mass of all jets in the event]
{\includegraphics[width=.32\linewidth]{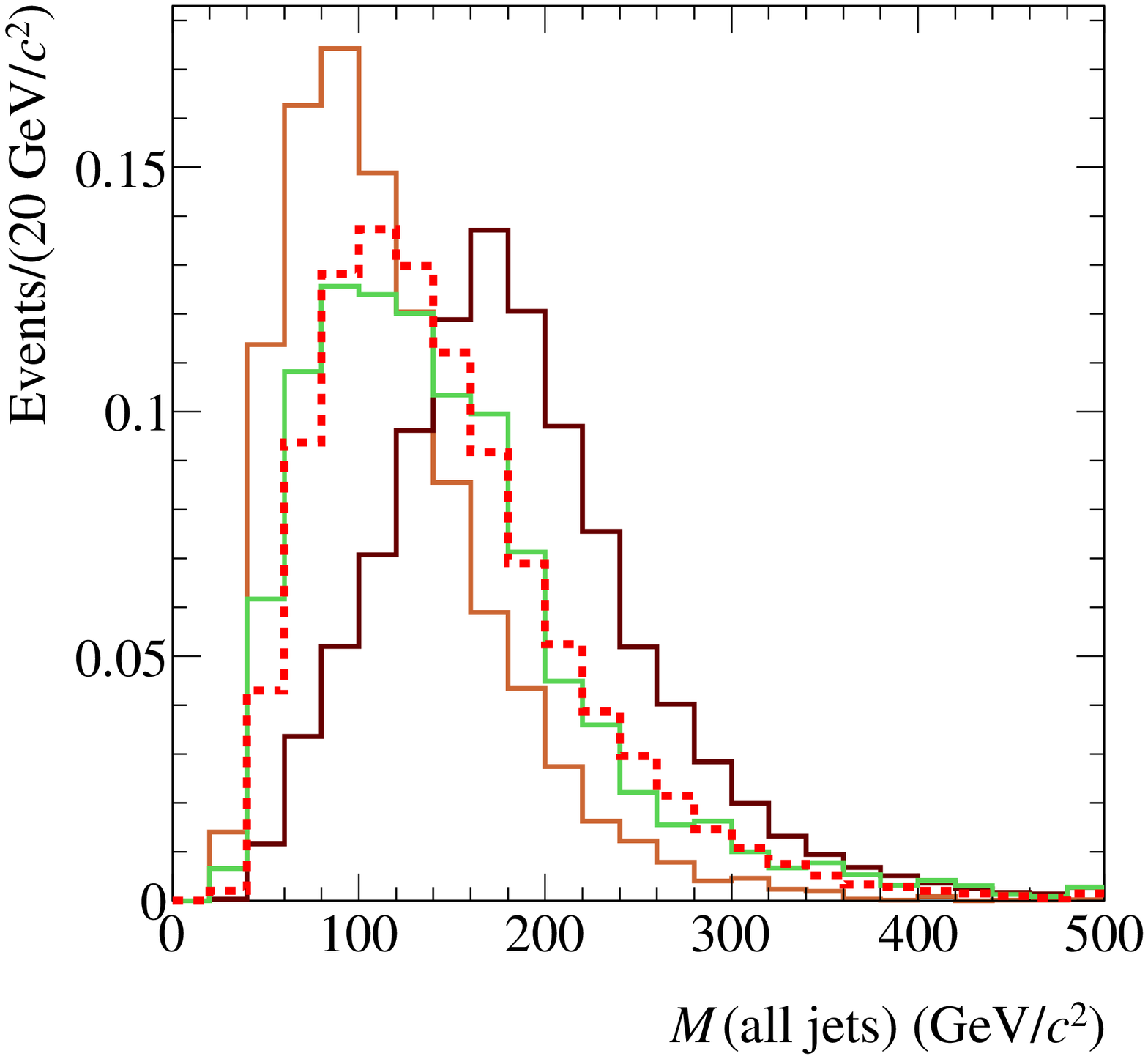}}
\subfigure
{\includegraphics[width=.32\linewidth]{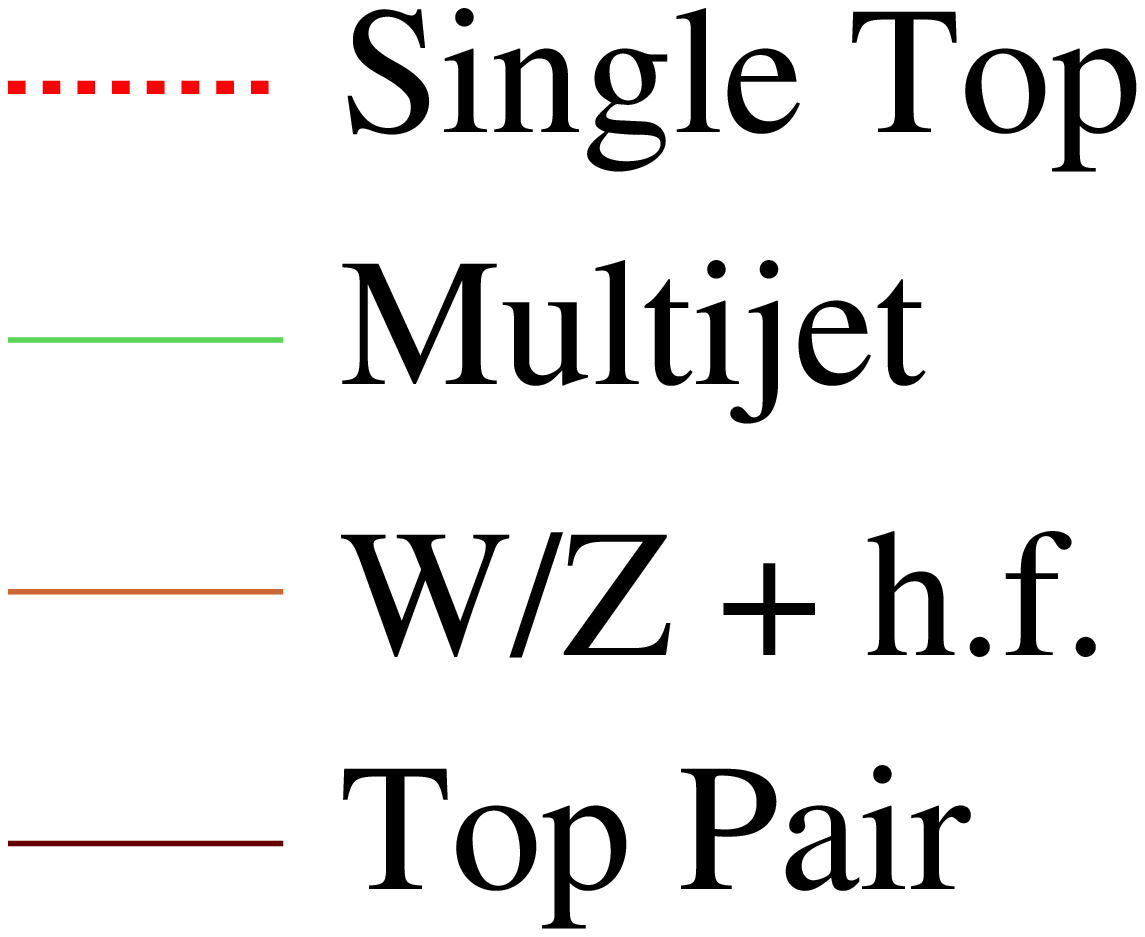}}
\caption{Kinematic distributions for the signal and background events in the signal region (NN$_{\rm QCD} > -0.1$). The three subsamples are summed together in their respective proportions. All histograms are normalized to unit area.} 
\label{fig:fNNinputs_1}
\end{figure*}
The distributions of the input variables for events in the signal region, for all $b-$tagged subsamples, are shown in Fig.\,\ref{fig:fNNinputs_1}, where the shapes of the distributions for each group of physics processes are compared. 
\begin{figure*}[htbp]
\centering
\subfigure
%[Invariant mass of $\met$ and 2nd jet]
{\includegraphics[width=.32\linewidth]{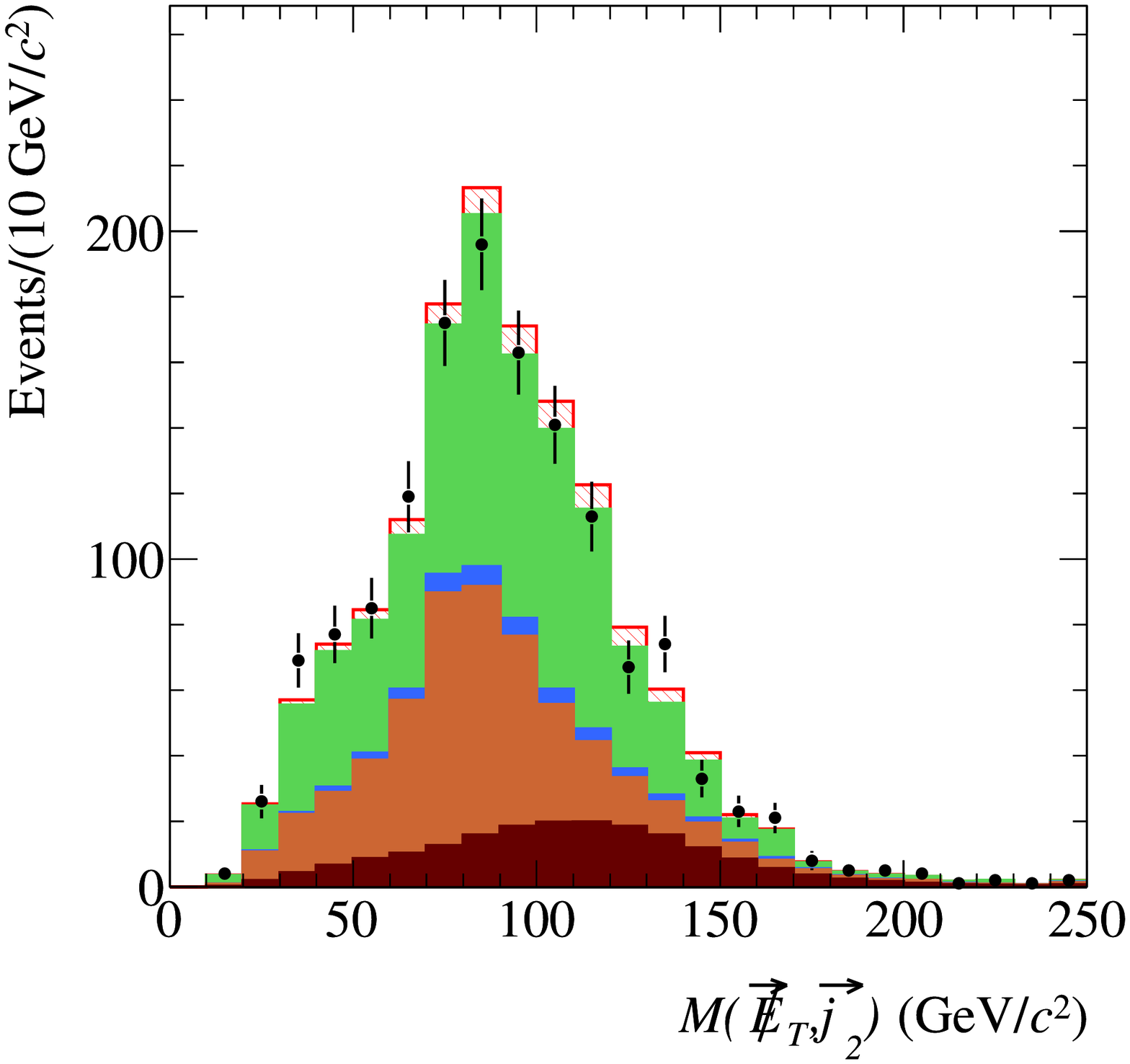}}
\subfigure
%[Scalar sum of jet $\Pt$'s, $\Ht$]
{\includegraphics[width=.32\linewidth]{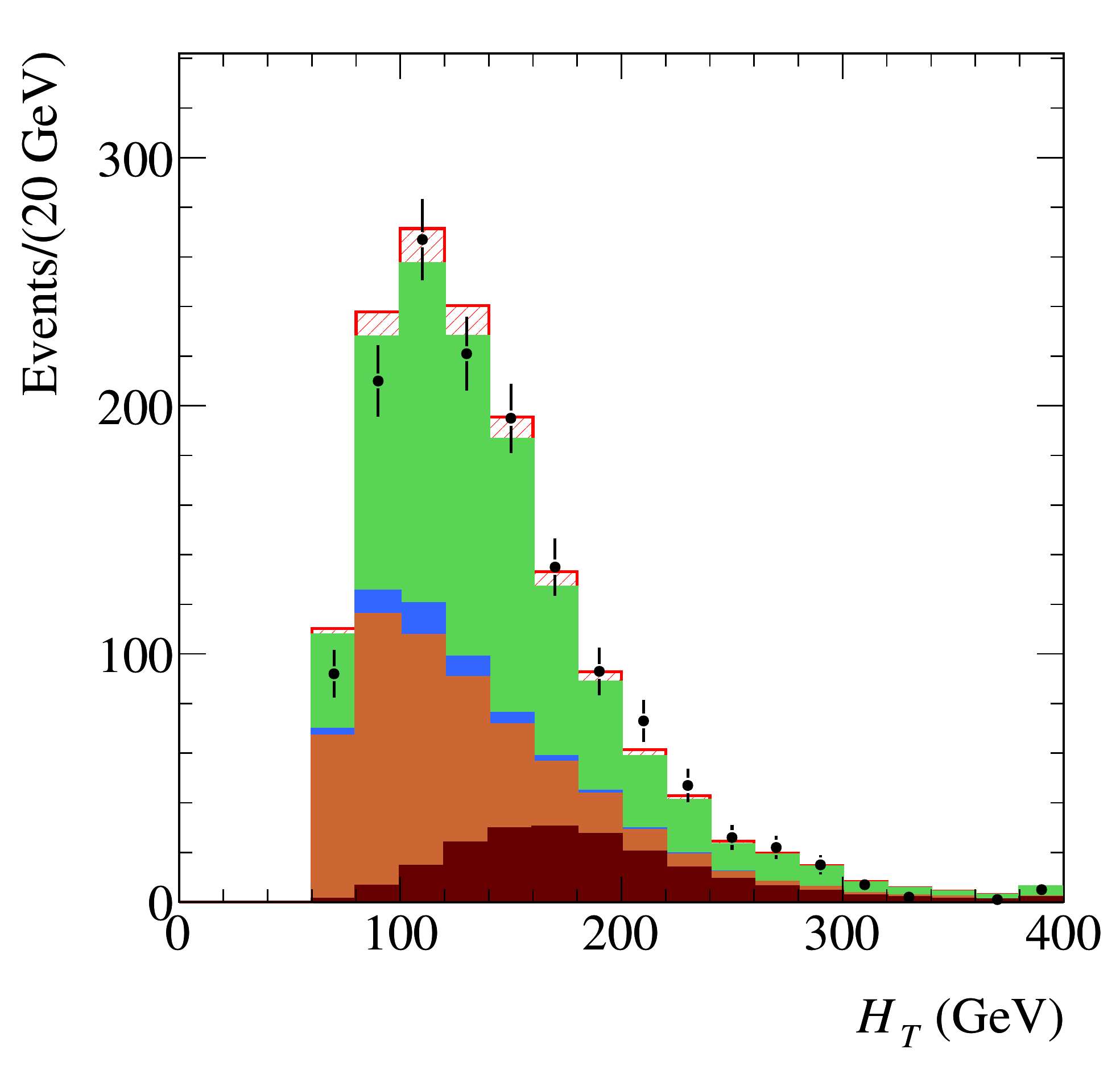}}
\subfigure
%[Min($\Delta \phi(\met,jet_i)$)]
{\includegraphics[width=.32\linewidth]{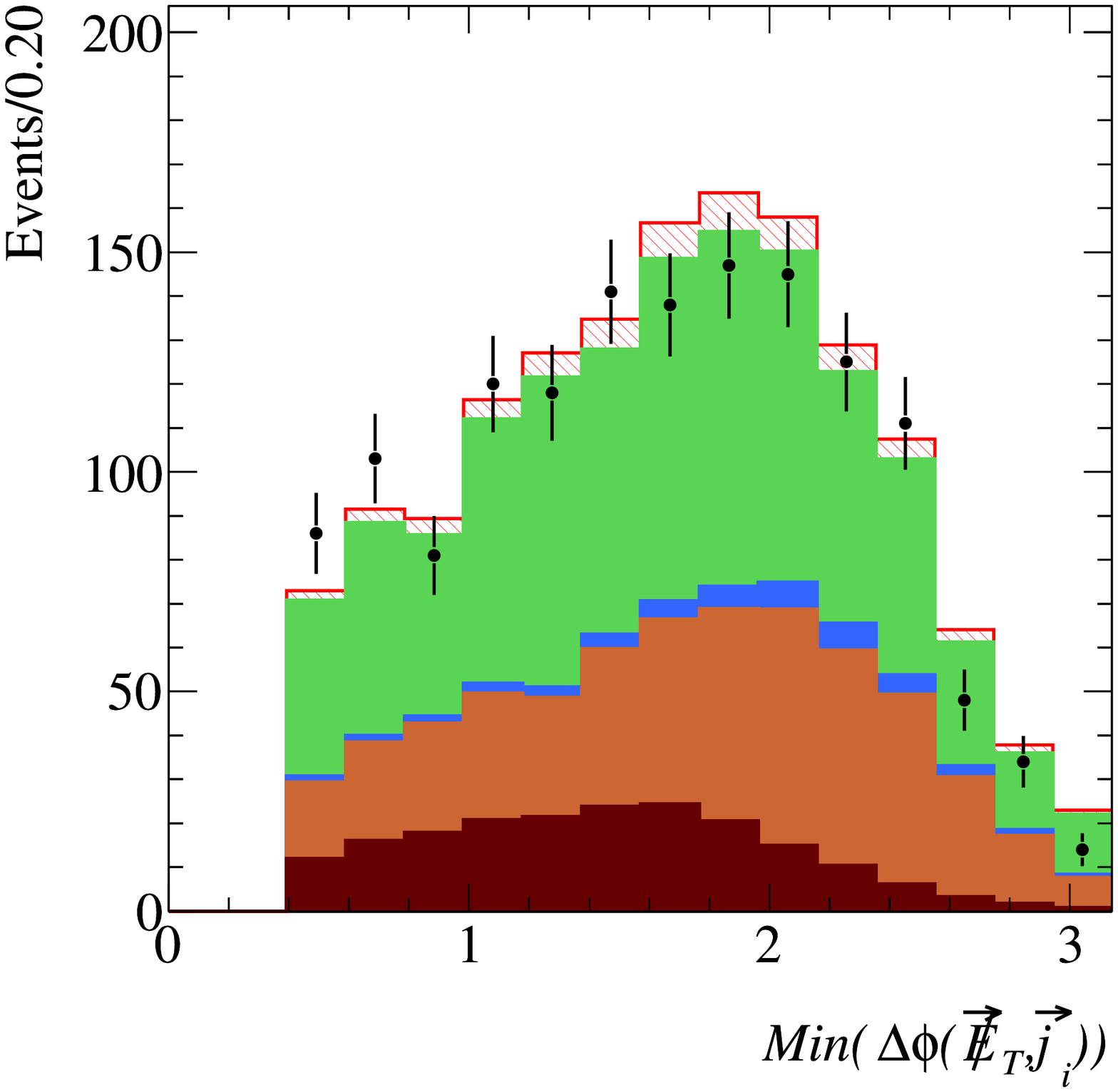}}
\subfigure
%[$\sum \Pt^{chgd}/\Pt^{jet}$ for the leading jet]
{\includegraphics[width=.32\linewidth]{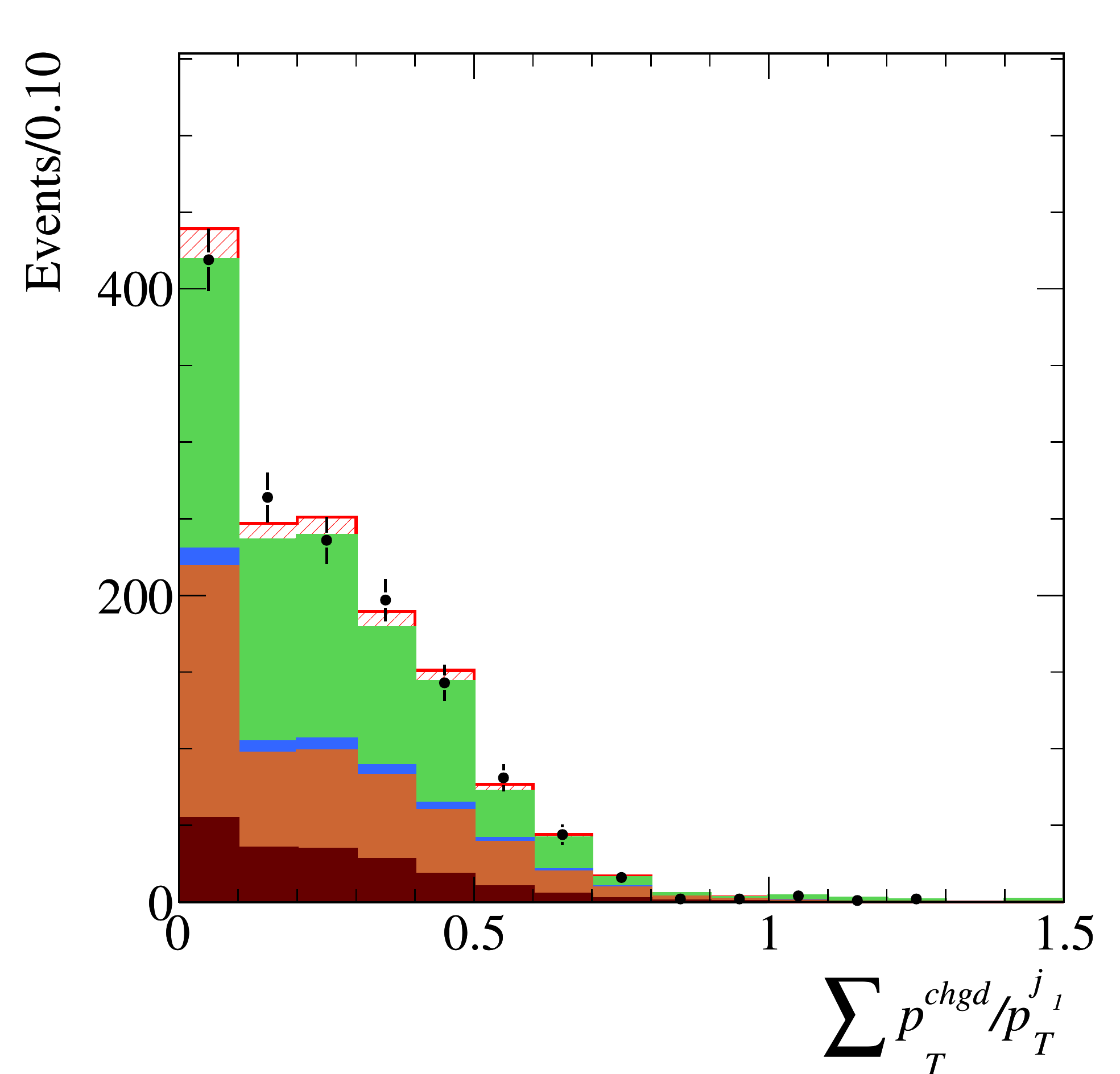}}
\subfigure
%[$\sum \Pt^{chgd}/\Pt^{jet}$ for the leading jet]
{\includegraphics[width=.32\linewidth]{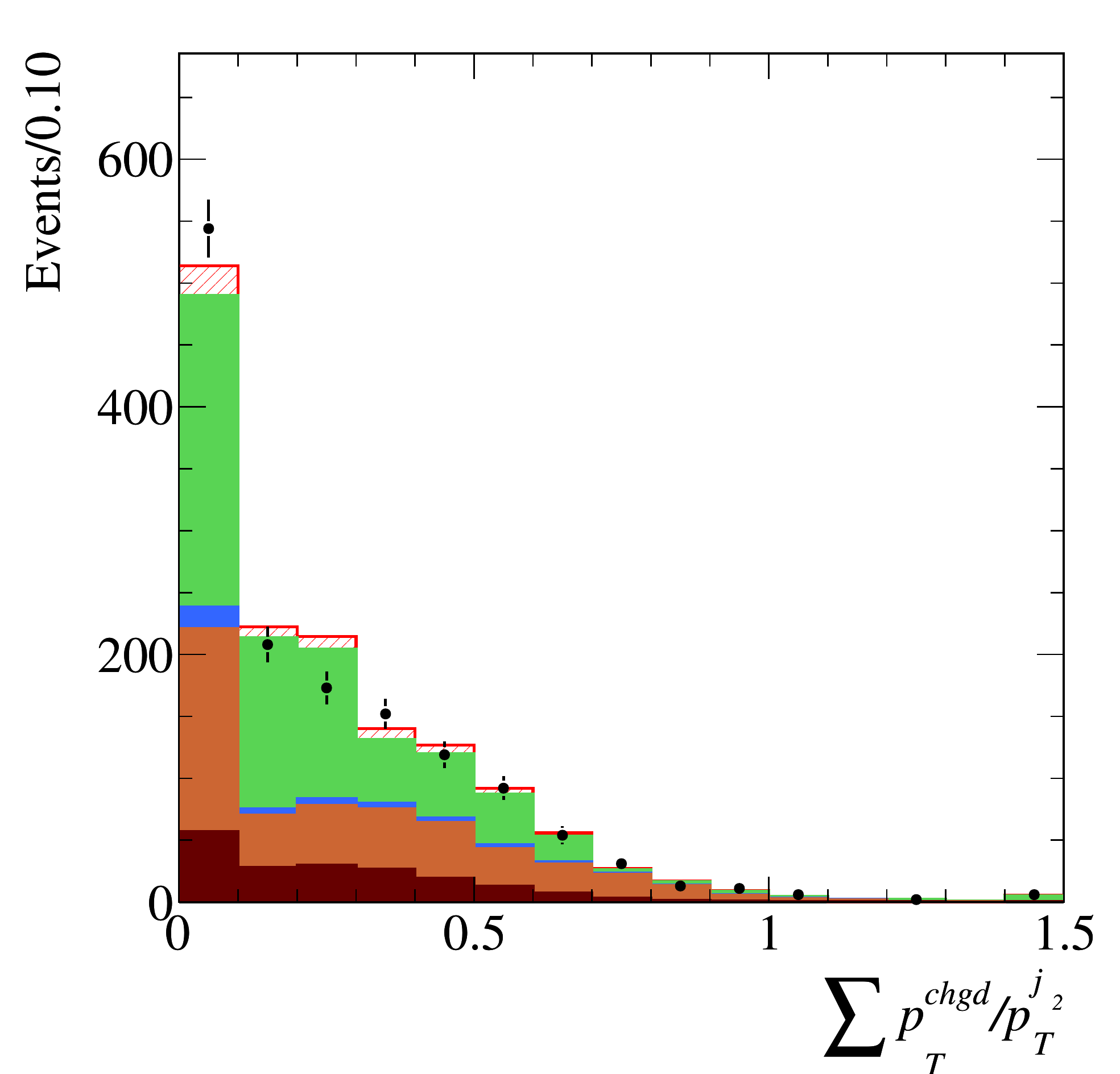}}
\subfigure
%[$\met$]
{\includegraphics[width=.32\linewidth]{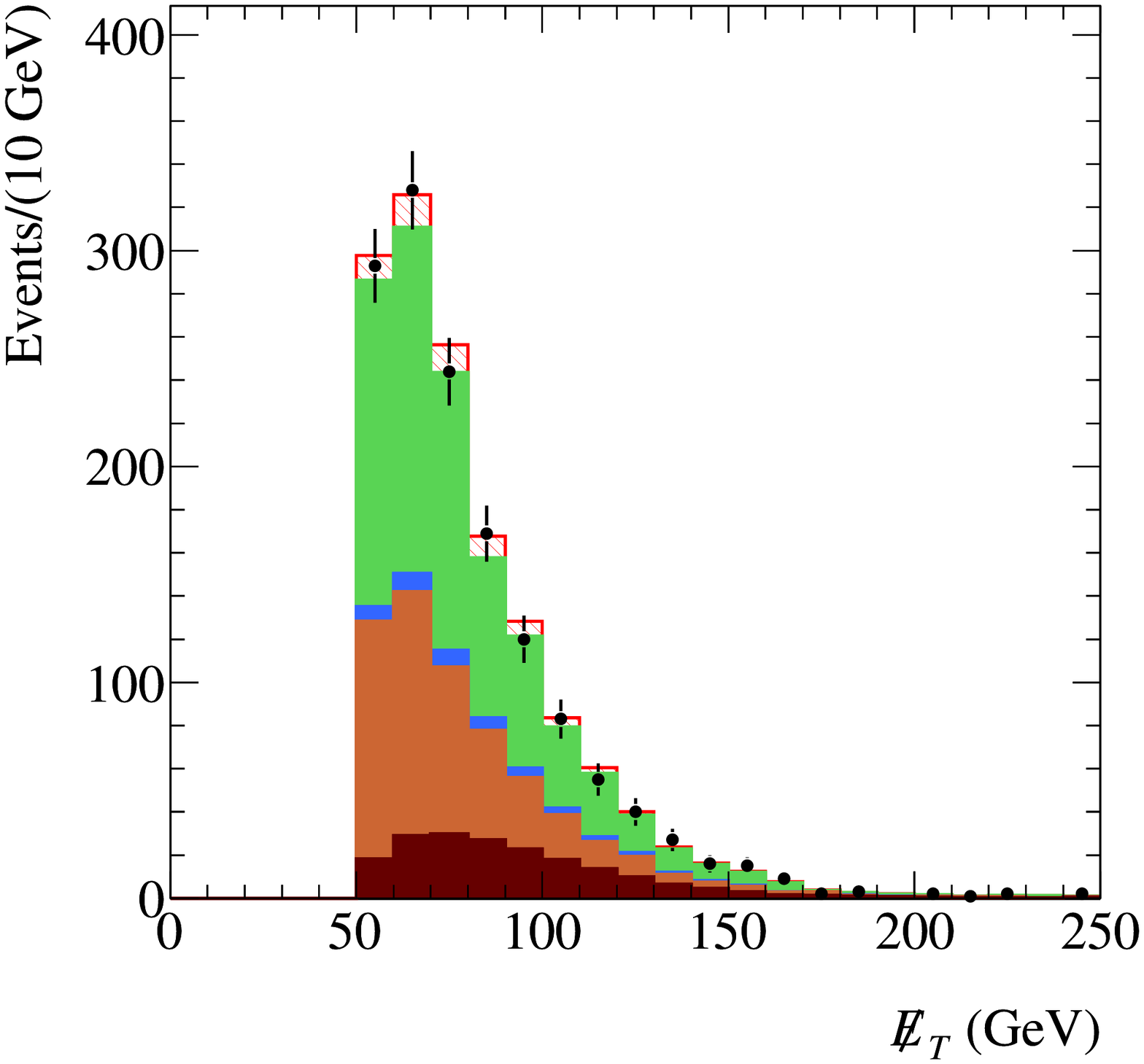}}
\subfigure
%[$\mpt$]
{\includegraphics[width=.32\linewidth]{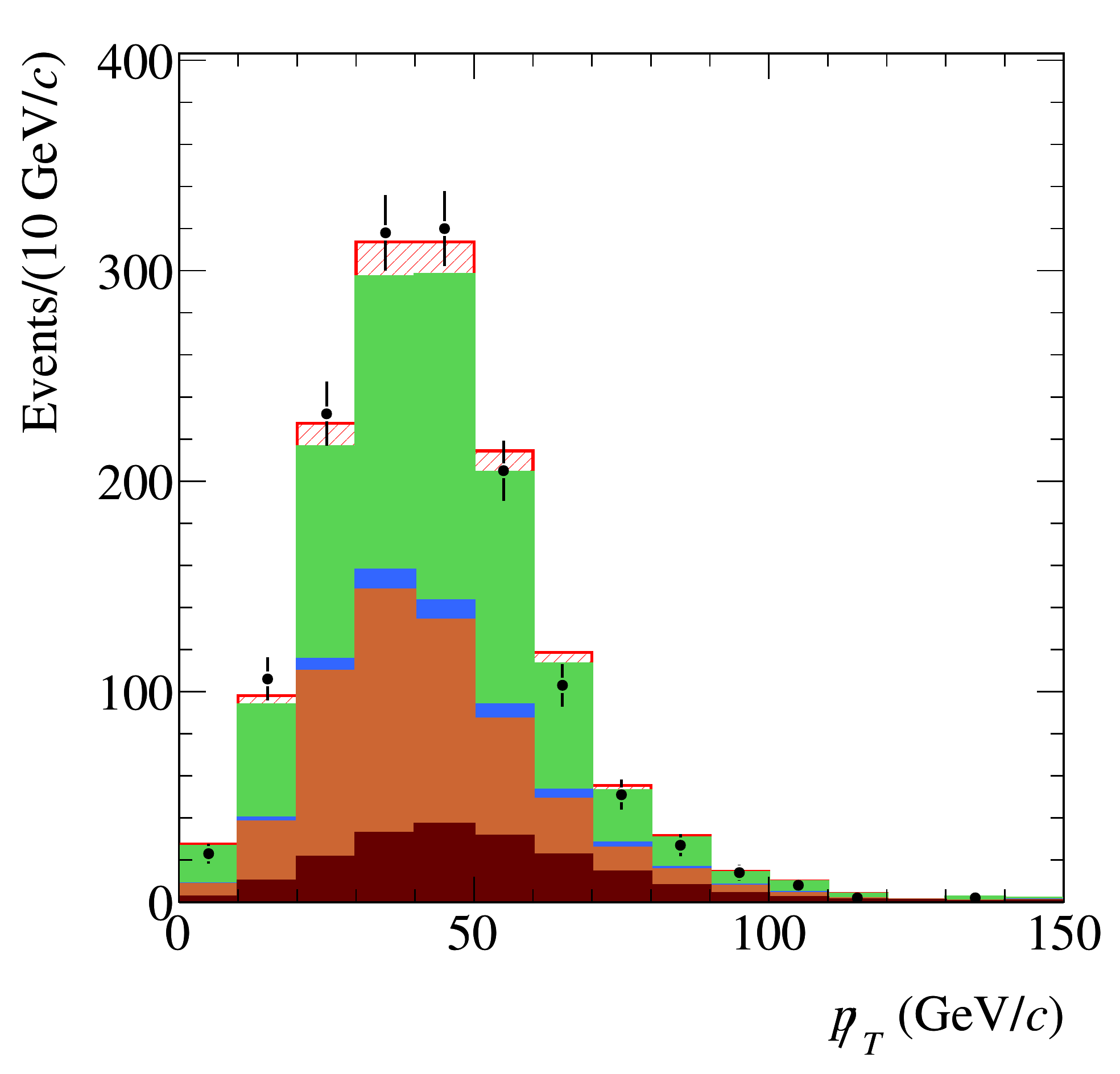}}
\subfigure
%[$\Delta \phi(jet_1,jet_2)$ with respect to the dijet system direction]
{\includegraphics[width=.32\linewidth]{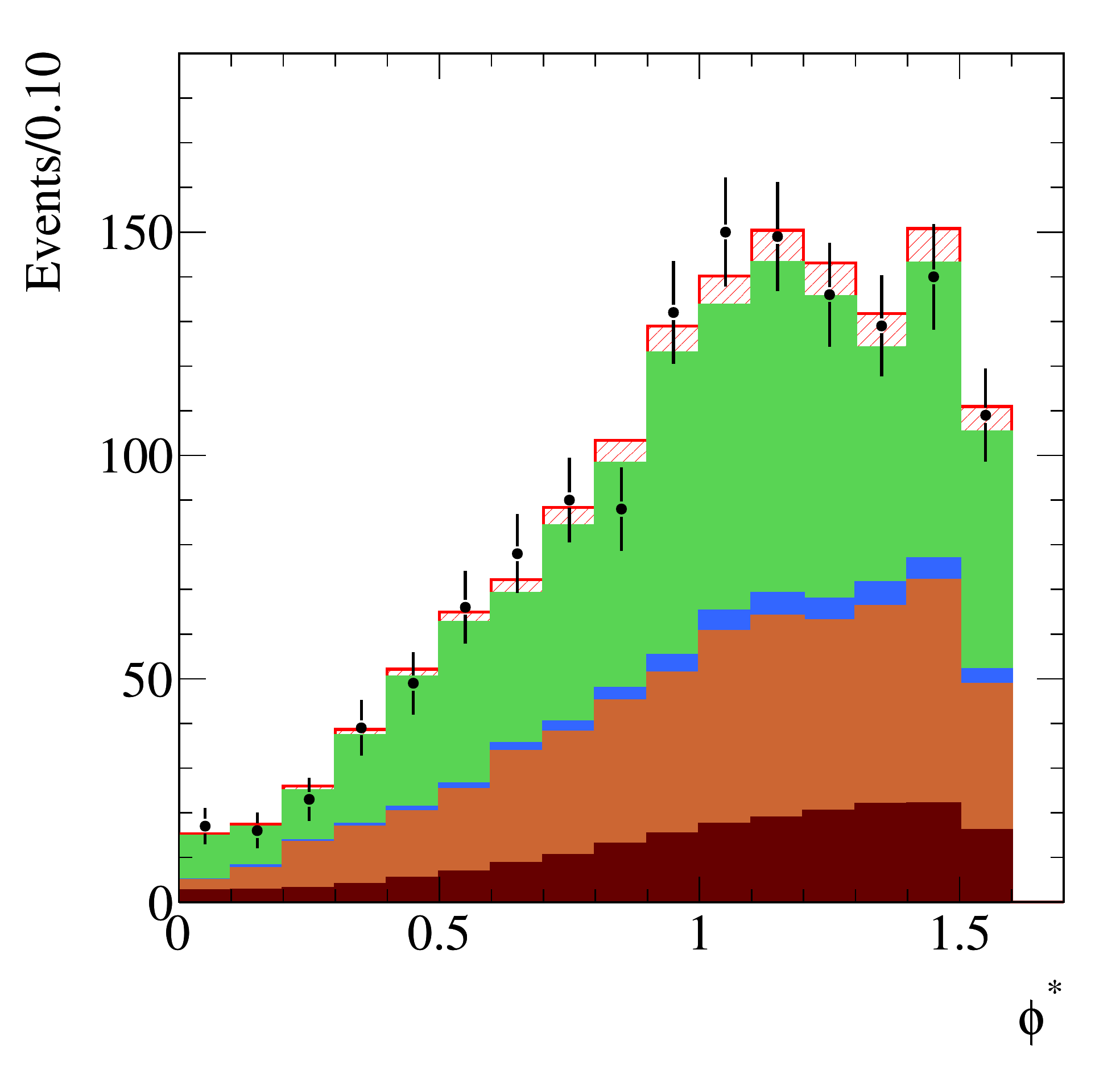}}
\subfigure
%[$\mht/\met$]
{\includegraphics[width=.32\linewidth]{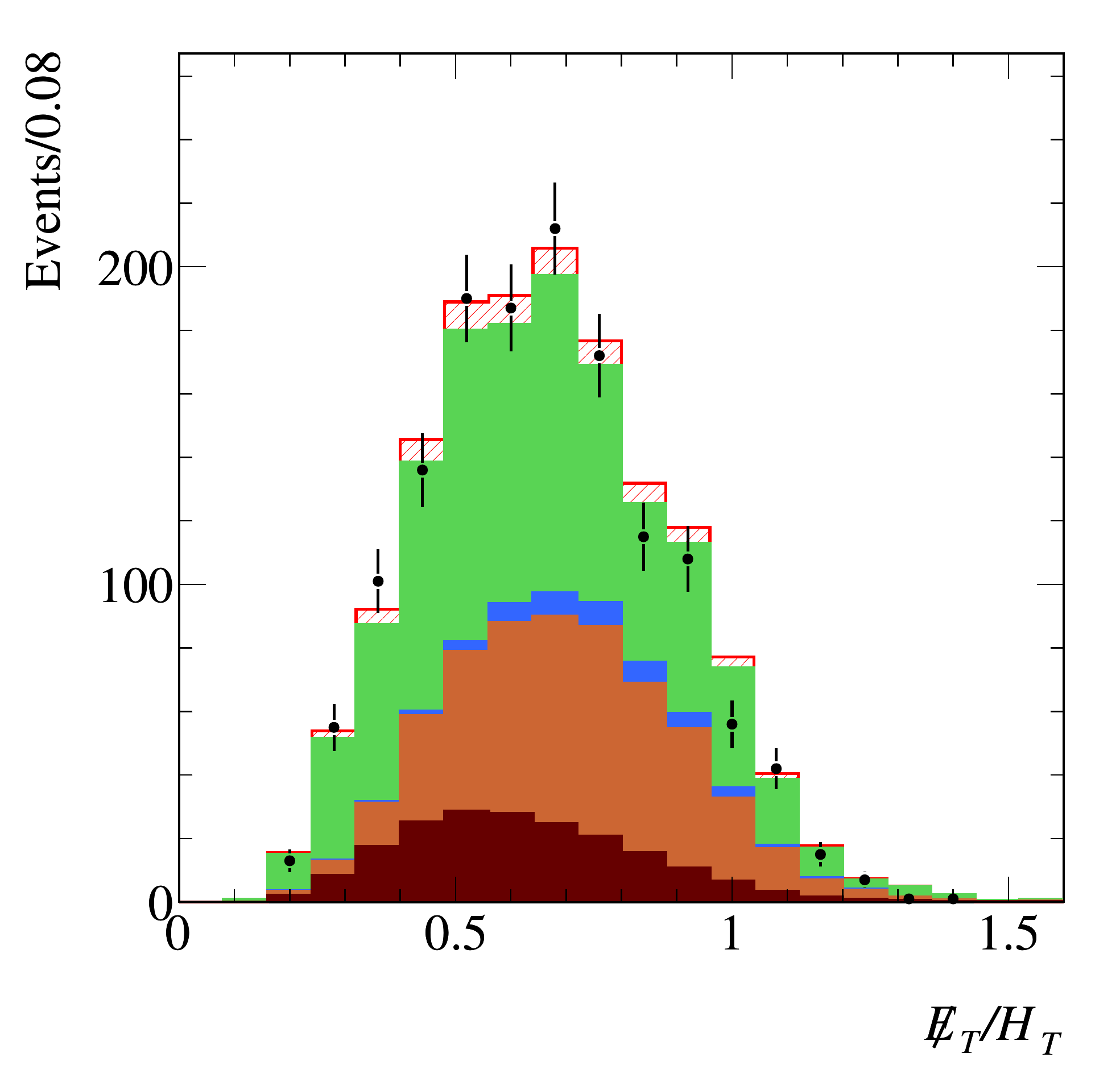}}
\subfigure
%[Invariant mass of $\met$ , $jet_1$ and $jet_2$]
{\includegraphics[width=.32\linewidth]{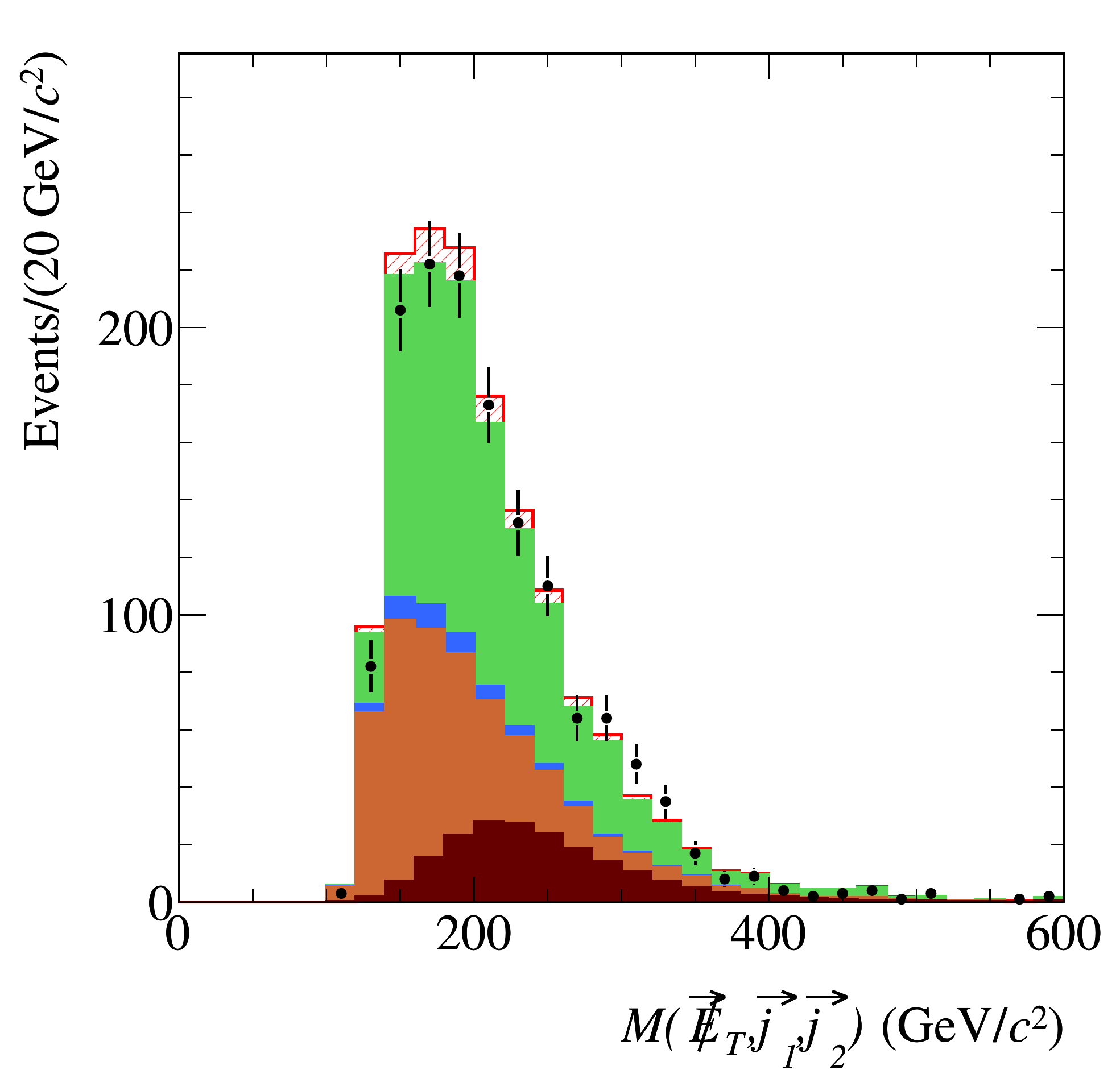}}
\subfigure
%[Invariant mass of all jets in the event]
{\includegraphics[width=.32\linewidth]{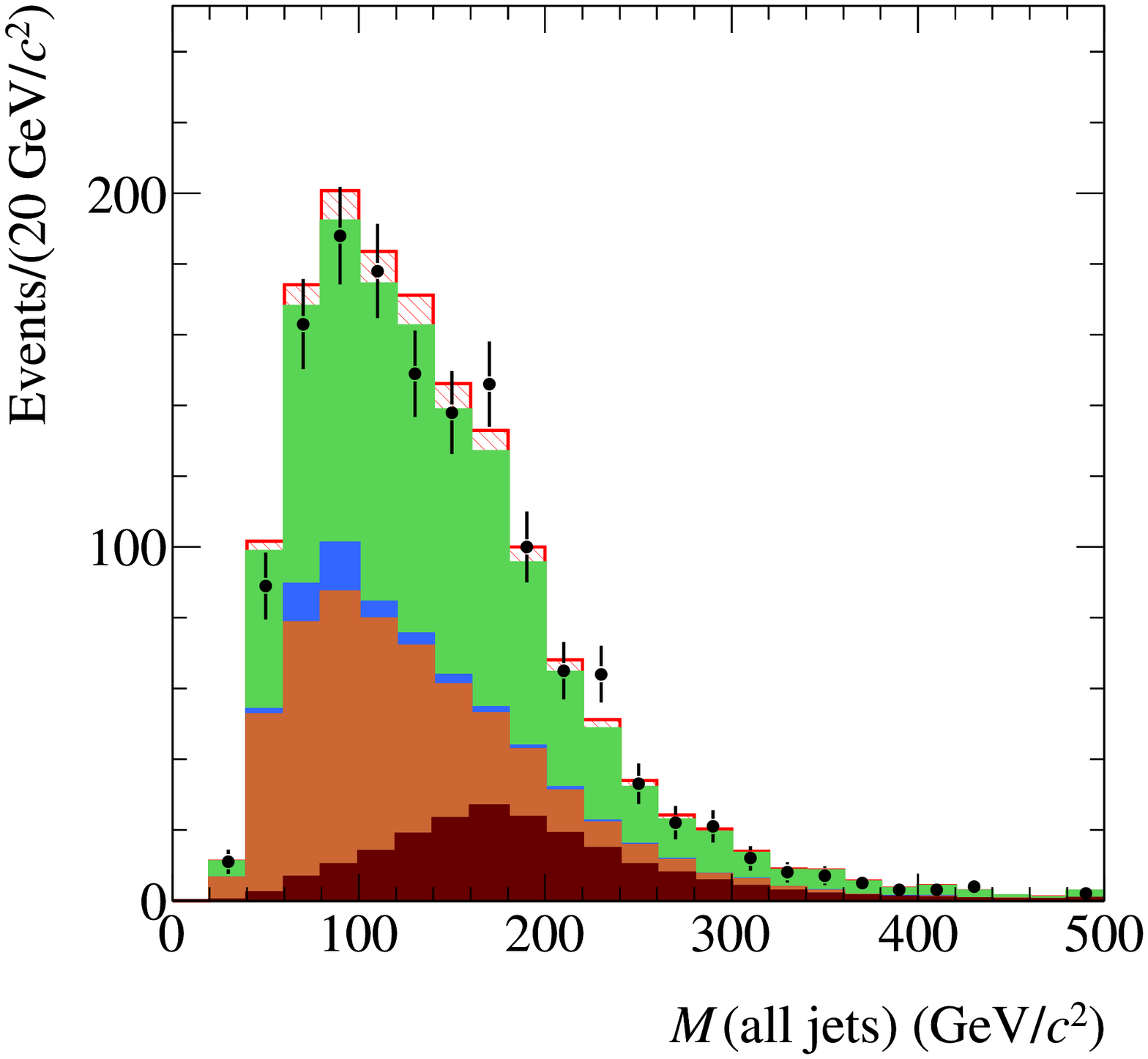}}
\subfigure
{\includegraphics[width=.32\linewidth]{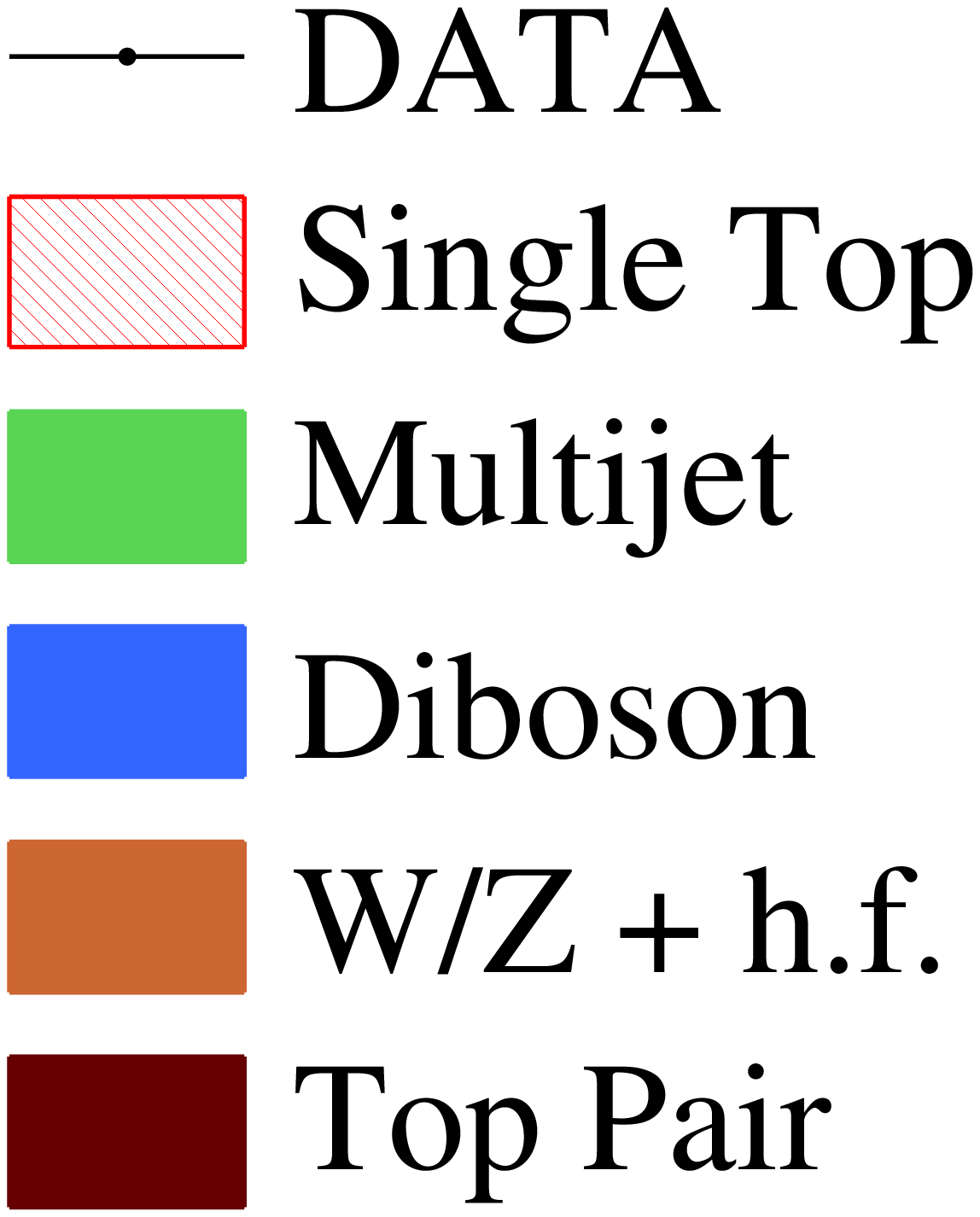}}
\caption{Kinematic distributions for the signal and background events in the signal region (NN$_{\rm QCD} > -0.1$). The three subsamples are summed together in their respective proportions. All physics processes contributions are normalized to the expected amount of events, as described in Sec.\,\ref{sec:SandBmodel}.} 
\label{fig:CRs}
\end{figure*}
Figure\,\ref{fig:CRs} shows that the predictions agree well with the observed data. The output of NN$_{\rm sig}$ is shown in Fig.\,\ref{fig:outFinalMLPshape}, where the contributions from the signal and main backgrounds are normalized to unit area. The single top quark signal events populate mostly the region of NN$_{\rm sig}$ around 0.3, while background events populate mostly the region with NN$_{\rm sig}$ around -0.3.
Finally, the same distribution is shown in Fig.~\ref{fig:outFinalMLP}, where the signal and backgrounds contributions have been normalized according to their estimates in the three subsamples, and the data are superimposed.
%The separation is helpful in the measurement of the cross section and the determination of the significance because backgrounds with large systematics are moved away from the bins where the signal peaks.

\section{\label{sec:sys}Systematic Uncertainties}

%Searching for single top production and measuring its cross section requires prior knowledge of what to expect.
%This knowledge is obtained from Monte Carlo simulations or extrapolations from control samples of the data in the form
%of predictions for signals and backgrounds. However, these differ from the ``true'' values by unknown amounts. This issue is
%addressed by assigning systematic uncertainties to our predictions and propagating their effects on the measured cross sections and on
%the significance of the signal.

%To address the fact that our predictions for various observables differ from their ``true" values, we assign systematic uncertainties and propagate their effects on the measured cross sections and on the significance of the signal.

%The search for single top quark production and the measurement of the cross section
%require substantial input from theoretical models, Monte Carlo simulations and 
%extrapolations from control samples in the data.
%The preditions of the expected signals and background from these models and extrapolations
%differ from their true values by unknown amounts.  We assign systematic uncertainties to our
%predictions and include the effects of these uncertainties on the measured cross sections as well as on
%the significance of the signal we quote.

Systematic uncertainties are grouped by their sources, where  a given source of uncertainty may affect several
background and signal distributions. The various systematic uncertainties originating from the same source are considered 100\% correlated. There are two categories of systematic uncertainties. Rate uncertainties are related to the predicted production rates, efficiency and acceptance of the various signal and background processes. Shape uncertainties express differences in the distributions due to a given systematic source.
% and, finally, the bin-by-bin uncertainties are statistical fluctuations arising from the limited size of the Monte Carlo or data samples (where applicable).
%We consider three categories of systematic uncertainties: uncertainties
%in the predicted rates of the signal and background processes, uncertainties in the shapes of the histogram
%templates, and independent bin-by-bin uncertainties arising from the limited size of the available Monte Carlo
%samples, and where applicable, the data samples used to fill the templates.  The systematic uncertainties
%are organized by their source.  Sources of uncertainty may affect multiple signal and background templates.
%The effects of systematic uncertainties from the same source are considered to be 100\% correlated, for example,
%the integrated luminosity is an ingredient of the predictions of both the Monte Carlo based backgrounds 
%and the expected signals.  The uncertainty on the integrated luminosity affects all of these in a correlated way.
%The effects of different sources of systematic uncertainty are
%considered to be uncorrelated.  A source of uncertainty may have correlated rate and shape effects on the
%predicted contributions. 
%\subsection{Rate Uncertainties}
Some sources of systematic uncertainty affect both rates and shapes. All rate uncertainties are assigned
a truncated Gaussian prior, preventing negative predictions. For shape uncertainties, if the prediction for a given bin is negative, it is set to zero.
%To reduce the impact of limited Monte Carlo statistics, the distributions uncertainties are smoothed bins. 
%More precisely, the template histograms for the varied values  of the uncertain parameters are divided by the template histograms for the central values. 
%Those ratio histograms are then median smoothed, assigning to a given bin the median of the rations in a window of five bins centered on that bin. 
%All but the first and last two bins are altered by this procedure. 
% The several sources of systematic uncertainties affecting this measurement are described in the following:
\subsection{Theoretical cross sections}  
%the Monte Carlo predictions for the single top quark $s$- and $t$-channel processes are made using {\sc madevent} for evaluating the tree-level diagrams and {\sc pythia} for the parton-shower. These Monte Carlo simulations predict the kinematic shapes rather well, but under-predict the total cross sections. 
%The predictions are scaled to NLO or better theoretical models and the associated uncertainties applied. The effects of the top quark mass are separated from the other sources of systematic uncertainties that affect theoretical predictions.
For all physics processes modeled by Monte Carlo simulation, we normalize to the most up-to-date theoretical computation of the cross section, and corresponding uncertainties. We use 12\% uncertainty for top quark pair production\,\cite{cacciari-2004-0404}, 40\% uncertainty  for the $W$ and $Z$ background processes\,\cite{abulencia:032008}, and 11\% for the diboson prediction\,\cite{DIBOSONS}.
\subsection{Integrated luminosity}
This systematic source accounts for the uncertainty in the $p \bar p$ inelastic cross section and for the uncertainty in the acceptance of the luminosity monitor of CDF to inelastic $p \bar p$ collision events\,\cite{CLC} and it is estimated as $6\%$. This uncertainty is applied to the rate predictions based on Monte Carlo simulation, {\it i.e.} all processes apart from multijet production. 
%The requirement that the primary vertex position in $z$ is within $\pm$60 cm of the origin has an uncertainty that is included in the luminosity uncertainty.
\subsection{Trigger efficiency}
Since we are using data below the fully efficient region of the trigger, we apply a parametrization of the trigger efficiency to the Monte Carlo simulated backgrounds. We assign a systematic uncertainty to both Monte Carlo simulated backgrounds and  signal acceptances by varying the trigger efficiency parameters\,\cite{artur}.
\subsection{$b$-tagging efficiency}
The $b$-tagging efficiency affects the predicted rates of signal and backgrounds estimates for which we use Monte Carlo simulations.  Known differences between the data and the simulation are corrected by scaling the simulation, and uncertainties on these scale factors are collected together in one source of uncertainty  (they affect the predictions in the same way). We assign an uncertainty of 4.3\% for the 1S prediction, 8.6\% for the 2S and and 12\% for SJ.
\subsection{Lepton veto}
The uncertainty in the efficiency of the cuts used to veto leptons was determined to be 2\%.
\subsection{Initial and final state radiation (ISR/FSR)}
To evaluate the ISR, {\sc pythia} uses a model of ``backwards evolution"\,\cite{PYTHIA,Sjo85}. The model used by {\sc pythia} for gluon radiation from partons emitted from the hard-scattering interaction has been tuned with LEP data. Monte Carlo simulated samples are generated for single top quark signals and \ttbar\ with $\Lambda_{\mathrm{QCD}}$ doubled (more ISR) or divided in half (less ISR) and with the initial transverse momentum scale and the hard scattering of the shower both multiplied (more ISR) or divided (less ISR) by four.
%Even with this tune to high precision data, uncertainties remain in the radiation from beam remnants [CHECK] 
The parameters for the final-state showering are also adjusted in {\sc pythia}, except for the hard-scattering scale. The uncertainties are then computed by comparing the efficiencies and kinematics of the varied ISR/FSR events to the nominal ones.
%
%The model used for ISR is {\sc pythia}'s ``backwards evolution'' 
%method~\cite{PYTHIA}.  This uncertainty is evaluated by generating new Monte Carlo samples for \ttbar\ and
%single top signals with $\Lambda_{QCD}$ doubled or divided in half, for samples with more ISR and less ISR,
%respectively.  
%Simultaneously, the initial transverse momentum scale is multiplied by four or divided by four,
%and the hard scattering scale of the shower is multiplied by four or divided by four, for more ISR and less ISR
%modeling, respectively.  
%
%These variations are checked with Drell Yan Monte Carlo and data samples.  
%The \Pt\ distributions
%of dileptons are compared between data and Monte Carlo, and the ISR more/less prescriptions generously bracket
%the available data, as can be seen in Figure~\ref{fig:isr}.  The ranges are generous because of the extrapolation
%required to predict ISR in higher-$Q^2$ events like \ttbar\ and single top events. {\sc pythia}'s model of gluon radiation from partons emitted from the hard-scattering interaction has been tuned with high precision to LEP data~\cite{PYTHIA}.  Nonetheless, uncertainty remains in the radiation from beam remnants, and analogous parameters are adjusted in {\sc pythia} for the final-state showering, except for the hard-scattering scale parameter.  
The effects of variations in ISR and FSR are treated as 100\% correlated with each other.
\subsection{Jet energy scale (JES)} \
Each step in the correction of the calorimeter response to particle jets involves an uncertainty, which is propagated to the final JES\cite{Bhatti:2005ai}. 
The effect of JES uncertainties are estimated by varying the jet energy scale in all Monte Carlo simulated samples twice: one upwards, one downwards. They are evaluated for each background and signal contribution, and both rate and shape uncertainties are taken into account.
%The calibration of the calorimeter response to jets is a multi-step process,
%and each step invovles an uncertainty which is propagated to the final jet-energy scale.  Raw jet energies
%are corrected for test-beam scales, detector non-uniformity, multiple interactions, and energy that is not
%assigned to the jet because it lies outside of the jet cone.  The uncertainties in the jet energy scale are incorporated
%by processing all events in all Monte Carlo samples with the jet energy scale varied upwards and again downwards. \cite{Bhatti:2005ai}
\subsection{Parton distribution functions (PDF)}
Lack of precise knowledge of the PDFs is a source of theoretical uncertainty for the amount of signal produced. The uncertainty is estimated using different sets of PDF eigenvectors. The default PDF set used in this analysis is the {\sc cteq5l} set\,\cite{Lai:1999wy}. The uncertainty is determined comparing:
\begin{itemize}
\item two different LO PDF parametrization sets, CTEQ5L\,\cite{Lai:1999wy} and MRST72\,\cite{Martin:1998sq};
\item MRST72 and MRST75 with different $\Lambda_{\mathrm{QCD}}$;
\item the variation within their uncertainties of each of the 20 signed eigenvectors of the NLO PDF set CTEQ6M\,\cite{Pumplin:2002vw} with the default PDF set.
\end{itemize}
The total PDF uncertainty is obtained adding the larger of the 20 eigenvectors' uncertainty (all added in quadrature) or the MRST72 and CTEQ5L PDF sets uncertainty in quadrature with the $\Lambda_{\mathrm{QCD}}$ uncertainty. A 2\% uncertainty was found to be sufficient for all the backgrounds. The PDF uncertainty on the signal acceptance range from 1\% to 2\% depending to the subsample. Shape variations induced by PDF systematic changes are considered only for the single top quark process.
\subsection{Multijet model}
The data-driven model for multijet production predicts the shapes of the distributions. The rates are obtained from the NN$_{\rm QCD} < -0.1$ control region. In this region, we assign a scale factor associated with the difference between the data rates and the Monte Carlo simulation prediction, which we then multiply to the multijet predicted rate. We then obtain the uncertainty on this scaling using propagation of errors. Depending on the subsample under study, we assign an uncertainty between 4.5\% and 13\%. 
%These values were obtained from the uncertainties on the scaling factor for QCD multijet production determined .
The variations in the tag rate probability parametrization used to estimate the multijet background also modify the shapes of the distributions. The shape uncertainty is obtained by varying the tag rate probability by the uncertainty in its estimation.
%by $\pm 1\sigma$.
%Those numbers are obtained from CR3, propagating the errors:
%\begin{eqnarray}
%\text{SF} &=& \frac{\text{DATA - Exp. (MC) Signal - Exp. MC Background}}{\text{QCD (before SF)}} \\
%\frac{\Delta \text{SF}}{\text{SF}} &=& \sqrt{\left(\frac{\text{MC}_{Err}}{\text{DATA - MC}}\right)^2+\left(\frac{\text{QCD}_{Err}}{\text{QCD}}\right)^2},
%\end{eqnarray}
%where MC is the number of expected MC events (including signal), QCD is the QCD prediction before applying the scale factor, QCD$_{Err}$ is the statistical error on the QCD prediction (before SF) and MC$_{Err}$ is is computed adding the errors (stat. + syst.) on its components in quadrature. The values fo these for each tagging category are shown in table\,\ref{tab:SFcomp}.
%
% \begin{table}[!ht]
%  \begin{center}
%    \begin{tabular}{|l|c|c|c|}
%      \hline
%      Process		&Excl. ST        & ST+ST              & ST+JP               \\
%      \hline\hline
%
%DATA & 2343 & 73 & 166\\
%\hline\hline
%QCD before SF&1992 +/- 65&86.9 +/- 11.3&207 +/- 20\\
%Exp. (MC) Signal & 5.9 +/- 0.2 & 0.46 +/- 0.05 & 0.65 +/- 0.06\\
%Exp. MC Background&188 +/- 66&4.4 +/- 1.2&8.0 +/- 2.1\\
%
%\hline\hline
%    \end{tabular}
%    \caption{Number of expected and observed events in the CR3 (QCD, signal like) in all tagging categories used to derive the QCD normalization scale factor.}
%    \label{tab:SFcomp}
%  \end{center}
%\end{table}
We also take into account the normalization uncertainty on the processes which are part of the background in the region from which we get the QCD normalization. Those uncertainties are anti-correlated with respect to the normalization on these processes, and are weighted appropriately. In the NN$_{\rm QCD} < -0.1$ control sample used to derive the normalization, the biggest contamination sources come from $W$ + heavy flavor jets ($5\%$ of the 1S and $<1.5\%$ of the 2S and SJ samples), $\ttbar$ (2.7\% of the 2S sample and $<1.5\%$ of the two other samples), and $Z$ + heavy flavor jets ($<1.8\%$ of the three $b-$tagged subsamples). Diboson contamination is negligible ($<0.5\%$).
%Table\,\ref{tab:NormAntiCorr} shows the proportion of the MC processes in the CR3.
%  Additionaly, the variations in Tag Rate Matrix are taken into account (see TRF systematic).  
%  
%\begin{table}[!ht]
%  \begin{center}
%    \begin{tabular}{|l|c|c|c|}
%      \hline
%      Process		& ST         & ST+ST        & ST+JP \\
%      \hline\hline
%Top Pair & 0.9\% & 2.7\% & 1.63\% \\
%W + h.f. & 5.1\% & 1\% & 1.66\% \\
%Diboson & 0.3\% & 0.5\% & 0.25\% \\
%Z + h.f. & 1.7\% & 1.8\% & 1.26\% \\
%\hline
%    \end{tabular}
%    \caption{Contribution of MC processes in CR3.}
%    \label{tab:NormAntiCorr}
%  \end{center}
%\end{table}
To avoid double counting non-multijet events in our estimation of the multijet background, we apply the tag rate parametrization to our Monte Carlo simulation predictions and subtract the output from the data. The single top quark signal as predicted by Monte Carlo simulation is also subtracted from the data. We associate a shape systematic uncertainty to this removal by varying the amount of single top quark we subtract by 50\%, more than three times the theoretical uncertainty on the single top quark cross section.

%[CHECK say that we have signal contamination and remove parenthesis] We associate another systematic to the removal of single top Monte Carlo from data (applying the tag rate parametrization to both) to estimate the QCD multijet background. We vary the amount of single top we subtract by 50\%, more than three times the theoretical uncertainty on the single top cross section.
\subsection{Background scaling}
A small fraction of the data events analyzed in this paper pass the event selection requiring identified charged leptons in the final state\,\cite{Aaltonen:2009jj}. The fraction has been computed using single top quark Monte Carlo simulation to be 2\%. To maintain a 100\% orthogonality with Ref.\,\cite{Aaltonen:2009jj}, these events are discarded from this analysis.  We scale down the predicted amount of background events by 2\%, and assign an additional uncertainty of 2\% to the background yields.

%[CHECK journalistic, merge] The systematic assigned to the backgrounds after scaling them down by 2\%, to remove overlap with the single top analyses with leptons in the final state\,\cite{PRDobs}. A few data events pass the event selection of the latter analyses as well as ours. To keep a 100\% orthogonality, these were discarded, requiring the scaling of the signal and background. There is no uncertainty assigned to the signal scaling because the same Monte Carlo is used by this and the lepton analyses, allowing a precise identification of the overlap. We assign an uncertainty of 2\% to the scaled down background yields. [Maybe I say too much?]
\subsection{Top quark mass dependence}
The most precise measurement of the top quark mass corresponds to M$_{top}=173.1 \pm 1.3\,$GeV/$c^2$\,\cite{:2009ec}. We consider for this analysis a nominal top quark mass of M$_{top}=175\,$GeV/$c^2$ for the acceptance computation and kinematic estimation, and use the two extreme values of 170 and 180\,GeV/$c^2$ to compute the systematic shifts. This uncertainty is considered for all processes producing top quarks when extracting the value of $V_{tb}$ and computing the significance of the measurement. \\

\par

The summary of the systematic sources, their effect on the rates of different processes and how they affect the kinematics can be found in Table~\ref{tab:SystTable}.

  \begin{table*}[!ht]
    \begin{center}
      \caption{Summary of systematic uncertainties and their treatment in the analysis. A range of values is listed for a systematic source to indicate that the effect is different in the three subsamples defined by the different heavy flavor jet content. The ``X" sign in the ``Shape column" means that the influence of the change in the systematic source on the kinematic distributions of the physics processes has been considered, while the ``-" sign means that it is not applicable, or not considered. The ``comment" column describes whenever a systematic is considered only for some physics processes, or treated differently for some physics processes, or considered only in certain computations.}
      \begin{tabular}{lccc}
       \hline\hline
      Systematic source & Rate & Shape & Comment \\
        \hline 
Top quark pair production cross section& $\pm$ 12\% & - & \\
%\hline
$W$/$Z$ + heavy flavor jets cross section & $\pm$ 40\% & - & \\
%\hline
Diboson cross section & $\pm$ 11\% & - & \\
%\hline
Luminosity & 6\% & - & Not for multijet \\
Trigger efficiency & $< 2.6\%$ & X & \\
%\hline
% \multirow{2}{*}{B tagging scale factors} & 4.3\% to \multirow{2}{*}{-} & \\
$b-$tagging efficiency & 4.3\% to 12\% & - &  \\
%\hline
% &  12\% (ST + JP) &  & \\\hline
Lepton veto & 2\% & - & \\
%\hline
ISR/FSR & -4.5\% \ldots +16\% & X & Only for top quark processes \\
%\hline
JES & -14\% \ldots +23\% & X & \\
%\hline
PDF & $\pm 1\% \ldots \pm 2\%$ & X & Shape for signal only\\
%\hline
%\multirow{3}{*}{PDF}& $\pm 0.5\%$ ($s$-chan) ; $\pm 2.2\%$ (t-chan) & X &\\\cline{2-3}
% & $ \pm 1.1\% (\ttbar)$& - & \\\cline{2-3}
% & $ 2\%$ (Monte Carlo backgrounds) & - & \\\hline
Multijet model & 4.5\% \ldots 13\% & X & \\
%\cline{1-4}
%\multirow{2}{*}{QCD multijet normalization} & 4.5\% (ST) ; 13\% (ST + ST) & \multirow{2}{*}{-} & \\
% &  10\% (ST + JP) & & \\\hline
%Signal contamination & - & X &  \\\hline
Background scaling & 2\% & - &\\
\hline
Single top quark cross section & $\pm$ 12\% & - &  \multirow{2}{*}{Only for $p$-value and $V_{tb}$ computation}\\
Top quark mass dependence & -16\% \ldots +7.5\% & X & \\
\hline\hline
    \end{tabular} \\
%{\small$^{(1)}$PDF shape systematic fon single top only}
	%     \caption{Summary of systematic uncertainties and their treatment in the analysis. A range of values is listed for a systematic source to indicate that the effect is different in the three subsamples defined by the different heavy flavor jet content. The ``X" sign in the ``Shape column" means that the influence of the change in the systematic source on the kinematic distributions of the physics processes has been considered, while the ``-" sign means that it is not applicable, or not considered. The ``comment" column describes whenever a systematic is considered only for some physics processes, or treated differently for some physics processes, or considered only in certain computations.}
     \label{tab:SystTable}
   \end{center}
  \end{table*}

\section{\label{sec:likelihood}  Cross section and significance extraction}

%We use the binned-likelihood technique to fit the data of Fig. 1, in order to gauge the significance of a possible signal-like effect and to measure the corresponding single top cross section
%As seen in Fig.\,\ref{fig:outFinalMLP} there is an excess of signal-like events over the standard model background expectation. Assuming the excess originates from single top quark production, 
We scan the NN$_{\rm sig}$ distribution using a binned likelihood technique to measure its cross section, as well as to determine the significance of the excess itself. 
The likelihood function $L$ is given by the product of the likelihood for each of the different subsamples $L_c$, $L= \prod_{c=1}^{N_c} L_c$ where $N_c = 3$ are the three subsamples subdivided according to the number of $b-$tagged jets and the tagging algorithm used (1S, 2S, SJ). The likelihood $L_c$ for each subsample to observe the data in the final NN distribution is defined as:
\begin{equation}
L_c =  \prod_{i=1}^{n_{\rm{bins}}} P(n_i | \mu_i)   =  \prod_{i=1}^{n_{\rm{bins}}} \frac{\mu_i^{n_i}e^{-\mu_i}}{n_i!}.
\end{equation} 
where $n_i$ is the data count in that particular bin and $n_{\rm bins}$ is the number of bins in the distribution which is scanned to look for an excess of signal-like events.
The prediction in each bin is a sum over signal and background contributions:
\begin{equation}
\mu_i = \sum_{k=1}^{n_{\rm bkg}} b_{ik} + s_{i} 
\end{equation}
where $b_{ik}$ is the background prediction in bin $i$ for background source $k$ given the number of background sources $n_{\rm bkg}$ and $s_i$ is the signal prediction in bin $i$ for the $s-$ and $t-$channel single top quark production summed according to the standard model proportions.
Uncertain nuisance parameters $\mbox{\boldmath$\theta$}$ affect the signal and background predictions and kinematics. The induced effect on the event rates and shapes of the kinematic distributions can be correlated with each other. The likelihood $L$ is then a function of the observed data $\mbox{\boldmath$D$}$, the signal cross section $\sigma_{s+t}$ and of the nuisance parameters $\mbox{\boldmath$\theta$}$ which affect the signal and background predictions, $ L(\mbox{\boldmath$D$}|\sigma_{s+t}, \mbox{\boldmath$\theta$})$. We use Bayes' theorem to convert the likelihood into a posterior density function in $\sigma_{s+t}$. 
We use the posterior density function to quote the measured value of the production rate\,\cite{Amsler:2008zzb}.
%The likelihood is a function of the observed data $\mbox{\boldmath$D$}$, the signal cross section $\sigma_{s+t}$ and of the nuisance parameters $\mbox{\boldmath$\theta$}$ which affect the signal 
%$\mbox{\boldmath$s$} = \{s_{i}\}$ 
%and background 
%$\mbox{\boldmath$b$} = \{b_{ik}\}$ 
%predictions:
%
The posterior density function is:
\begin{equation}
p(\sigma_{s+t}|\mbox{\boldmath$D$}) =  \frac{1}{N} \int L(\mbox{\boldmath$D$}|\sigma_{s+t}, \mbox{\boldmath$\theta$}) \pi(\mbox{\boldmath$\theta$}, \sigma_{s+t}) d\mbox{\boldmath$\theta$}
\end{equation}
where $N$ is an overall normalization obtained from the requirement $\int p(\sigma_{s+t}|\mbox{\boldmath$D$}) d\sigma = 1$.
The function $\pi(\mbox{\boldmath$\theta$}, \sigma_{s+t})$ is the prior probability density, which encodes our knowledge of the parameters $\mbox{\boldmath$\theta$}$ and $\sigma_{s+t}$. Since our knowledge of the cross section is independent of our prior knowledge of the nuisance parameters, we can write the prior density as $\pi(\mbox{\boldmath$\theta$},\sigma_{s+t}) = \pi(\mbox{\boldmath$\theta$}) \cdot \pi(\sigma_{s+t})$. 
The ignorance about the true single top quark production cross section is encoded through the prior probability density function, which is set equal to the Heaviside function in $\sigma_{s+t}$, $\pi(\sigma_{s+t})=1$ if $\sigma_{s+t} \ge 0$ and $\pi(\sigma_{s+t})=0$ otherwise. 
%The priors on the nuisance parameters $G(\tilde{\nu},\sigma_{\nu})$ are set to be Gaussian densities constraints using the estimated central value $\tilde{\nu}$ and the associated uncertainty $\sigma_{\nu}$.
The prior probability density function on the nuisance parameters are set to be Gaussian distributions, characterized by the estimated central value of the systematic source and the associated uncertainty. The posterior density function can thus be written as 
\begin{equation}
p(\sigma_{s+t},\mbox{\boldmath$D$}) =  \frac{1}{N} \int L(\mbox{\boldmath$D$}|\sigma_{s+t},\mbox{\boldmath$\theta$}) \pi(\mbox{\boldmath$\theta$}) \pi(\sigma_{s+t}) d\mbox{\boldmath$\theta$}
\end{equation}
The marginalization of the posterior density function is done using Monte Carlo integration, 
by generating a large number of points in the nuisance parameters space, according to their priors probability density functions.

In doing the generation, we take into account the correlations between nuisance 
parameters. Shape and rate uncertainties due to a given nuisance parameter are treated as 100\% correlated. 
%Since some of the uncertainties may result in shape variations of the relevant distributions 
%(such as the JES uncertainty),  the histograms are interpolated and extrapolated within their shape uncertainties.
We define the measured cross section $\sigma^{\rm{meas}}_{s+t}$ as the value corresponding to the mode of the $p(\sigma_{s+t}|\mbox{\boldmath$D$})$ distribution, and its uncertainty as the smallest interval enclosing $68\%$ of the integral.
In order to measure the single top quark production cross section and its uncertainty, we do not include the $m_t$ uncertainty in the \ttbar\ background or in the signal, but rather quote the measurement at the assumed top quark pole mass of $m_t=175$\,\gevcc.  We assume the standard model ratio between $s$- and $t$-channel production.

\begin{figure}[h]
  \centering
\includegraphics[width=8cm]{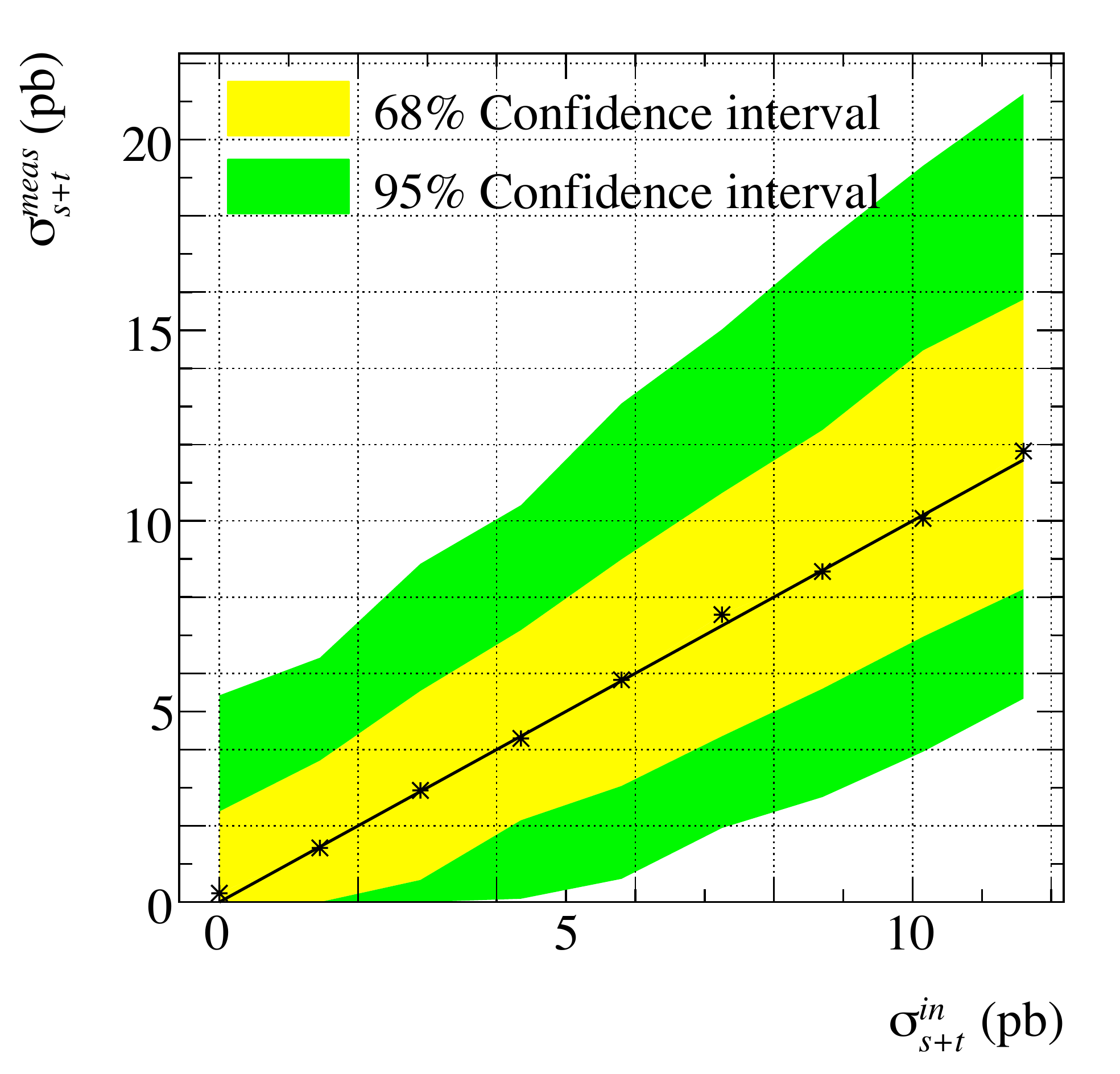}
    \caption{Distributions of 68\% and 95\% of extracted cross sections centered on the medians are shown as a function of a set of input cross section. The line represents $\sigma_{s+t}^{\rm{meas}} = \sigma_{s+t}^{in}$. The plot shows that the fit technique does not introduce bias. }
  \label{fig:Linearity_Test}
\end{figure}

The measured cross section %and $p$-value
depends on the true cross section but also the random outcome of the data. The sensitivity of the analysis is evaluated using the expected distribution of outcomes, assuming a signal is present. 
The event selection cut values and the final discriminant used to scan for the signal have been chosen to maximize the expected sensitivity to the signal.
We check our cross section fit method using pseudoexperiments generated varying the input signal cross section and systematic uncertainties, which are then fit to measure the signal cross section to check for possible biases. The procedure used cannot produce a negative cross section measurement, since the priors are zero for negative values. For an input cross section of zero, half of the measured cross sections then are exactly zero, and the other half form a distribution of positive fit cross sections. 
We therefore use the median fit cross section for our linearity check to avoid the bias which would be introduced by using the average instead. Distributions of 68\% and 95\% of extracted cross sections centered on the medians are then shown as a function of the input cross section in Fig.\,\ref{fig:Linearity_Test}. It can be deduced from the same plot that the fit technique used here does not introduce a bias.

\subsection{Significance calculation}

In addition to measuring its cross section, it is also important to estimate the significance of the measurement itself. To do so, we use the $p$-value, which is the probability of observing an outcome of our experiment at least as signal-like as the one observed or more, assuming that a signal is absent. By convention, an observed $p$-value of less than $1.35\times 10^{-3}$ constitutes evidence for a signal, and an observed $p$-value less than $2.87\times 10^{-7}$ constitutes a discovery. These are the one-sided integrals of the tails of a unit Gaussian
distribution beyond $+3\sigma$ and $+5\sigma$, respectively.
The experimental outcome is ranked on a one-dimentional scale using the likelihood ratio test statistic\,\cite{Neyman}:
%%%%%%%%%
%shortcut
%%%%%%%%%%
%$-2 \ln Q$
%%%%%%%%%%%%%%%%
%FULL EXPLANATION
%%%%%%%%%%%%%%%%
\begin{equation}
-2\ln Q = -2\ln\frac{L(\mbox{\boldmath$D$} | \sigma_{s+t} = \sigma^{\prime}_{s+t}) } {L(\mbox{\boldmath$D$} | \sigma_{s+t} = 0) }
\end{equation}
%\begin{equation}
%-2\ln Q = -2\ln \frac{ \int L(\mbox{\boldmath$D$}|\sigma_{s+t}) \pi(\mbox{\boldmath$\theta$}) \pi(\sigma_{s+t}) d\mbox{\boldmath$\theta$}} { L(\mbox{\boldmath$D$}|\sigma_{s+t}=0)}
%\end{equation}
In the computation of the $-2 \ln Q$ and thus the $p$-value, we include all sources of systematic uncertainties, including the
theoretical uncertainty on the single top quark production cross section and on the top quark mass itself. 
We perform two sets of a large number of pseudoexperiments and compute corresponding $-2 \ln Q$ distributions. 
In the first one, pseudodata are generated in the hypothesis that single top quark production is present in the SM-predicted amount (S+B). In the second, 
pseudodata are generated according to the background-only hypothesis (B). The $p$-value is the probability that $-2 \ln Q < -2 \ln Q_{\rm meas}$ in the B hypothesis.
%\begin{equation}
%p = p(-2\ln Q\le -2\ln Q_{\rm{meas}}| \sigma_{s+t}=0).
%\%end{equation}
To compute the expected $p-$value, we set $Q_{\rm{meas}}$ as the mode of the $Q$ distribution assuming that $\sigma^{\prime}$ is equal to the theoretical prediction for the SM single top quark production cross section. To compare the data with the SM predictions, we then set $\sigma^{\prime}$ to be equal to $\sigma^{\rm meas}_{s+t}$ and compute  the corresponding $Q_{\rm{meas}}$ and then the observed $p$-value.

\subsection{Constraining $V_{tb}$}

We can also use the knowledge of the standard model prediction for the single top quark production cross section to compute the $V_{tb}$ element of the Cabibbo-Kobayashi-Maskawa matrix. Under the standard model hypothesis, with the assumption that $|V_{td}|^2+|V_{ts}|^2 \ll |V_{tb}|^2$ and that only $|V_{tb}|$ incorporates new physics contributions, one can measure $V_{tb}$ using the relation $|V_{tb}|^2 =  \frac{\sigma_{s+t}^{\rm{meas}}}{\sigma_{s+t}^{\rm{SM}}}$. The theoretical uncertainty on $\sigma_{s+t}^{\rm{SM}}$ is taken into account when setting the $V_{tb}$ constraints, together with the uncertainty on the top quark pole mass measurement.

\section{Results}
\label{sec:results}

\begin{figure}[h]
  \centering
\includegraphics[width=8cm]{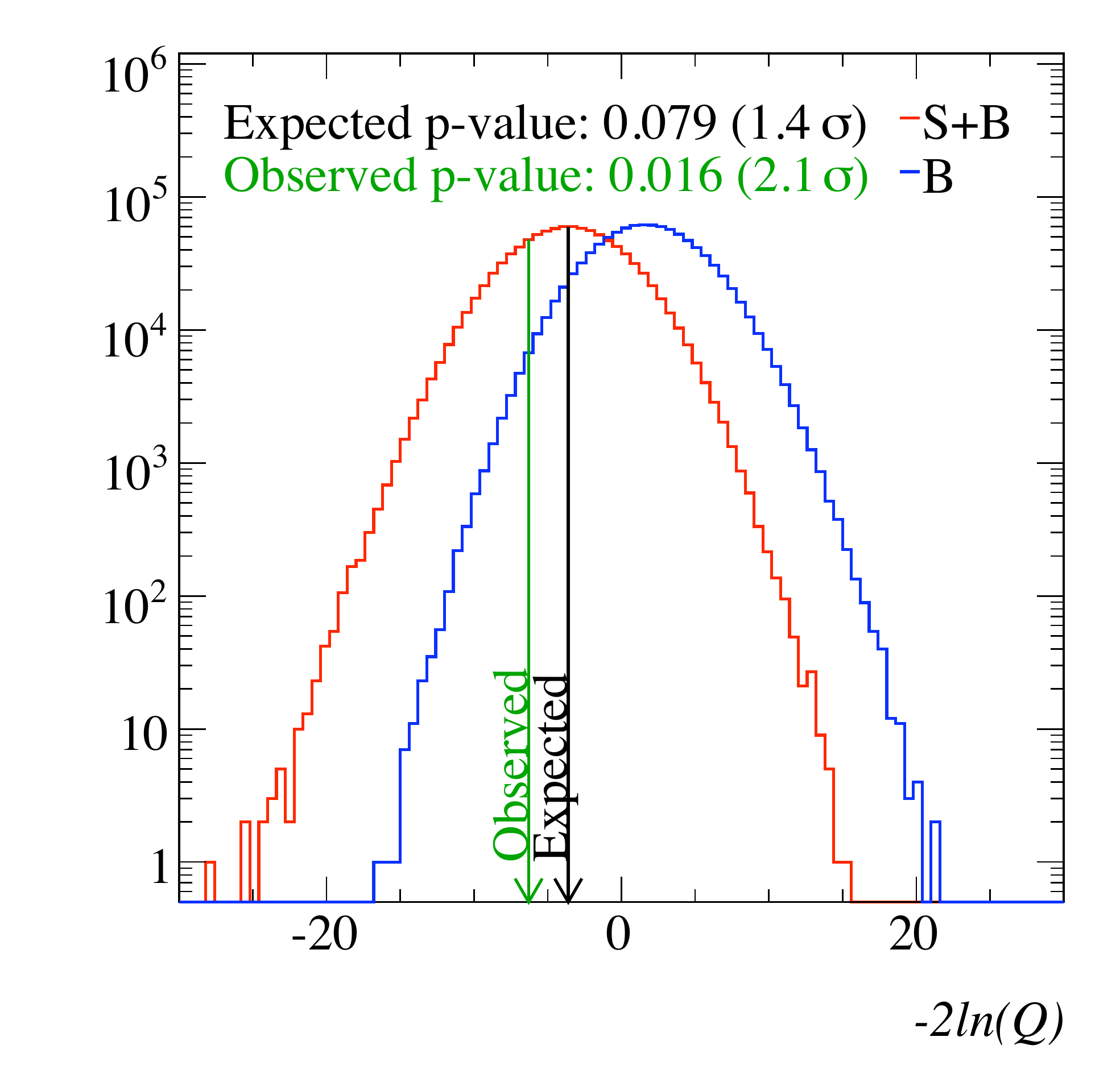}
    \caption{Distribution of  $-2 \ln Q$ for the B or S+B hypothesis. The dark vertical arrow shows the mode of the  $-2 \ln Q$ distribution in the S+B hypothesis using pseudoexperiments,
    while the bright vertical arrow shows the outcome in data. The expected (observed) $p$-value is the integral of the $-2 \ln Q$ distribution in the B hypothesis at the left of the dark (bright) arrow.}
  \label{fig:PValue}
\end{figure}

%We apply our analysis to the first 2.1 fb$^{-1}$ of data recorded by the CDF II experiment. 
Using the signal and background modeling described in Sec.\,\ref{sec:SandBmodel}, scanning the multivariate discriminant described in Sec.\,\ref{sec:NNdisc} 
%using the likelihood described in Sec.\,\ref{sec:likelihood}, we expect to measure the single top quark cross section as $\sigma_{s+t}^{exp} =  2.7^{+2.3}_{-2.1} \text{ pb}.$ 
and using the statistical test described above, we compute the probablility that the background (B) 
%fluctuated equal or up to the observed value in the data 
looks at least as signal-like as the data (observed $p$-value) or as the median of signal plus background (S+B) pseudo-experiments outcomes (expected $p$-value). 
%The purely statistical sensitivity for the analysis amounts to almost $3 \sigma$ excess over the background. IS THIS TRUE??
Once including all systematic sources, we obtain an expected $p$-value of $7.9\times10^{-2}$ (1.4\,$\sigma$) and an observed $p$-value of $1.6\times10^{-2}$ (2.1\,$\sigma$). The distributions of $-2 \ln Q$ for the B or S+B hypothesis are shown in Fig.\,\ref{fig:PValue}, together with the observed outcome in the data.
\par
\begin{figure}[h]
  \centering
\includegraphics[width=8cm]{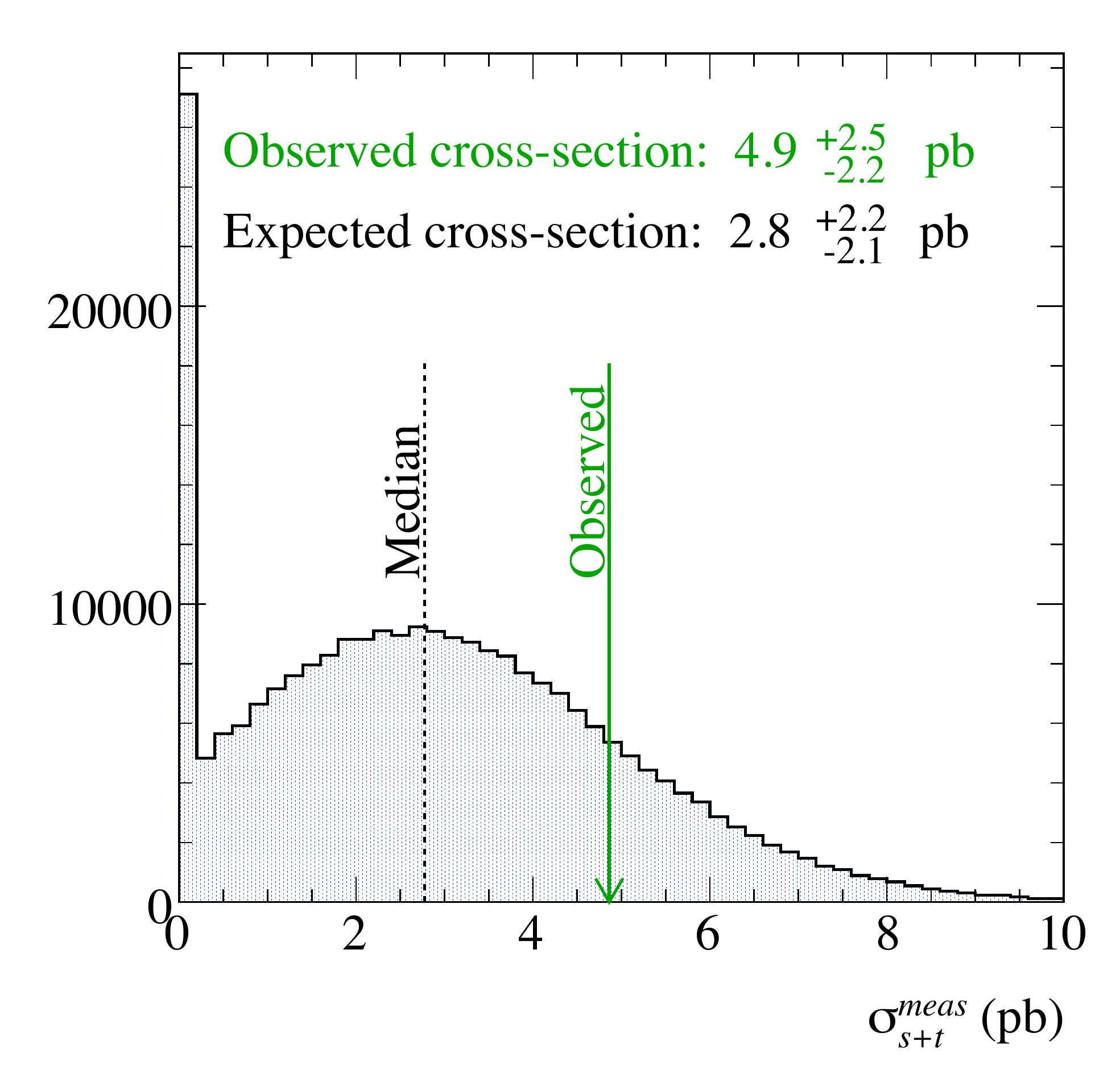}
   \caption{Distribution of cross section measurement outcomes using pseudoexperiments. The green arrow shows the cross section measured in the data. The probability to measure a cross section higher than the median expected cross section shown as a dotted line is 18$\%$.}
  \label{fig:PExs}
\end{figure}

Interpreting the 2.1\,$\sigma$ excess as originating from single top quark production, we measure a single top quark production cross section of $$\sigma_{s+t}^{\rm{meas}} = 4.9^{+2.5}_{-2.2} \text{ pb.}$$ 
The value measured in the data is compared to the measurement outcomes from 150\,000 pseudoexperiments, as shown in Fig.\,\ref{fig:PExs}. 
The single top quark production cross section measurement presented here is consistent with a $+1\,\sigma$ statistical upward fluctuation with respect to the standard model cross section. The probability to measure a cross section higher than 4.9\,pb has been estimated to be 18\%. 

\begin{figure}[h]
  \centering
\includegraphics[width=8cm]{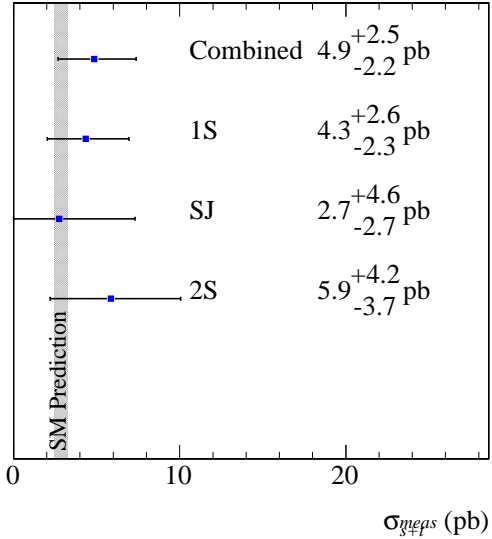}
    \caption{Measurement of the single top quark cross section production. We show the combined result in the whole dataset analyzed, and the result obtained in each tagging category. All measurements are consistent with the standard model theoretical cross section within uncertainties.}
  \label{fig:ResultsWData}
\end{figure}

As a cross-check, we perform the measurement separately in the three subsamples. The results are shown in Fig.\,\ref{fig:ResultsWData} and indicate that, while the precision is low, the three orthogonal measurements are in agreement with each other.

%\begin{table}[hbtp]
%\begin{center}
%\begin{tabular}{l|c|c|c|c}
%\hline\hline
%Data sample & 1S & SJ & 2S & All\\
%\hline
%$\sigma_{s+t}^{\rm{meas}}$ (pb) & $4.3^{+2.6}_{-2.3}$ & $2.7^{+4.6}_{-2.7}$  & $5.9^{+4.2}_{-3.7}$ & $4.9^{+2.5}_{-2.2}$ \\
%\hline\hline
%\end{tabular}
%\end{center}
%\caption{Measurement of the single top quark cross section production. We show the combined result in the whole dataset analyzed, and the result obtained in each tagging category. All measurements are consistent with the standard model theoretical cross section within uncertainties.}
% \label{tab:ResultsWData}
%\end{table}

\begin{figure}[h]
  \centering
\subfigure[]{\label{fig:VTB}\includegraphics[width=8cm]{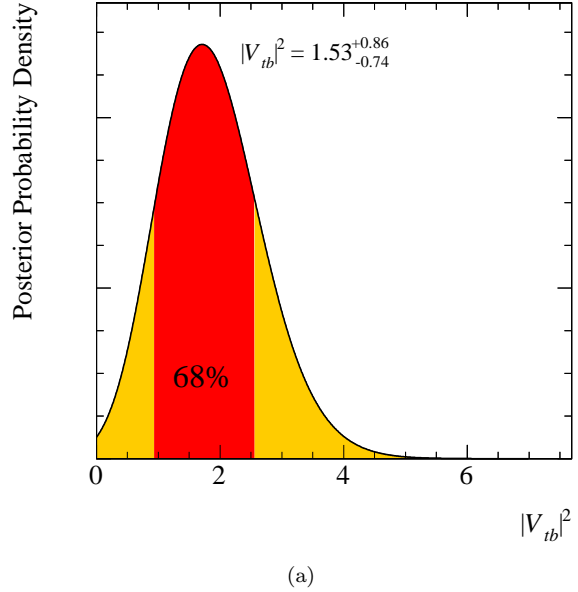}}
\subfigure[]{\label{fig:VTB_CL}\includegraphics[width=8cm]{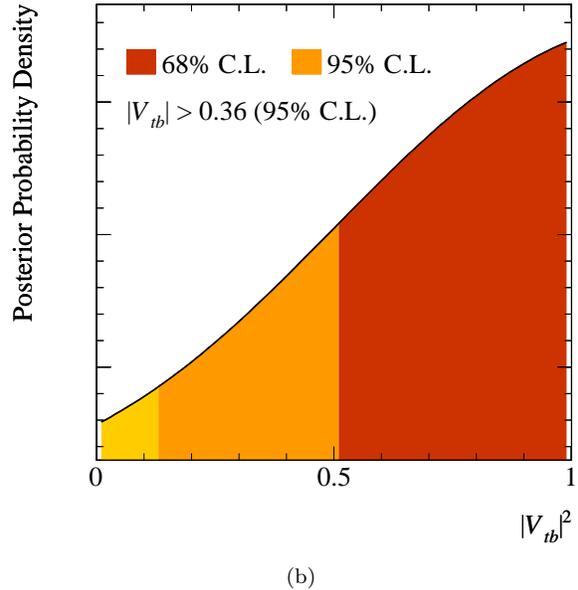}}
    \caption{Posterior probability density of the square of the $V_{tb}$ element of the CKM matrix, assuming an unconstrained flat prior on $V_{tb}^2$ (a) and assuming a flat prior on $V_{tb}^2$ constrained between zero and one (b).}
%  \label{fig:VTB}
\end{figure}

Finally, we measure the $V_{tb}$ element of the CKM matrix. Using an unconstrained flat prior on $|V_{tb}|^2$, we find $|V_{tb}| = 1.24^{+0.34}_{-0.29}\pm 0.07 \text{(theory)}$ as shown in Fig.\,\ref{fig:VTB}. The theoretical uncertainty is due to the uncertainty on the standard model theoretical cross section for single top quark production\,\cite{singletops,singletopt}. Assuming a flat prior on $|V_{tb}|^2$ between 0 and 1, Fig.\,\ref{fig:VTB_CL} shows that $V_{tb} > 0.36$ at 95\% credibility level.

%\begin{figure}[h]
%  \centering
%\includegraphics[width=.6\linewidth]{./newfinalPlots8/VTB_CL}
%    \caption{Determination of the $V_{tb}$ element of the CKM matrix assuming a flat prior between 0 and 1. }
%  \label{fig:VTB_CL}
%\end{figure}

\section{\label{sec:Summary}Summary}
\label{sec:summary}

We have presented the first search for $s$- and $t$-channel electroweak single top quark production in the $\met$+jets signature. This dataset is orthogonal to the one used to achieve the evidence level at CDF\,\cite{Aaltonen:2008sy} and D0\,\cite{Abazov:2006gd,Abazov:2008kt}, and it is sensitive to the $W \to \tau \nu$ decays. 
Using an optimized neural network-based kinematic selection and $b$-jet identiÞcation techniques, we are able to improve the $s/b$ of the initial sample obtained with a $\met$+jets trigger from about 1/10\,000 to about 1/20.
We have analyzed 2.1\,fb$^{-1}$ of integrated luminosity recorded with the CDF II detector and observed an excess of signal-like events with respect to the standard model background prediction. The probability that the background-only hypothesis would produce the observed data is $1.6\times10^{-2}$ (2.1\,$\sigma$).  
Assuming that the excess originates from single top quark production through $s-$ and $t-$channel and a top quark mass of 175 GeV/$c^2$ we measure $$\sigma_{s+t}^{\rm{meas}} = 4.9^{+2.5}_{-2.2} \text{ pb}.$$ 
%Given the SM signal-plus-background hypothesis we estimate the probability to measure a cross section as large or larger to be 18$\%$.
We use the theoretical computation of the signal cross section to measure the $V_{tb}$ element of the CKM matrix: $$|V_{tb}| = 1.24^{+0.34}_{-0.29}\pm 0.07 \text{(theory)}.$$ Assuming $0 \le V_{tb} \le 1$, we set the lower limit of $V_{tb} > 0.36$ at 95$\%$ confidence level. This analysis has been combined with the search performed by the CDF collaboration in events with an identified charged lepton plus $\met$ plus jets signature\,\cite{Aaltonen:2009jj}. The combination of the two searches observes an excess of signal-like events over the background expectations at the 5$\sigma$ level, thus estabilishing the existence of this rare process. Finally, the combination of the measurement presented here with the measurements in the charged lepton plus $\met$ plus jets signatures by CDF and D0 has been performed to obtain the most precise direct measurement of $V_{tb}$ to date\,\cite{Bestvtb}.

\begin{acknowledgments}
We thank the Fermilab staff and the technical staffs of the participating institutions for their vital contributions. This work was supported by the U.S. Department of Energy and National Science Foundation; the Italian Istituto Nazionale di Fisica Nucleare; the Ministry of Education, Culture, Sports, Science and Technology of Japan; the Natural Sciences and Engineering Research Council of Canada; the National Science Council of the Republic of China; the Swiss National Science Foundation; the A.P. Sloan Foundation; the Bundesministerium f\"ur Bildung und Forschung, Germany; the World Class University Program, the National Research Foundation of Korea; the Science and Technology Facilities Council and the Royal Society, UK; the Institut National de Physique Nucleaire et Physique des Particules/CNRS; the Russian Foundation for Basic Research; the Ministerio de Ciencia e Innovaci\'{o}n, and Programa Consolider-Ingenio 2010, Spain; the Slovak R\&D Agency; and the Academy of Finland. 

\end{acknowledgments}


\begin{thebibliography}{99}
%%%%%%%%        TTBAR theoretical REFERENCES      %%%%%%%%%%%%%%%%

%\bibitem{Kidonakis:2008mu}
%  N.~Kidonakis and R.~Vogt,
%  Phys.\ Rev.\  D {\bf 78} (2008) 074005
%  arXiv:hep-ex/0805.3844.

%\bibitem{Moch:2008ai}
%  S.~Moch and P.~Uwer,
%  Nucl.\ Phys.\ Proc.\ Suppl.\  {\bf 183} (2008) 75
%  arXiv:hep-ex/0807.2794.

\bibitem{cacciari-2004-0404}
  M.~Cacciari, S.~Frixione, M.~L.~Mangano, P.~Nason, and G.~Ridolfi,
  J.\ High Energy Phys. 04 (2004) 068;  
  arXiv:hep-ph/0303085v1.
  
%%%%%%%%%%%%%    CKM  %%%%%%%%%%

\bibitem{Cabibbo}
  N.~Cabibbo,
  %``Unitary Symmetry and Leptonic Decays,''
  Phys.\ Rev.\ Lett.\  {\bf 10} (1963) 531.
  
%\cite{Kobayashi:1973fv}
\bibitem{KM}
  M.~Kobayashi and T.~Maskawa,
  %``CP Violation In The Renormalizable Theory Of Weak Interaction,''
  Prog.\ Theor.\ Phys.\  {\bf 49}, 652 (1973).

%%%%%%%%  MOTIVATIONS %%%%%%

\bibitem{Alwall:2006bx}
  J.~Alwall {\it et al.},
  %``Is V(tb) ~= 1?,''
  Eur.\ Phys.\ J.\  C {\bf 49} (2007) 791;
  arXiv:hep-ph/0607115.

\bibitem{Tait:2000sh}
  T.~M.~P.~Tait and C.~P.~P.~Yuan,
  %``Single top quark production as a window to physics beyond the Standard
  %Model,''
  Phys.\ Rev.\  D {\bf 63}, 014018 (2001).  

%%%%%%%%%% SINGLE TOP THEORETICAL %%%%%%%%%%%

\bibitem{singletops} %Sullivan:2004ie
Z.~Sullivan, Phys.\ Rev.\ D {\bf 70}, 114012 (2004).

\bibitem{singletopt}
B.W.~Harris {\it et al.}, Phys.\ Rev.\ D {\bf 66}, 054024 (2002).

\bibitem{Campbell:2009ss}
  J.~M.~Campbell, R.~Frederix, F.~Maltoni, and F.~Tramontano,
  %``$t^-$ channel single top production at hadron colliders,''
  arXiv:hep-ex/0903.0005.

%%%%%%%%%%     SINGLE TOP EXPERIMENTAL REFERENCES   %%%%%%%%%%%%%%

%D0 evidence PRL
\bibitem{Abazov:2006gd}
  V.~M.~Abazov {\it et al.}  (D0 Collaboration),
  Phys.\ Rev.\ Lett.\  {\bf 98} (2007) 181802;
  arXiv:hep-ex/0612052.

%D0 evidence PRD
\bibitem{Abazov:2008kt}
  V.~M.~Abazov {\it et al.}  (D0 Collaboration),
  Phys.\ Rev.\  D {\bf 78} (2008) 012005;
  arXiv:hep-ex/0803.0739.

%CDF evidence
\bibitem{Aaltonen:2008sy}
  T.~Aaltonen {\it et al.}  (CDF Collaboration),
  Phys.\ Rev.\ Lett.\  {\bf 101}, 252001 (2008);
  arXiv:hep-ex/0809.2581.
 
%D0 observation
\bibitem{Abazov:2009ii}
  V.~M.~Abazov {\it et al.}  (D0 Collaboration),
  %``Observation of Single Top-Quark Production,''
  Phys.\ Rev.\ Lett.\  {\bf 103} (2009) 092001;
  arXiv:hep-ex/0903.0850.
  %%CITATION = PRLTA,103,092001;%%
  
 %CDF observation PRL
\bibitem{Aaltonen:2009jj}
  T.~Aaltonen {\it et al.}  (CDF Collaboration),
  %``First Observation of Electroweak Single Top Quark Production,''
  Phys.\ Rev.\ Lett.\  {\bf 103} (2009) 092002;
  arXiv:hep-ex/0903.0885.
  %%CITATION = PRLTA,103,092002;%%

%%CDF observation PRD
%\bibitem{PRDobs}
%A.~Aaltonen, {\em et. al.}
%Phys.\ Rev. D.\ {\bf ZZZ}, CDF Observation PRD (2009).

%%%%%%%%         HIGGS PAPER                 %%%%%%%%%%%%

\bibitem{HiggsPRL} 
T.~Aaltonen {\it et al.} (CDF collaboration), submitted to Phys. Rev. Lett.
arXiv:0911.3935;

%%%%% MET

%\bibitem{met} The missing transverse energy $\vec{\met}$ is calculated as the vector sum of the energy in each calorimeter tower multiplied by a unit vector in the azimuthal direction of the tower. If isolated high momentum muons are found in the event, the $\vec{\met}$ is corrected by subtracting the muon energy in the calorimeter and adding the muon \Pt\ to the vector sum.  $\met$ is defined as the magnitude of  $\vec{\met}$.

%%%%%%%%%%% DETECTOR CITATIONS   %%%%%%%%%%%

\bibitem{CDFdetector} D.~Acosta {\it et al.} (CDF Collaboration), Phys. Rev. D {\bf 71}, 032001 (2005).

\bibitem{coordinate} We use a cylindrical coordinate system where $\theta$ is the polar angle to the proton beam direction at the event vertex, $\phi$ is the azimuthal angle about the beam axis, and pseudorapidity is defined $\eta = - \ln \tan(\theta/2)$. We define transverse energy as $\Et = E \sin\theta$ and transverse momentum as $p_T = p\sin\theta$ where $E$ is the energy measured in the calorimeter and $p$ is the magnitude of the momentum measured by the spectrometer.

\bibitem{L00} C.~S.~Hill (CDF Collaboration), Nucl.\ Instrum.\ Methods A {\bf 530}, 1 (2004).

\bibitem{SVXII} A.~Sill (CDF Collaboration), Nucl.\ Instrum.\ Methods A {\bf 447}, 1 (2000). 

\bibitem{ISL} A.~A.~Affolder et al. (CDF Collaboration), Nucl.\ Instrum.\ Methods A {\bf 453}, 84 (2000). 

\bibitem{COT} T.~Affolder {\it et al.} (CDF Collaboration), Nucl.\ Instrum.\ Methods A {\bf 526}, 249 (2004).

\bibitem{ecal} L.~Balka {\it et al.}, Nucl. Instrum. and Methods A {\bf 267}, 272 (1988).

\bibitem{pem} M.~Albrow {\it et al.}, Nucl. Instrum. and Methods A {\bf 480}, 524 (2002).

\bibitem{cha} S.~Bertolucci {\it et al.}, Nucl. Instrum. and Methods A {\bf 267}, 301 (1988).

\bibitem{TDR} R.~Blair {\it et al.}, {\it The CDF-II detector: Technical Design Report}, FERMILAB-PUB-96-390-E.

%\bibitem{jets} F.~Abe {\it et al.} (CDF Collaboration), Phys. Rev. D {\bf 45}, 1448 (1992).

\bibitem{CMU} G.~Ascoli {\it et al.}, Nucl. Instrum. Methods A {\bf 268}, 33 (1988).

\bibitem{CMX} A.~Abulencia {\it et al.} (CDF Collaboration), J. Phys. G {\bf 34},  2457 (2007).

\bibitem{WZprl} F.~Abe {\it el al.} (CDF Collaboration), Phys. Rev. Lett. {\bf 94}, 091803 (2005).


%%%%%%%%%%%%%%%%%%%%%%%%%%%%%%%%%%%%%%%%%%%%%%%%%%%%%%%%%%

\bibitem{artur}
  A.~Apresyan, Ph.~D.~Thesis, Purdue University, 2009,  FERMILAB-THESIS-2009-09.

%%%%%%%   JETS      %%%%%%%%%%%%%%%%%%%%%%%%%%%%%%%%%%%%%

\bibitem{Bhatti:2005ai}
  A.~Bhatti {\it et al.},
  Nucl.\ Instrum.\ Methods A {\bf 566} (2006) 375;
  arXiv:hep-ex/0510047.

\bibitem{H1} C. Adloff {\it et al.} (H1 collaboration), Z. Phys. C {\bf 74} (1997) 221.

%%%%%%%%%    B-TAGGING          %%%%%%%%%%%%%%%%%%%%%

\bibitem{secvtx}
D. Acosta {\it et al.} (CDF Collaboration),
Phys.\ Rev.\ D {\bf 71}, 052003 (2005).

\bibitem{jetprob}
A. Abulencia {\it et al.} (CDF Collaboration), Phys.\ Rev.\ D {\bf 74}, 072006 (2006)

%%%%%%       MC GENERATORS           %%%%%%%%%%

\bibitem{madevent} %
J.~Alwall {\it et al.}, J.\ High Energy Phys. 09 (2007) 028.

\bibitem{Lai:1999wy}
H.L.~Lai {\em et~al.}, Eur. Phys. J. C {\bf 12}, 375 (2000);
arXiv:hep-ph/9903282. 

\bibitem{PYTHIA} T.~Sjostrand {\it et al.}, Comput. Phys. Commun. {\bf 135}, 238 (2001).

\bibitem{Boos:2006af}
E.~E.~Boos {\it et al.},
% ``Method for simulating electroweak top-quark production events in the NLO approximation: SingleTop event generator",
  Phys.\ Atom.\ Nucl.\  {\bf 69} (2006) 1317
  [Yad.\ Fiz.\  {\bf 69} (2006) 1352].
  %%CITATION = YAFIA,69,1352;%%

\bibitem{DGLAP}
V.N. Gribov {\it et al.}, Sov.~J.~Nucl.~Phys. {\bf 15}, 438 (1972).
G. Altarelli {\it et al.}, Nucl.~Phys. {\bf B126}, 298 (1977).
Yu.~L.~Dokshitzer, Sov.~Phys.~JETP {\bf 46}, 641 (1977).

\bibitem{Jan}
J.~Lueck, M.~S.~thesis, Karlsruhe University, 2009, FERMILAB-MASTERS-2006-01.

%%%%%%%%%%%%%%%%%%%%%%%%%%%%%%%%%%%%%%%%%%%%%%%%%%%%%%%%

\bibitem{Abulencia:2005ix} A.~Abulencia {\it et al.} (CDF Collaboration), J. Phys. G {\bf 34}, 2457 (2007);
arXiv:hep-ex/0508029. 

\bibitem{abulencia:032008} A.~Abulencia {\it et al.} (CDF Collaboration), Phys. Rev. D {\bf 74}, 032008 (2006),
D.~Acosta {\it et al.} (CDF Collaboration), Phys.\ Rev.\ Lett.\ {\bf 94}, 091803 (2005).

\bibitem{MCFM}
J.~M.~Campbell and R.~K.~Ellis, Phys.\ Rev.\ D {\bf 62}, 114012 (2000).

\bibitem{DIBOSONS}
J.~M.~Campbell and R.~K.~Ellis, Phys.\ Rev.\ D {\bf 60}, 113006 (1999).

%%%%%%        ROOT REFERENCES      %%%%%%%%%%

\bibitem{mlp}
K.~Hornik {\em et al.}, Multilayer Feedforward Networks are Universal Approximators, Neural Networks, Vol. 2, pp 359-366 (1989), 
D.~W.~Ruck {\em et al.}, The Multilayer Perceptron as an Approximation to a Bayes Optimal Discriminant Function, IEEE Transactions on Neural Networks, Vol. 1, nr 4, pp296-298 (1990).

\bibitem{TMVA}
  A.~Hocker {\it et al.},
  PoS A {\bf CAT}, 040 (2007);
  arXiv:physics/0703039.

\bibitem{root}
  R.~Brun and F.~Rademakers,
  Nucl.\ Instrum.\ Methods  A {\bf 389} (1997).

\bibitem{sphericity}
%A momentum tensor is defined as $M_{ij} = \frac{\sum_o p^o_i p^o_j}{\sum_o | \vec{p^o}|}$, where $\vec{p^o}$ is the momentum of a reconstructed object $o$, and $i$ and $j$ are Cartesian coordinates. The sum over $o$ includes up to three jets of leading $p_T$. The sphericity in an event is defined as $S = \frac{3}{2} (\lambda_2 + \lambda_3)$, where $\lambda_2$ and $\lambda_3$ are the smallest two eigenvalues of the normalized momentum tensor.
A momentum tensor is defined as $M_{lm} = \frac{\sum_o j^o_l j^o_m}{\sum_o | \vec{j^o}|}$, where $\vec{j^o}$ is the momentum of a reconstructed jet, and $l$ and $m$ are Cartesian coordinates. The index $o$ runs over the number of jets in the event. The sphericity in an event is defined as $S = \frac{3}{2} (\lambda_2 + \lambda_3)$, where $\lambda_2$ and $\lambda_3$ are the smallest two eigenvalues of the normalized momentum tensor.

\bibitem{CLC}
D.~Acosta {\it et al.}, 
 Nucl.\ Instrum.\ Methods\ A {\bf 494}, 57 (2002).

% for right order
\bibitem{Sjo85} 
T.~Sjostrand, Phys. Lett. B {\bf 157}, 321 (1985); 
M.~Bengtsson {\it et al.}, Z. Phys. C {\bf 32}, 67 (1986).

%%%%%%%%%    SYSTEMATICS        %%%%%%%%%%%%%%%%%%%%%

\bibitem{Martin:1998sq}
  A.~D.~Martin, R.~G.~Roberts, W.~J.~Stirling, and R.~S.~Thorne,
  Eur.\ Phys.\ J.\  C {\bf 4}, 463 (1998);
  arXiv:hep-ph/9803445.

\bibitem{Pumplin:2002vw}
  J.~Pumplin, D.~R.~Stump, J.~Huston, H.~L.~Lai, P.~M.~Nadolsky, and W.~K.~Tung,
  JHEP {\bf 0207}, 012 (2002);
  arXiv:hep-ph/0201195.

\bibitem{:2009ec}
  Tevatron Electroweak Working Group and CDF and D0 Collaborations,
  arXiv:hep-ex/0903.2503.

\bibitem{Amsler:2008zzb}  C.~Amsler {\it et al.}  (Particle Data Group),
  Phys.\ Lett.\  B {\bf 667} (2008).

\bibitem{Neyman}
J.~Neyman and E.~Pearson, Phil.\,Trans. of the Royal Soc.\, of London A {\bf 31}, 289 (1933).

\bibitem{Bestvtb} 
  Tevatron Electroweak Working Group and CDF and D0 Collaboration,
  arXiv:hep-ex/0908.2171.

%%%%%%%%%%%%%   CROSS-SECTIONS %%%%%%%%%%

%\bibitem{CDF7780:ZHF}
%A.~Mehta and B.~Heinemann,  
%%``Measurement of the b jet cross section for events with a $Z^0$ bosonÓ, 
%CDF/ANAL/JET/CDFR/7780, 2006 

%\bibitem{metjetPRL} D.~Acosta {\it et al.} (CDF Collaboration), Phys. Rev. Lett {\bf96}, 202002 (2006). 

%\bibitem{BFGS} S. McLoone {\it et al.}, IEEE Trans. Neural Networks {\bf 9} (1998) 669

%\bibitem{Herwig} G.~Marchesini {\it et al.}, Comput. Phys. Commun. {\bf 67}, 465 (1992); G.~Corcella {\it et al.}, J. High Energy. Phys. {\bf 0101}, 010 (2001).
%21

%\bibitem{Sullivan:2004ie}
%Z.~Sullivan,
%\newblock Phys. Rev. {\bf D70}, 114012 (2004), hep-ph/0408049.
%%%CITATION = HEP-PH/0408049;%%

%\bibitem{limits}
%J.~Heinrich, C.~Blocker, J.~Conway, L.~Demortier, L.~Lyons, G.~Punzi, and P.K.~Sinervo
%[arXiv:physics/0409129v1] [physics.data-an].

\end{thebibliography}
\end{document}